\documentclass[
reprint,
superscriptaddress,
amsmath,amssymb,
aps,
prb,
]{revtex4-2}

\usepackage{graphicx}
\usepackage{dcolumn}
\usepackage{bm}
\usepackage{xcolor}
\usepackage{hyperref}

\begin{document}

\title{Thermoelectric transport of strained CsK$_2$Sb: The role of electron velocities and scattering within extended Fermi surfaces
}

\author{Øven A. Grimenes}
\email{oven.andreas.grimenes@nmbu.no}
\affiliation{
 Department of Mechanical Engineering and Technology Management, \\ Norwegian University of Life Sciences, NO-1432 Ås, Norway}

\author{G. Jeffrey Snyder}
\affiliation{Materials Science and Engineering, Northwestern University, Evanston, Illinois 60208, USA}

\author{Ole M. Løvvik}
\affiliation{SINTEF Sustainable Energy Technology, Forskningsveien 1, NO-0314 Oslo, Norway}

\author{Kristian Berland}
\email{kristian.berland@nmbu.no}
\affiliation{
 Department of Mechanical Engineering and Technology Management, \\ Norwegian University of Life Sciences, NO-1432 Ås, Norway}

\date{\today}

\begin{abstract}
\noindent 
In this first-principles study, we investigated the thermoelectric properties of the full-Heusler compound CsK$_2$Sb at different compressive strains. This material exhibits a valence band structure with significant effective mass anisotropy, forming tube-like energy isosurfaces below the band edge, akin to that of two-dimensional (2D) systems. Such systems can have a large number of high-mobility charge carriers and a beneficial density of states profile. In the calculations, we predicted a maximum p-type figure of merit ($zT$) of 2.6 at 800\;K, in line with previous predictions of high $zT$. This high $zT$ arises from the low lattice thermal conductivity of 0.35 Wm$^{-1}$K$^{-1}$ and the beneficial electronic band structure. The high density of states significantly increased the electron-scattering space, but this effect was largely compensated by reduced scattering rates of electrons with large momentum ${\mathbf{q}}$. We further explored the effect of enhancing the low-dimensionality through compressive strain. This increased the p-type power factor by up to 66\%; partly due to more strongly pronounced 2D features of the valence band, but primarily due to increased Fermi velocities. However, compressive strain also increased phonon velocities and hence the lattice thermal conductivity. The maximum p-type $zT$ thus only increased slightly, to 2.7 at 1\%\ compressive strain. In the conduction band, strain aligned the $\Gamma$- and X-centered valleys, resulting in the optimal n-type $zT$ increasing from 0.9 to 2.3 at 2\%\ compressive strain. Thus, highly strained CsK$_2$Sb has the potential for both good p- and n-type thermoelectricity. 
\end{abstract}

\maketitle

\section{Introduction}
\label{sec: Introduction}

The ability to convert heat to electricity and vice versa makes thermoelectric (TE) materials useful for energy harvesting and distributed cooling \cite{Snyder_2008,Vining_2009}. With higher efficiency, they could see more widespread use, such as for waste-heat recovery, geothermal systems, or combined electricity and heat generators\cite{Schwab_2022,Bell_2008}, which could be key in the ongoing green-energy transition. Enhancing the TE efficiency is, however, highly nontrivial, as expressed by the TE figure of merit of a material, $zT = S^2\sigma T/(\kappa_\mathrm{e} +\kappa_\ell)$. Here, $S$ is the Seebeck coefficient, $\sigma$ the electrical conductivity, $T$ the temperature, and $\kappa_\mathrm{e}$ and $\kappa_\ell$ the electron and lattice thermal conductivity, respectively. These parameters are highly interdependent; increasing $\sigma$ typically decreases $S$ and increases $\kappa_\mathrm{e}$. Reducing $\kappa_\ell$ through increased phonon scattering can also inadvertently reduce $\sigma$.

A wide range of strategies to reach higher $zT$ have been proposed and investigated. Starting from promising materials discovered through experimental and computational studies, $\kappa_\ell$ can be reduced by isovalent alloying that selectively scatters phonons more than electrons\cite{Tranas_2022}, or through nanostructuring that impedes phonons with long mean free path (MFP) compared to short-MFP electrons\cite{Minnich_2009}. Electronic transport properties can be enhanced by increasing the number of degenerate conduction or valence band valleys through strain\cite{Zhang_2016,Li_2018} and alloying\cite{Pei_2011,Zhu_2022}. Resonance states\cite{Bilc_2004,Heremans_2008,Heremans_2012} and interfacial effects described as energy filtering \cite{Faleev_2008,Gayner_2020, Lin_2020} are other possibilities. Low-dimensional materials have also attracted interest as they can exhibit an increased density of states (DOS) near the Fermi level, greatly enhancing $S$\cite{Hicks_1993,Hicks_1993a} while also reducing $\kappa_\ell$. However, in systems where multiple sub-bands are present, low-dimensionality has been shown to reduce the PF \cite{Cornett_2011,Cornett_2011a}. The practical realization of low-dimensional band structures has typically been associated with low-dimensional materials such as superlattice structures, which require complex processing methods such as molecular beam epitaxy\cite{Yang_2018}. Interestingly, several recently identified materials possess band structures with features similar to those of low-dimensional materials, despite being bulk 3D isotropic materials. Such materials can have band structures with nearly flat energy dispersions in one or more directions similar to those found in low-dimensional quantum wells. A band that is flat in one direction will exhibit Fermi surfaces (charge carrier energy isosurfaces) consisting of tubes (2D quantum well), as seen in Figure \ref{fig: Fermi}, while a band flat in two directions will form sheets (1D quantum well) as opposed to spheres that arise in traditional 3D parabolic band structures \cite{Parker_2013,Bilc_2015,Dylla_2019,Park_2021,Brod_2021}.

Computational transport calculations are nowadays a key tool in the quest for materials with high $zT$, both in the optimization of individual material properties, but also for identifying novel promising materials. These calculations are typically based on the semi-classical Boltzmann transport equation, with electronic properties obtained from density functional theory (DFT). A key input in these calculations is the electronic relaxation time $\tau$. Computing this relaxation time, however, is complex and the constant relaxation time approximation (CRTA) therefore remains widely used, particularly in screening studies \cite{Madsen_2006,Wang_2011,Bhattacharya_2015,Ricci_2017,Berland_2019,Berland_2021}.

For complex band structures like those with extended energy isosurfaces, high $zT$ predictions are often attributed to simultaneously obtaining a high DOS and high group velocities\cite{Bilc_2015,Park_2019}. However, earlier computational studies have usually been based on the CRTA or energy-dependent approximation for $\tau = \tau(E)$, the latter counteracting the beneficial effect of high DOS\cite{Parker_2013,Bilc_2015,Brod_2021}. Recent studies have highlighted the importance of fully accounting for the scattering, with findings such as significantly reduced inter-valley compared to intra-valley scattering\cite{Park_2021a,Li_2024}. Only a handful of studies have discussed the effect of low-dimensional extended band energy isosurfaces close to the Fermi level on the electronic transport properties beyond the CRTA or other simple approximations\cite{Park_2019,Ji_2022,Park_2021}. Thus, it is critical to establish the role of electron scattering in materials with such extended energy isosurfaces. 

Among the A$_3$Sb group of materials, where A is an alkali metal, several exhibit low-dimensional band structures\cite{Ettema_2000,Kalarasse_2010,Liu_2024a}. One such material is the full-Heusler CsK$_2$Sb (sometimes denoted K$_2$CsSb), which is a known photocathode\cite{Sommer_1963,Ghosh_1978}. So far, four DFT-based TE studies beyond the CRTA have been published for this compound\cite{Yue_2022,Yuan_2022,Singh_2022,Sharma_2023}. While they all reported ultralow values of $\kappa_\ell$, their electron transport results were somewhat contradictory. Sharma et al.\cite{Sharma_2023} predicted the material to be superior as n-type TE, while Yue et al.\cite{Yue_2022}  and Yuan et al.\cite{Yuan_2022} found stronger p-type performance.

In this study, we computed the transport properties of CsK$_2$Sb and performed an in-depth analysis of the electron scattering of the compound. In particular, to accentuate the extended energy isosurfaces, we investigated the effect of compressive strain on the electronic and lattice TE properties. Up to 5\%\ compressive strain was considered, which is not necessarily representative of what can be engineered in a typical device, though quite extreme strains can be achieved with a diamond anvil. However, this range provided a rough representation of chemical pressure, i.e., alloying with smaller elements. This allowed an investigation of the interplay between the electronic band structure, the electronic scattering, and $\kappa_\ell$.

\section{Methods}
\label{sec: Methods}

\subsection{Electron transport}
\label{subsec: Electron transport theory}

\begin{figure}
    \centering
    \includegraphics[width=\linewidth]{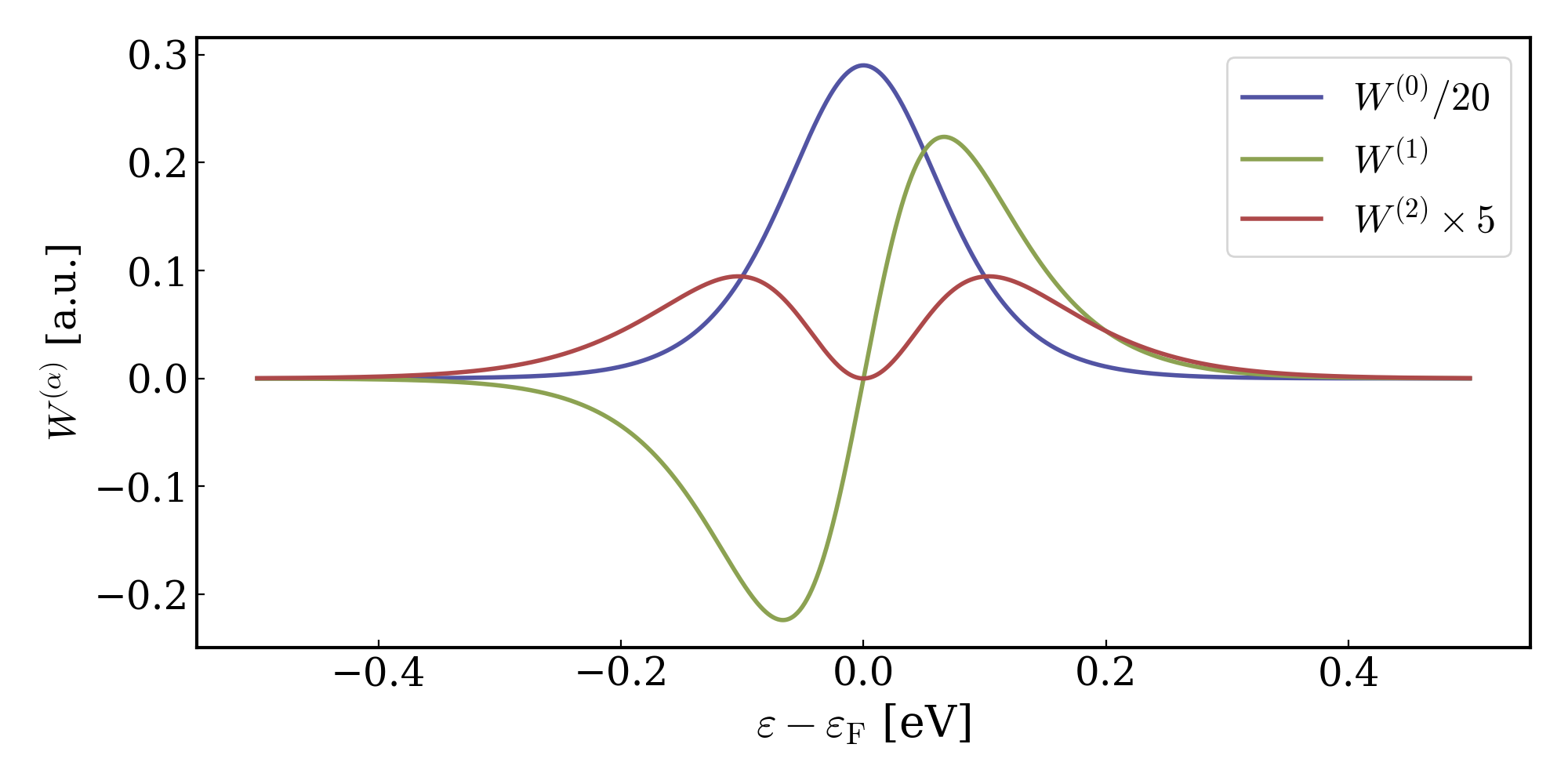}
    \caption{Selection functions for $\alpha = 0,1,2$ at T = 500\;K. $W^{(0)}$ and $W^{(2)}$ are scaled down and up, respectively, for easier comparison.
    \label{fig: selc_func}}
\end{figure}

Electron transport properties were calculated with the \textsc{AMSET} code\cite{Ganose_2021}, based upon \textsc{BoltzTraP2}\cite{Madsen_2018}, which solves the Boltzmann transport equation in the relaxation time approximation. At this level of theory, the transport properties depend on the transport distribution function, as follows
\begin{equation}
    \label{eq: transport_distribution}
    \Sigma(\varepsilon) = \int_\mathrm{BZ} 
    \frac{\mathrm{d}\mathbf{k}}{8\pi^3}
    \sum_n \mathbf{v}_{n\mathbf{k}} \otimes \mathbf{v}_{n\mathbf{k}} 
    \tau_{n\mathbf{k}} \delta(\varepsilon - \varepsilon_{n\mathbf{k}}) \,,
\end{equation}
where $\delta(\varepsilon - \varepsilon_{n\mathbf{k}})$ is the Dirac delta function, $\mathbf{v}_{n\mathbf{k}}$ is the electron group velocity, and $\tau_{n\mathbf{k}}$ is the relaxation time for a given band $n$ and wave vector $\mathbf{k}$. $\varepsilon$ indicates carrier energies. For electron transport, $\Sigma(\varepsilon)$ is weighted by the (Fermi) selection functions,
\begin{equation}
    \label{eq: sel_func}
    W^{(\alpha)}(\varepsilon) = (\varepsilon - \varepsilon_\mathrm{F})^\alpha\left(\frac{-\partial f(\varepsilon)}{\partial \varepsilon}\right)\,,
\end{equation}
where $\varepsilon_\mathrm{F}$ is the chemical potential, and $f$ is the equilibrium Fermi-Dirac distribution function around $\varepsilon_\mathrm{F}$ at a given temperature $T$. The selection functions for $\alpha = 0,1,2$ are shown in Fig.\ \ref{fig: selc_func}. The thermoelectric properties can be obtained from the generalized transport coefficients 
\begin{equation}
    \label{eq: general_transport_coeff}
    \mathcal{L}^{(\alpha)}(\varepsilon_\mathrm{F}) = q^2 \int_{-\infty}^\infty \Sigma(\varepsilon) W^{(\alpha)}(\varepsilon) \mathrm{d}\varepsilon \,,
\end{equation}
where $q$ is the elementary charge. In turn, $\sigma$, $S$, and $\kappa_\mathrm{e}$ are given as
\begin{align}
    &\sigma = \mathcal{L}^{(0)} \,, \label{eq: conductivity}   \\
    &S = \frac{\mathcal{L}^{(1)}}{q T \mathcal{L}^{(0)}} \,, \label{eq: Seebeck} \\
    &\kappa_\mathrm{e} = \frac{1}{q^2 T}\left[\mathcal{L}^{(2)} - \frac{(\mathcal{L}^{(1)})^2}{\mathcal{L}^{(0)}}\right]. \label{eq: kappa_e}
\end{align}

The electronic relaxation time is given by the inverse of the total scattering rate $\tau_{n\mathbf{k}} = 1/\Gamma_{n\mathbf{k}}^{\mathrm{Tot}}$ which is the sum of the acoustic deformation potential (ADP), polar optical phonon (POP), ionized impurity (IMP), and piezoelectric (PIE) scattering rates\cite{Ganose_2021}:
\begin{equation}
    \Gamma^\mathrm{Tot}_{n\mathbf{k}} = 
    \Gamma^\mathrm{ADP}_{n\mathbf{k}} + 
    \Gamma^\mathrm{PIE}_{n\mathbf{k}} +
    \Gamma^\mathrm{POP}_{n\mathbf{k}} + 
    \Gamma^\mathrm{IMP}_{n\mathbf{k}}.
\end{equation}
Expression for matrix elements of each scattering mechanism can be found in Appendix \ref{app: a} and further details about the methodology can be found in Ref.~\cite{Ganose_2021}.

\subsection{Computational details}
\label{subsec: Computational details}

DFT as implemented in \textsc{VASP}\cite{Kresse_1993,Kresse_1999} version 6.3.2
was used to compute input properties for the \textsc{AMSET} transport calculations. The vdW-DF-cx\cite{Berland_2014,Berland_2014a} functional was used due to its generally accurate bulk modulus and lattice constants of solids\cite{Berland_2014,Berland_2014a,Bjorkman_2014,Tran_2019}, with the HSE06\cite{Krukau_2006} hybrid functional used as a reference. The plane wave energy cutoff was set to 520\;eV and the number of valence electrons for the basis set was 9 (Cs), 9 (K), and 5 (Sb); corresponding to the standard Materials Project (MP) settings\cite{Jain_2013}. The initial crystal structure was retrieved from the MP database (mp-581024) and relaxed on a 10$\times$10$\times$10 k-points grid until forces fell below 5.0$\times 10^{-5}$\;eV/Å. A second ionic relaxation was subsequently performed to remove Pulay stress. Density functional perturbation theory (DFPT) calculations\cite{Gajdos_2006} were used to compute the static and high-frequency dielectric constants $\epsilon_\mathrm{s}$, $\epsilon_\infty$, Born effective charges, and $\Gamma$-point phonon frequencies $\omega_{\Gamma\nu}$. The elastic tensor $C_{ij}$ was calculated using finite differences\cite{LePage_2002}. For both of these calculations, a 12$\times$12$\times$12 k-point grid was used. For each strain, the relevant properties were computed at the equivalent external pressure (separately determined with DFT). A 20$\times$20$\times$20 k-point grid interpolated to a 53$\times$53$\times$53 grid in \textsc{AMSET} was used in the transport calculations, while the deformation potential was extracted using a 12$\times$12$\times$12 \textbf{k}-points grid.

The lattice thermal conductivity $\kappa_{\ell}$ was calculated using the stochastic temperature-dependent effective potential method (sTDEP)\cite{Shulumba_2017} as implemented in the TDEP code\cite{Hellman_2011,Knoop_2024}. Forces were calculated with \textsc{VASP} at the same level of theory and the same numerical settings as the electron transport calculations, but at a lower k-point density (2$\times$2$\times$2) for the supercell configuration calculations. Stochastic sampling of atomistic configurations was based on the Debye model and was used as the initial input to an iterative self-consistent calculation of force constants. Second- and third-order force constants were in each step extracted from displacement-force data from DFT. The force constants generated a new generation of configurations for new force calculations. The criterion for self-consistent force constants was a difference in the phonon free energy of less than 1 meV/atom between two consecutive steps. The configurations consisted of 152 atoms in non-diagonal supercells with a spatial filling ratio of 99.5{\%} of the ideal cube. The cutoff radius for extracting force constants was set to half of the supercell lattice constant. The $\mathbf{q}$-point grid used to calculate $\kappa_{\ell}$ was varied between 15$\times$15$\times$15 and 25$\times$25$\times$25 and extrapolated to infinity.

\section{Results}
\label{sec: Results}

\subsection{Materials properties}
\label{subsec: Materials properties}

The lattice constant of CsK$_2$Sb was relaxed to 8.57\;Å with DFT (at $T=0$\;K). With thermodynamic data from TDEP, we predicted the room temperature (300\;K) lattice constant to be 8.623\;Å, in good agreement with the experimental value of  8.615\;Å\cite{McCarroll_1965}. Table \ref{tab: Mat_prop} lists various materials properties obtained at different compressive strains, which were used to compute $\tau_{n \mathbf{k}}$. The unstrained material has rather low elastic coefficients, but they increase significantly under compressive strain as the interatomic distances shorten. Similarly, the effective $\omega_{\mathrm{po}}$ increases substantially with strain and reduces the ionic part of the dielectric tensor. This can be understood from the shortened interatomic distances, which make the lattice stiffer and the force constants larger; both these effects increase the (effective) phonon frequencies and hamper the ions' reaction to an electric field. The ionic part of the dielectric tensor is thus reduced, with a corresponding reduction in $\epsilon_\mathrm{s}$. The deformation potential for selected $\mathbf{k}$-points is given in Supplemental Material (SM)\cite{SM}.

\begin{table}[h!]
  \caption{Equivalent external pressure, elastic constants, dielectric tensors, and average optical phonon frequency of CsK$_2$Sb at different amounts of strain.}
\label{tab: Mat_prop}
\begin{ruledtabular}
\begin{tabular}{lllllll}

Strain             & 0\%    & 1\%        & 2\%       & 3\%    & 4\%    & 5\%\\ \hline
Pressure [kB]      & 0.0    & 4.46       & 9.70      & 15.74  & 22.69  &    30.66  \\ 
$C_{11}$ [GPa]     & 22.7   &  25.1      &  27.8     & 30.9   &  34.2  &    37.8   \\ 
$C_{12}$ [GPa]     & 9.8    &  10.9      &  12.2     & 13.7   &  15.4  &    17.2   \\ 
$C_{44}$ [GPa]     & 13.1   &  14.6      &  16.2     & 18.0   &  19.8  &    21.6    \\
$\epsilon_\mathrm{s}$ & 28.1&  26.2      &  23.2     & 19.2   &  19.0  &    16.5    \\
$\epsilon_\infty$  & 7.5    &  7.6       &  7.8      & 7.5    &  7.6   &    7.4     \\
$\omega_\mathrm{po}$ [THz]& 1.57   &  1.71      &  1.87     & 2.04   &  2.24  &    2.44    \\

\end{tabular}
\end{ruledtabular}
\end{table}

\subsection{Electronic structure}
\label{subsec: Electronic structure}

\begin{figure}[h!]
    \centering
    \includegraphics[width=\linewidth]{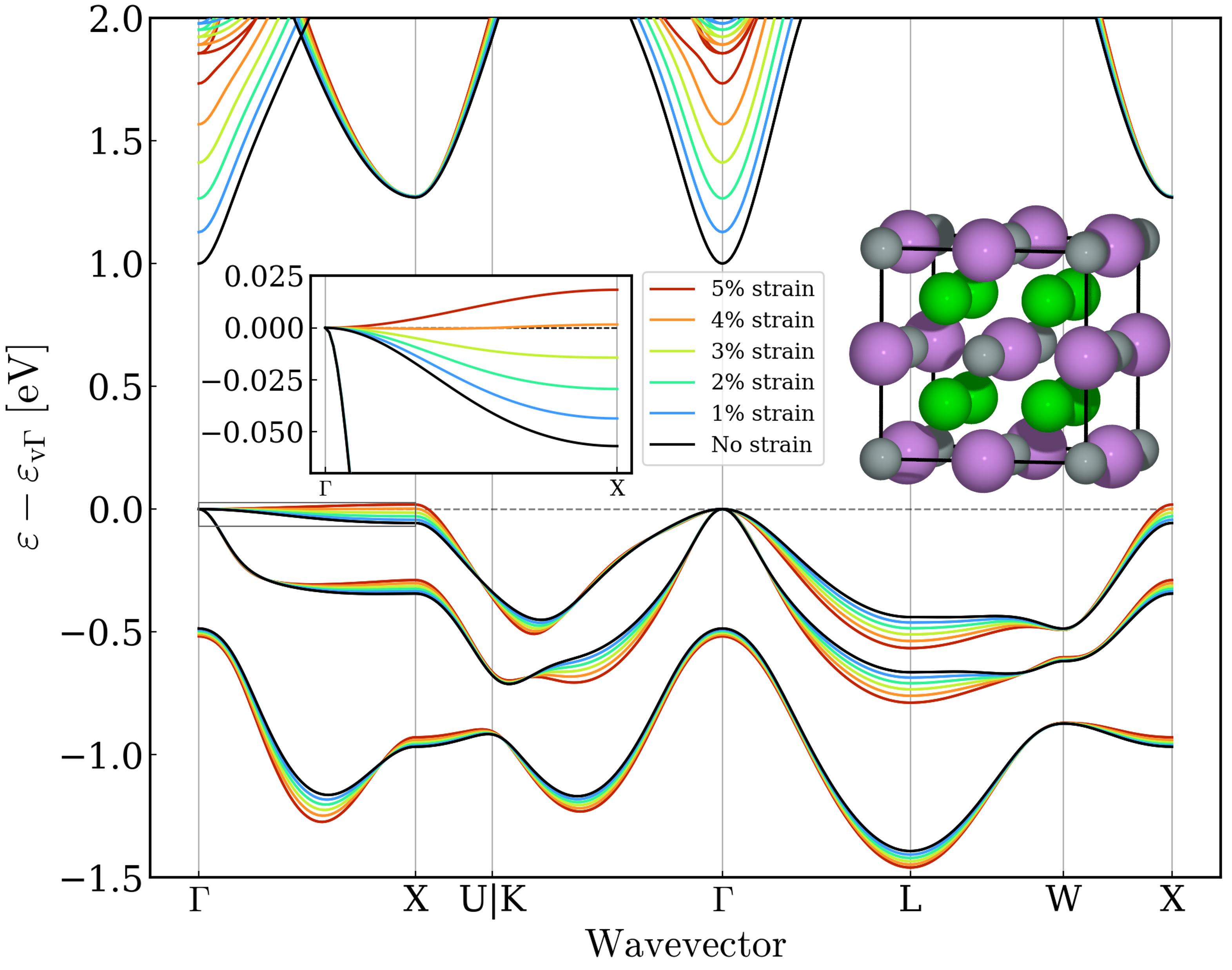}
    \caption{Electronic band structure at different amounts of compressive strain for CsK$_2$Sb calculated with the vdW-DF-cx functional. The black lines show the band structure of the equilibrium unit cell. A close-up of the energy dispersion between $\Gamma$ and X is shown in the inset. The conventional cell of the crystal structure is also shown.
    \label{fig: band_structure}}
\end{figure}
For unstrained CsK$_2$Sb, we found a band gap of 1.0\;eV with vdW-DF-cx\cite{Berland_2014} and 1.6\;eV with HSE06\cite{Krukau_2006}, in good agreement results from previous HSE06 calculations\cite{Wu_2023} and many-body perturbation theory\cite{Cocchi_2019}, but somewhat higher than results from experiments\cite{Ghosh_1978}. The HSE06 band structure is provided in SM\cite{SM} Fig. S2. Since the band structure shapes of the two functionals are very similar, we used vdW-DF-cx to obtain $\varepsilon_{n\mathbf{k}}$ on a dense $\mathbf{k}$-point grid.

\begin{figure*}[t]
    \centering
    \includegraphics[width=\linewidth]{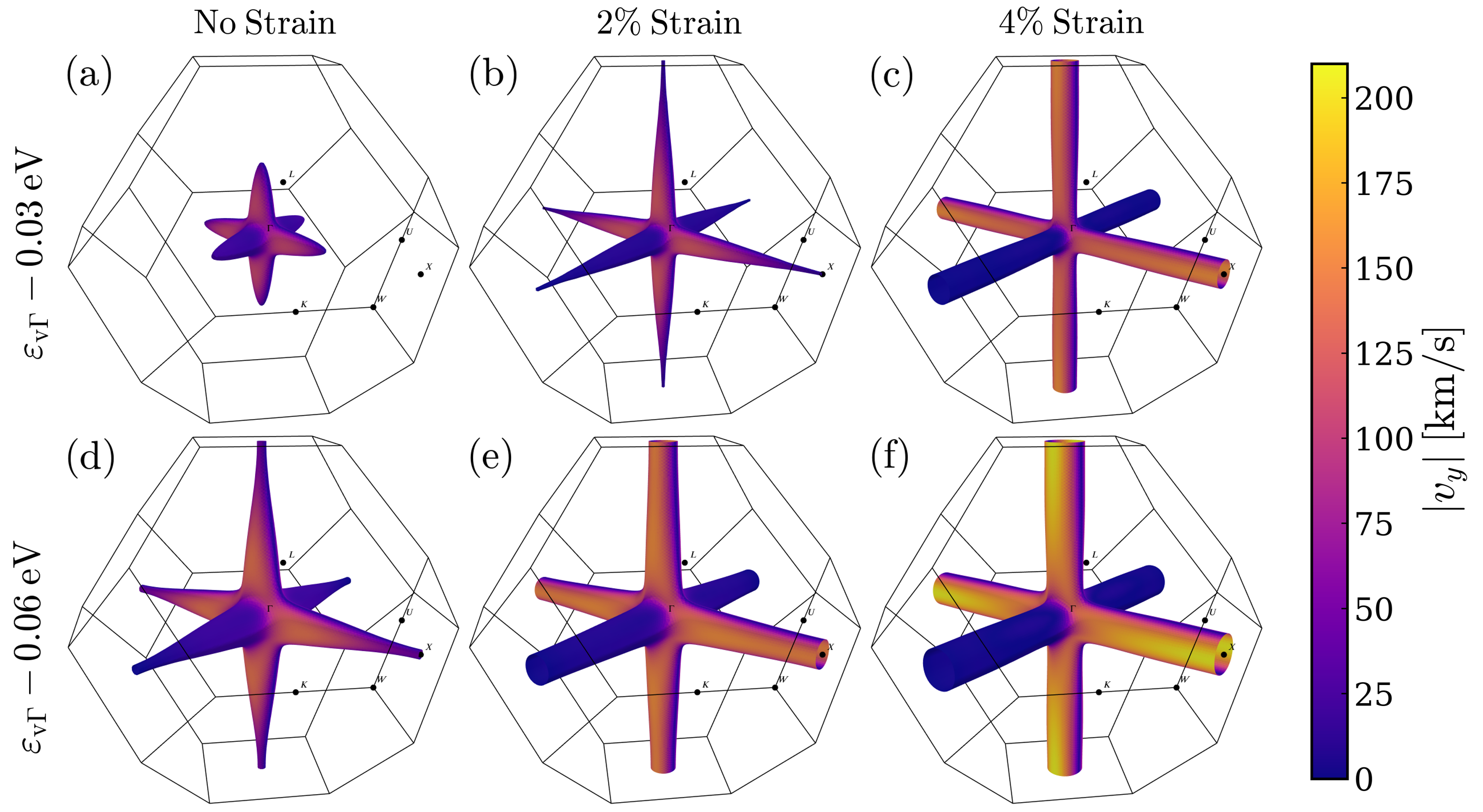}
    \caption{Energy isosurfaces of CsK$_2$Sb at 0\%, 2\%, and  4\% compressive strain. The top (bottom) row shows the surfaces at 0.03\;eV (0.06\;eV) below $\varepsilon_{\mathrm{v}\Gamma}$. Colors representing the absolute values of the $y$-component of the electron velocity are projected onto the surfaces. Surfaces were plotted with \textsc{IFermi}\cite{Ganose_2021a}.
    \label{fig: Fermi}}
\end{figure*}

Figure \ref{fig: band_structure} shows the electronic band structure at different strains. For ease of comparison, the band structures are aligned to the $\Gamma$ point valence band ($\mathrm{v}$) energy $\varepsilon_\mathrm{v\Gamma}$. Corresponding bands aligned to core levels, used for competing deformation potentials are provided in the SM\cite{SM}.

The conduction band at zero strain has a $\Gamma$-centered minimum with a dispersion similar to a Kane-band, i.e., parabolic-to-linear\cite{Kane_1957}. Under strain, the conduction band $\Gamma$ valley shifts upwards while the three-fold degenerate conduction band valley at X remains nearly constant in energy, relative to $\varepsilon_{\mathrm{v}\Gamma}$. At a strain of 2\%, the two valleys align, while for larger strains, the conduction band minimum (CBM) shifts to X.

The valence band between $\Gamma$ and X has a very flat slope. The band flattens further with increasing compressive strain, i.e., smaller unit cells, and becomes almost constant at 4\%\ strain. Beyond this point, the valence band maximum (VBM) shifts to X. In tandem with the band flattening, the curvature and hence also the electron group velocities, increases perpendicular to this line,  as seen by the valence band along the X--U, $\Gamma$--L, and X--W lines. Increasing velocities with decreasing volume can be explained by increasing wave function overlaps\cite{Ziman_1972}.

Fig.\ \ref{fig: Fermi} shows energy isosurfaces close to the VBM at different strains and energy values. The 2D-like cylinders result from the almost flat band structure along $\Gamma$--X. The flattening of the band structure at 4\%\ strain is reflected in energy isosurface almost extending to three intersecting cylinders even for very small energies below the VBM. The color shades of the isosurfaces indicate the absolute value of the $y$-component of the electron velocities $|v_y|$, showing that straining the structure not only enlarges the energy isosurfaces, but also enhances the corresponding effective velocities.

\begin{figure}[h!]
    \centering
    \includegraphics[width=\linewidth]{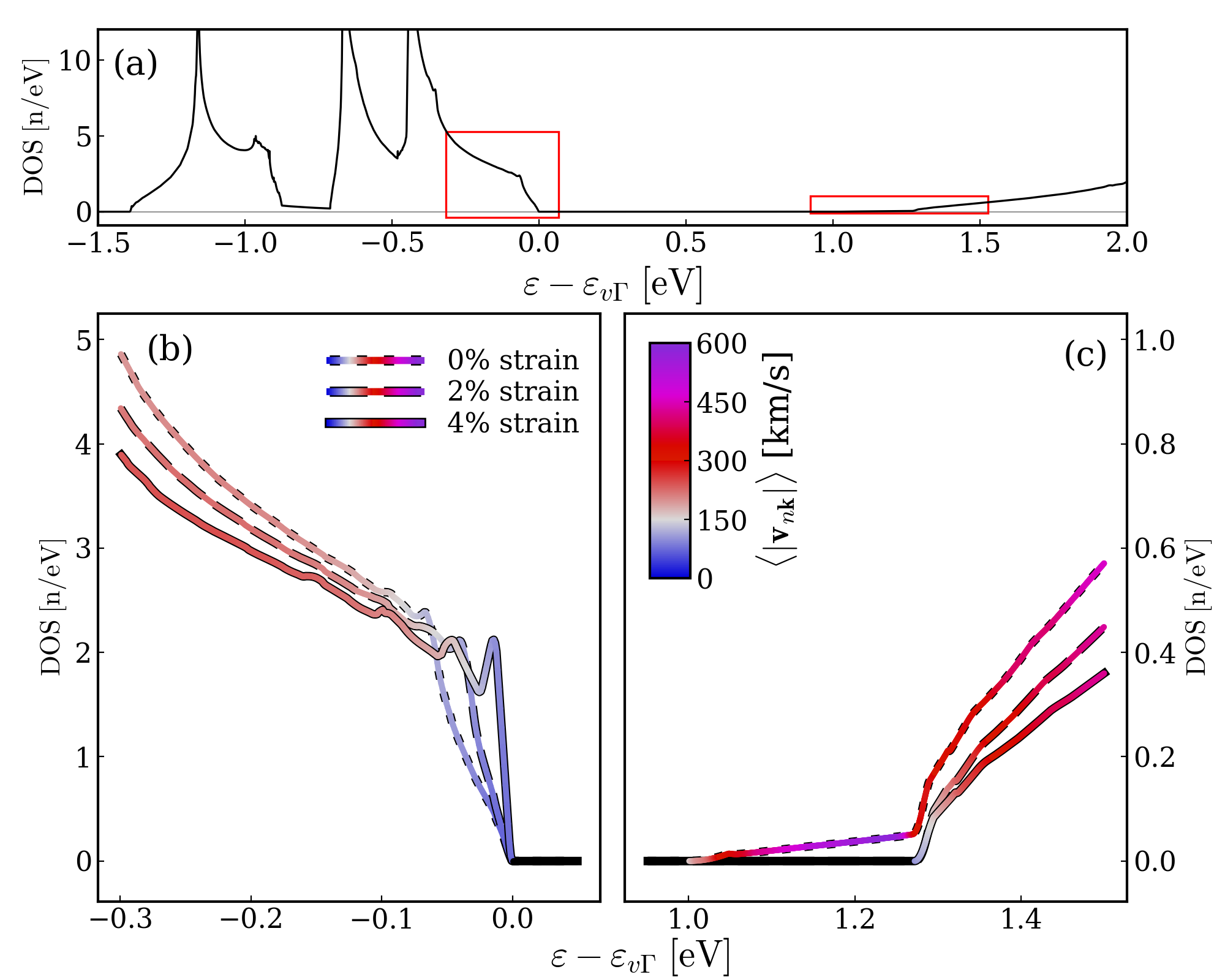}
    \caption{Density of states (DOS) of CsK$_2$Sb without strain (a). The red squares mark the close-ups around the VBM (b) and the CBM (c). In (b) and (c), the DOS is given at 0\%, 2\%, and 4\%\ strain. Each DOS was colored by the moving average absolute velocity $\langle | \mathbf{v}_{n\mathbf{k}}| \rangle$. When the DOS is zero, the color is black. 
    \label{fig: DOS}}
\end{figure}

Figure \ref{fig: DOS} shows the electronic DOS of CsK$_2$Sb without strain over a wide energy range (a). The near-gap VBM (b) and CBM (c) regions of the DOS are shown at 0\%, 2\%, and 4\%\ compressive strain. The coloring of the DOS in (b) and (c) is given by the value of running Gaussian average of absolute electron velocity $\langle | \mathbf{v}_{n\mathbf{k}}| \rangle$ (see SM\cite{SM} for details). The flattening of the valence band makes the sharp increase in the DOS close to the VBM more pronounced. The fact that both the DOS and velocities increased with strain is very distinct from the inherent trade-off between high DOS and velocities in single parabolic band structures\cite{Ziman_1972}.

\subsection{Electron transport}
\label{subsec: Electron transport}

\begin{figure}[h!]
    \centering
    \includegraphics[width=\linewidth]{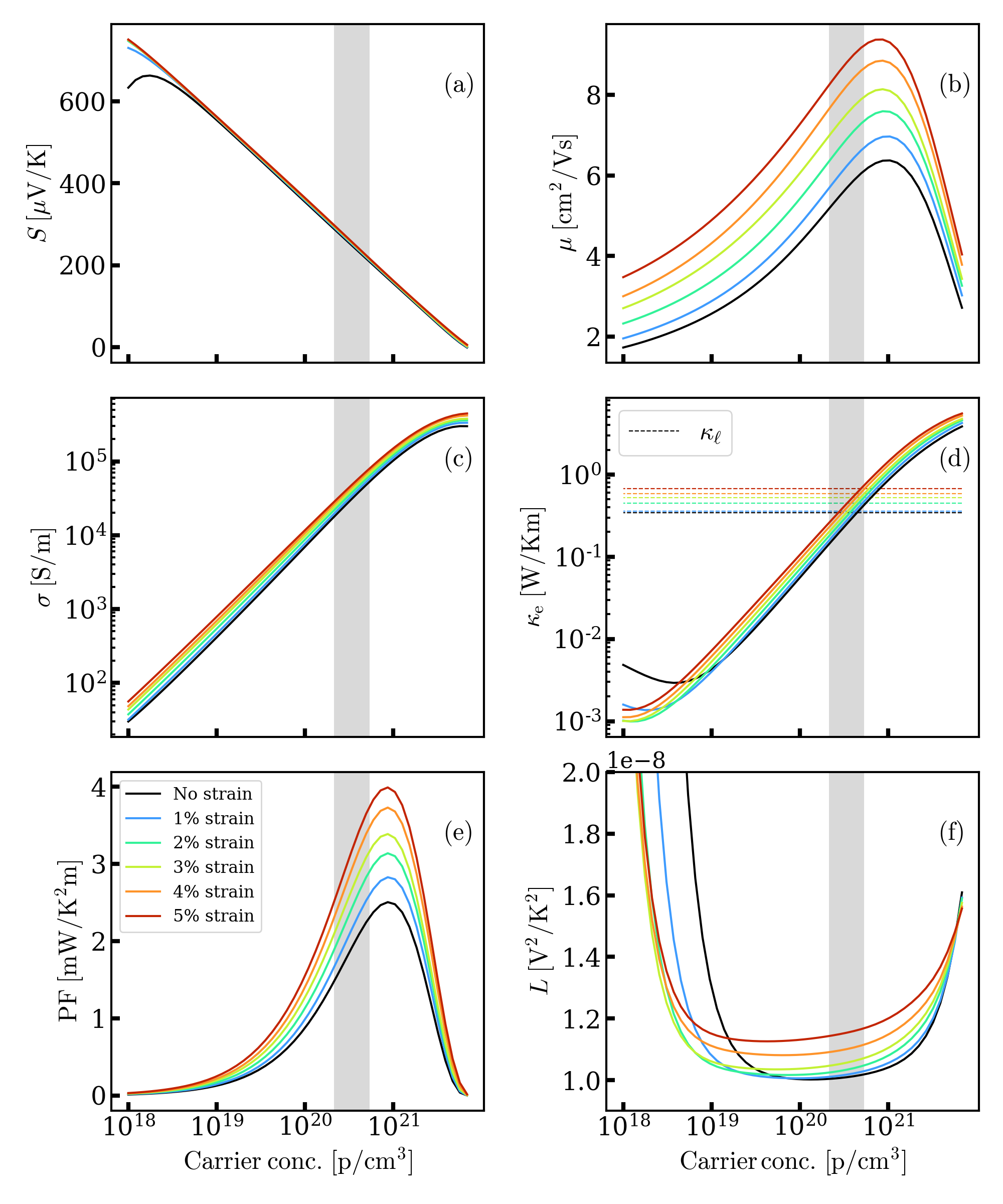}
    \caption{Electron transport properties: The Seebeck coefficient $S$ (a), mobility $\mu$ (b), electronic conductivity $\sigma$ (c), electronic part of the thermal conductivity $\kappa_\mathrm{e}$ (d), power factor PF (e), and Lorentz number $L$ (f) at 800\;K for p-type CsK$_2$Sb at varying compressive strain. The grey area indicates the carrier concentration region around maximum $zT$.
    \label{fig: transp_p}}
\end{figure}

Figure~\ref{fig: transp_p} shows the Seebeck coefficient $S$, mobility $\mu$, electrical conductivity $\sigma$, electronic thermal conductivity $\kappa_{\mathrm{e}}$, power factor (PF), and Lorentz number ($L$) for p-type CsK$_2$Sb at 800\;K as a function of carrier concentration, for different strains. The predicted $\mu$ of the unstrained material is low compared to several common TE materials\cite{Maeda_2015,Cha_2019,Dughaish_2002}, resulting in relatively low $\sigma$ and $\kappa_\mathrm{e}$. However, the high DOS allows for a high $S$ resulting in a high PF. Our predicted values agree well with those of Yuan et al.\cite{Yuan_2022}. For the doping level of 3.39$\times 10^{20}$ p/cm$^{-3}$ (corresponding to the optimal carrier concentration $N_\mathrm{opt}$ for $zT$, see Fig. \ref{fig: zT}), $L \approx 1.0\times 10^{-8}$, which is lower than the value typically used for semiconductors in the non-degenerate limit ($L_0 = 1.5\times 10^{-8}\;$V$^2$/K$^2$)\cite{Kim_2015,Flage-Larsen_2011}. As more and more materials with intrinsic low $\kappa_\ell$ are identified along with improvements in our ability to engineer reduced $\kappa_\ell$, it also becomes more important to identify low $L$ materials along with high $S$ at relevant doping concentrations.

Comparing the results for different strains shows that $\mu$ and hence both $\sigma$ and $\kappa_\mathrm{e}$ of the p-doped material increased with strain. The latter increased somewhat more, so that $L$ increased by 14\%\ to $L=$1.15$\times 10^{-8}$ at 5\%\ strain. Interestingly, there were only small changes to $L$ in the first two percentages of strain. The PF, however, despite only small improvements in $S$, increased continuously as a function of strain up to an increase of 66\%\ as a result of 5\%\ strain. The resulting $zT$ also taking into account $\kappa_\ell$ (Sec. \ref{subsec: Lattice thermal conductivity}) will be discussed in Sec. \ref{subsec: Figure of merit}.

\begin{figure}[h]
    \centering
    \includegraphics[width=\linewidth]{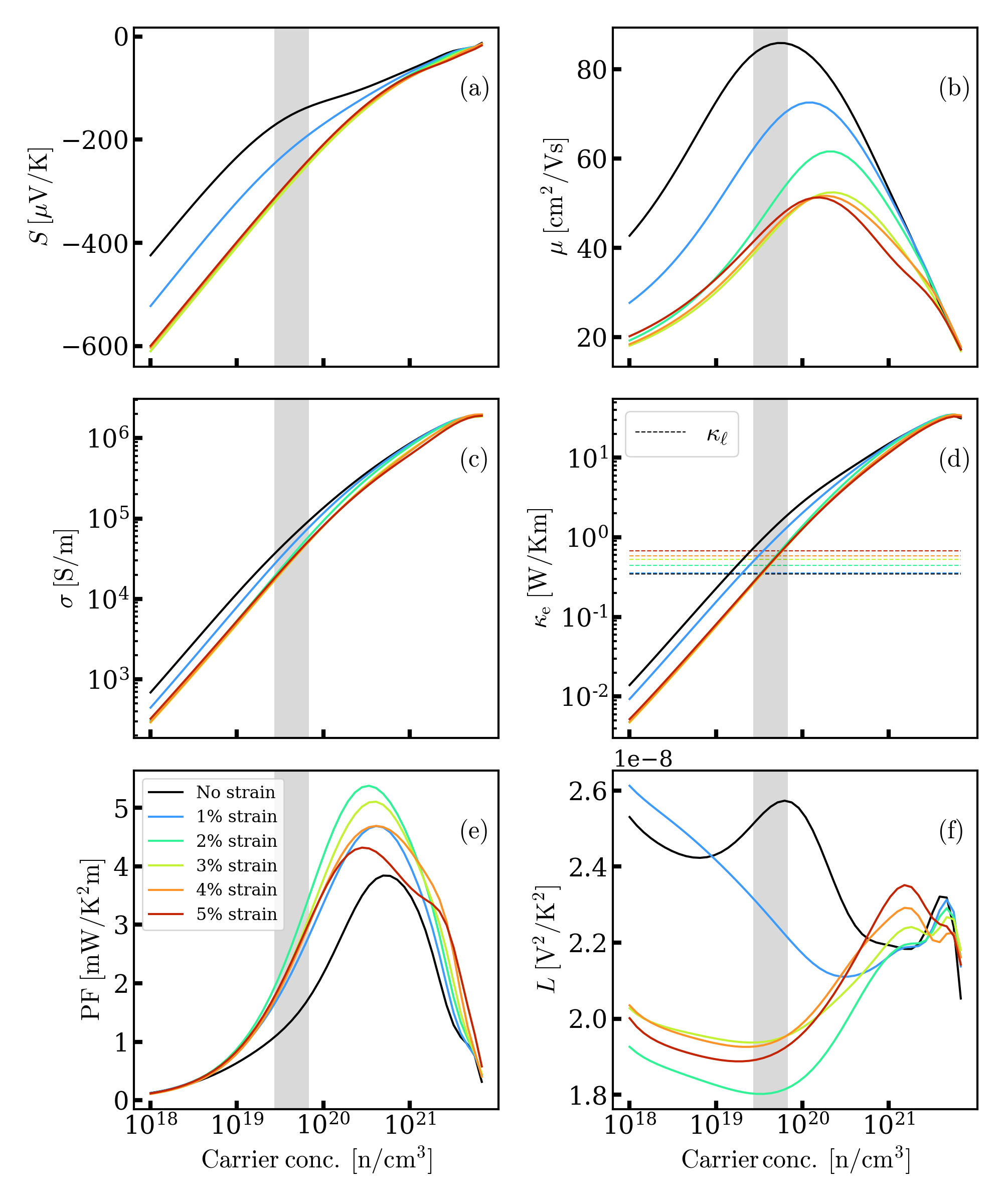}
    \caption{Electron transport properties: (a) $S$, (b) $\mu$, (d) $\sigma$, (d) $\kappa_\mathrm{e}$, (e) PF, and (f) $L$ at 800\;K for n-doped CsK$_2$Sb at varying levels of strain.
    \label{fig: transp_n}}
\end{figure}

Figure \ref{fig: transp_n} shows the TE properties of the n-type system. The $|S|$ values for the unstrained system are significantly lower than those of the p-type at their respective relevant doping concentrations. This result can be linked to a CBM at $\Gamma$ with low effective mass. However, the low effective mass, also makes $\mu$ more than an order of magnitude larger than for the p-type. The high $\mu$ results in a higher $\sigma$ and $\kappa_\mathrm{e}$, and an $L$ that is more than twice that of the p-type material. As strain aligns the $\Gamma$ and X valleys, with a corresponding increase in DOS, the $|S|$ nearly doubles. As the less dispersive X valley becomes dominant for transport, both $\sigma$ and $\kappa_\mathrm{e}$ are reduced at a fixed doping concentration, but $\kappa_\mathrm{e}$ falls off faster than $\sigma$, greatly reducing $L$. Despite the reduction in $\sigma$ for a given carrier concentration, the large increase in $|S|$ dominates and the peak PF still increases up to 2\%\ strain.

\subsection{Electronic scattering}
\label{subsec: Electron scattering scattering}

\begin{figure}[h!]
    \centering
    \includegraphics[width=\linewidth]{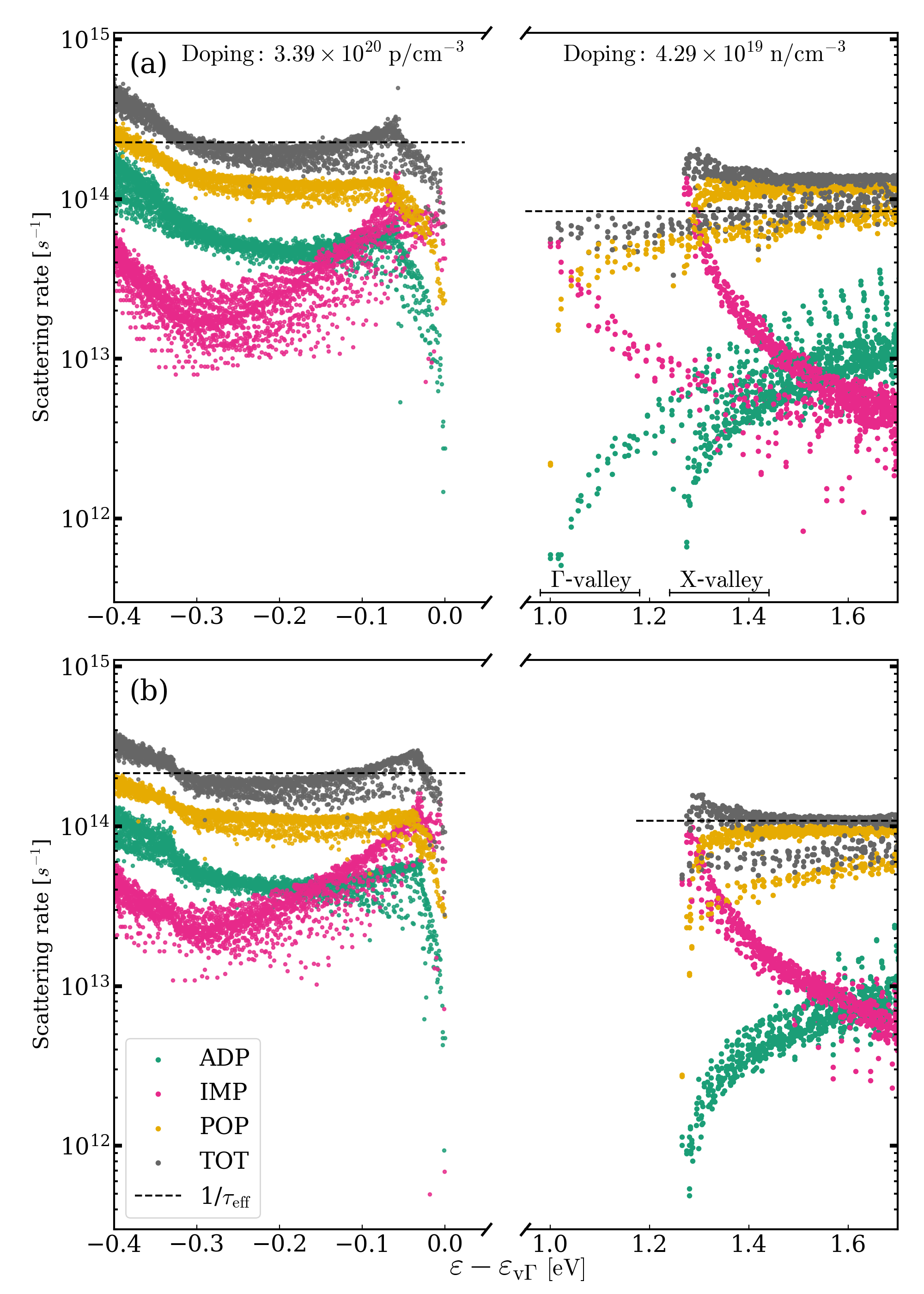}
    \caption{Scattering rates $\Gamma_{n \mathbf{k}}$ of ADP, IMP, POP, and total scattering shown for the valence band and conduction band of unstrained CsK$_2$Sb at 800\;K for (a) 0\%\ strain and (b) 2\%\ strain. For the conduction band at 0\%\ strain, the lowest 0.2 eV of the $\Gamma$-valley and X-valley are indicated.
    \label{fig: scattering}
    }

\end{figure}

The different scattering rates for CsK$_2$Sb at selected doping levels are shown for no strain in Fig.~\ref{fig: scattering} (a) and with 2\%\ strain (b). The colored dots indicate the scattering rate $\Gamma_{n\mathbf{k}}$ for the conduction and valence bands for the different scattering mechanisms. It shows that the POP scattering is the dominant scattering mechanism both for p- and n-type, a common feature in polar materials\cite{Zayachuk_1997b, Cao_2018,Ganose_2021,Hauble_2023,Li_2024}. The POP scattering is according to Eq. \ref{eq: POP} particularly strong when $\epsilon_\infty$ is low and $\epsilon_0$ is high relative to $\epsilon_\infty$, as is the case in this material. For n-type, the POP scattering rates of the $\Gamma$ valley are lower than p-type, while for the X valley the scattering rates are almost equal. IMP scattering is also important for low-energetic carriers, while ADP is more important than IMP for p-type and less important for n-type. The total scattering rates for most of the n- and p- type states are almost the same magnitude, especially at 2\%\ strain where the higher scattering rate X valley contributes more. 

From the commonly used approximation $\tau \propto 1/\mathrm{DOS}$\cite{Ziman_1960,Xu_2014,Xia_2019} assuming similar materials parameters (used in Eq. \ref{eq: ADP}-\ref{eq: pop_weight}), one would expect the p-type scattering rates to be vastly larger than the n-type. However, this is not the case. This result can be understood by the $\mathbf{q}$-dependence of the individual scattering mechanisms. For the dominant POP scattering, the $\propto 1/\mathbf{q}^2$ factor results in reduced scattering between the different cylinders of the energy isosurface due to the large distance in $\mathbf{k}$-space. For long-ranged ADP scattering, lacking this factor, the p-type ADP scattering is an order of magnitude larger than the n-type. Finally, for the short-ranged ($\propto 1/\mathbf{q}^4$) IMP scattering, both the n- and p-type materials scatter the most close to their respective band minimum or maximum. Further away from the band edge, the scattering rates quickly fall off due to the longer distances within the energy isosurface. Increasing the anisotropy of band structures has been shown to lead to a larger increase in momentum-dependent scattering mechanisms than that of momentum-independent\cite{Graziosi_2023}. This strong short-range scattering would also be present in cylinders and could be part of the reason why POP scattering dominates. Note that the n- and p-type scattering rates are shown for different carrier concentrations, each at their respective $N_\mathrm{opt}$. Higher carrier concentration tends to decrease POP scattering due to free-carrier screening, especially when the DOS is also high. However, high doping increases the number of ionized impurities and thereby IMP scattering. 

With more realistic scattering models like in \textsc{AMSET}, the relaxation time depends on the band and the $\mathbf{k}$-point and thus generally have a wide spread of values. To assist in the interpretation of the magnitude of different scattering mechanisms, we introduce an effective relaxation time $\tau_\mathrm{eff}$ defined as the CRTA value that reproduces the same $\sigma$ as the \textsc{AMSET} distribution of $\tau_{n\mathbf{k}}$ for a given $T$ and carrier concentration.

\begin{figure}[h!]
    \centering
    \includegraphics[width=\linewidth]{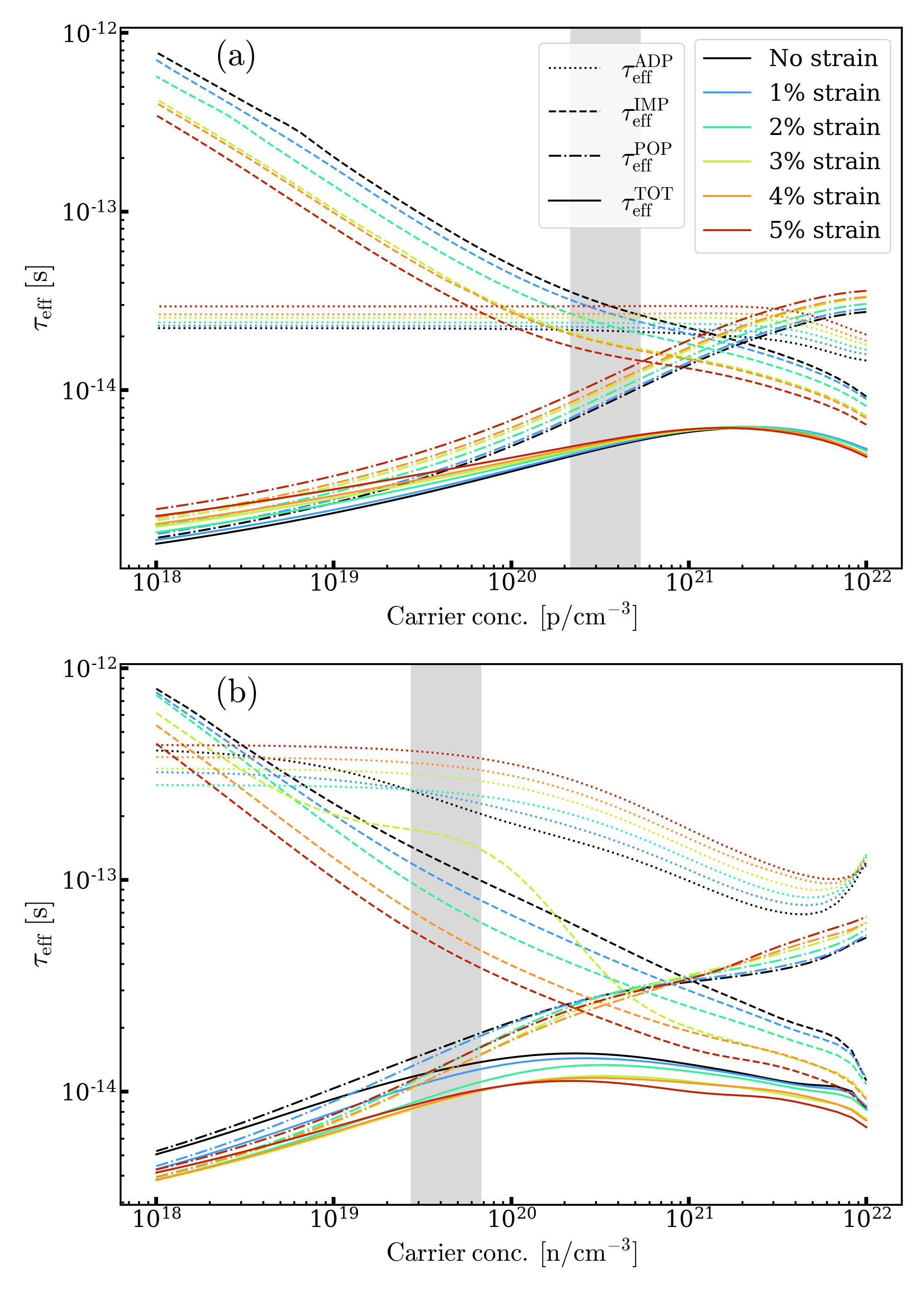}
    \caption{Effective relaxation time $\tau_\mathrm{eff}$ of the valence band at carrier concentrations and different strains at a temperature of 800\;K for (a) p-type and (b) n-type. Plots are shown for relaxation times due to single scattering mechanism as well as that of the total effective scattering rate $\tau^\mathrm{TOT}_\mathrm{eff}$.
    \label{fig: tau_eff}
    }
\end{figure}

Figure \ref{fig: tau_eff} shows $\tau_\mathrm{eff}$ for different scattering mechanisms with varying strain and doping concentrations. Each "bundle" of lines with the same line style represents the $\tau_\mathrm{eff}$ of a specific scattering mechanism at different strains indicated by the color. The total effective scattering rate $\tau^\mathrm{TOT}_\mathrm{eff}$ is given by the lower full lines. At low doping, the latter is almost entirely limited by POP scattering. With increasing doping, the free carrier concentration reduces the POP scattering. Simultaneously, increased doping also increases the ionized impurity scattering, making it the dominant scattering mechanism at high carrier concentrations. ADP scattering is independent of carrier concentration and $\tau^\mathrm{ADP}_\mathrm{eff}$ only changes at very high carrier concentrations due to the shift in Fermi level this doping represents.

The effect of strain on the different scattering mechanisms can be linked to the changing material properties listed in Table \ref{tab: Mat_prop}. Under compressive strain, with a reduction in the ionic contribution to the static dielectric tensor, the POP scattering decreases according to Eq. \ref{eq: POP}. Simultaneously, the IMP scattering increases due to the reduction in $\epsilon_\mathrm{s}$. For ADP, the major change in scattering rates is due to the changes to the elastic constants and the resulting changes to the sound velocity in Eq.~\ref{eq: ADP}. The net effect on $\tau^\mathrm{TOT}_\mathrm{eff}$, at $N_{\mathrm{opt}}$, is an increase of 10\%\ when going from 0 to 5\%\ strain. As this is much less than the increase in mobility, the effect of the changing band curvature must be substantial.

Figure~\ref{fig: tau_eff} shows the conduction band $\tau_\mathrm{eff}$ as a function of carrier concentration at varying strain. The general trends of $\tau_\mathrm{eff}$ are similar to those of the valence band but with some notable differences. Primarily, these differences stem from the two conduction band valleys ($\Gamma$ and X) contributing with different scattering rates, see Fig. \ref{fig: scattering}. Since the Fermi level is determined by doping and strain changes the relative energy minimum of the two valleys, both of these parameters change the relative contribution of the two valleys to $\tau_\mathrm{eff}$. For POP scattering at low carrier concentration, increasing strain decreases $\tau_\mathrm{eff}$ as the X-valley becomes the major contributor to transport with higher scattering rates, see Fig.~\ref{fig: scattering}. At high doping, both valleys contribute to the expected change in $\tau^\mathrm{POP}_\mathrm{eff}$, seen for p-type when strain is introduced. 

For $\tau^\mathrm{ADP}_\mathrm{eff}$, we observe something similar, but only up to 2\%\ strain. For ADP, the two valleys have similar scattering rates that increase with carrier concentration. At a given carrier concentration; however, the effective carrier concentration per band is reduced as the band aligns, resulting in a reduced $\tau^\mathrm{ADP}_\mathrm{eff}$. As $\tau^\mathrm{TOT}_\mathrm{eff}$ is almost entirely dominated by POP scattering at $N_\mathrm{opt}$, the changes to $\tau^\mathrm{TOT}_\mathrm{eff}$ under strain closely resembled the changes to $\tau^\mathrm{POP}_\mathrm{eff}$.

Although the lifetimes were substantially shorter for p-type at optimal doping, the n-type optimum is quite similar to the conventional choice of $\tau = 10^{-14}\;{\mathrm{s}}$.

\begin{figure}[h!]
    \centering
    \includegraphics[width=\linewidth]{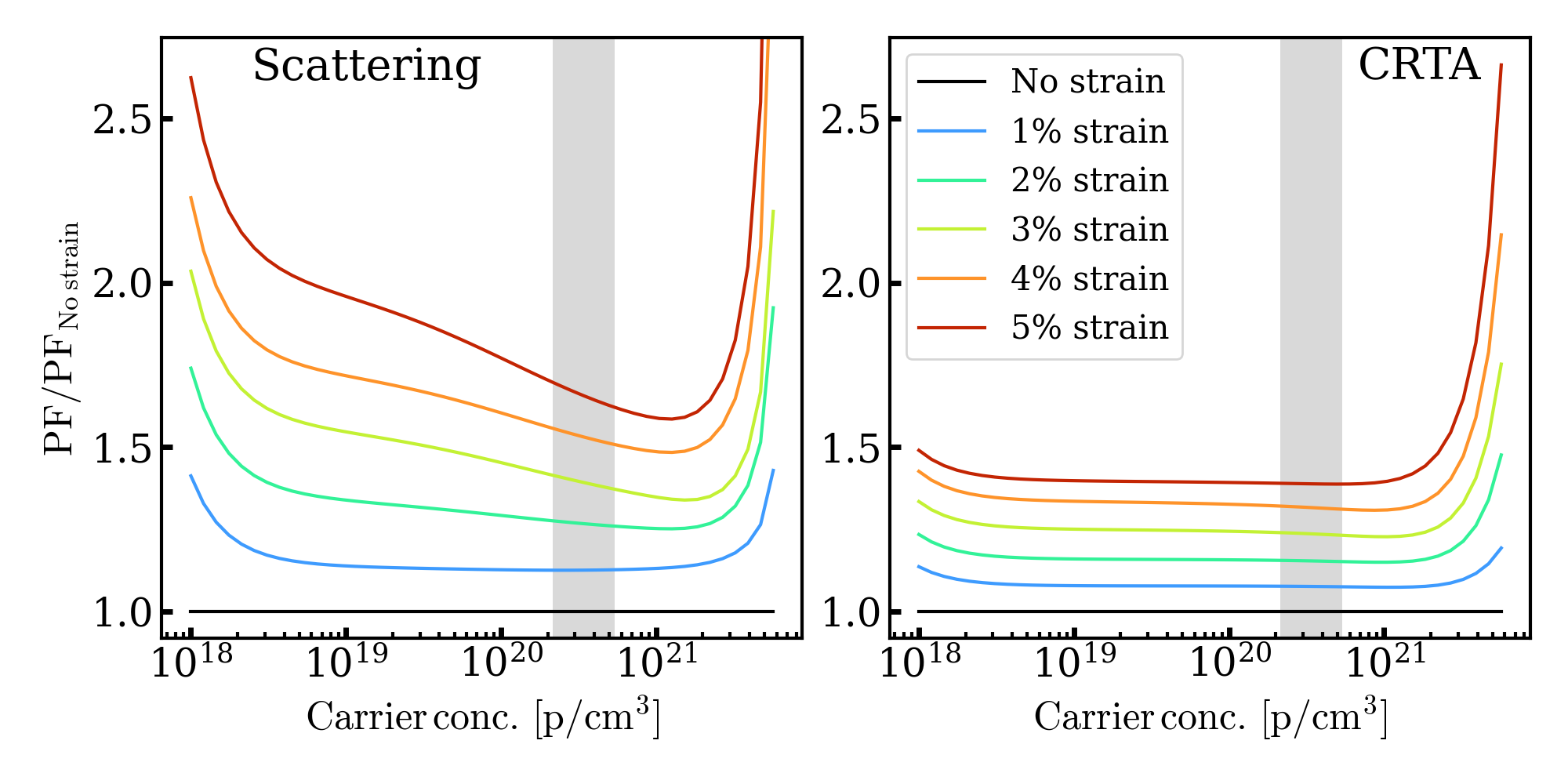}
    \caption{Power factor enhancement due to strain with electron scattering rates calculated by \textsc{AMSET} (left) and with constant relaxation time (right) for p-type CsK$_2$Sb at 800\;K.
    \label{fig: PF_const}
    }
\end{figure}

Fig.~\ref{fig: PF_const} shows the enhancement of the p-type PF due to strain with scattering rates from \textsc{AMSET} (left) and based on the CRTA (right). This allows us to gauge the role of the velocities and DOS (Figs.~\ref{fig: Fermi}, \ref{fig: DOS}) under strain, without the effect of the scattering rates. At $N_\mathrm{opt}$ (the grey shaded area), the CRTA PF increases by 39\%\ at 5\%\ strain, with an evenly distributed increase of each percentage of strain. This shows that the increased $v_\mathrm{g}$ and DOS are important for increasing the PF. However, with scattering, the PF increases by 66\%\ with the same strain rate, highlighting that reduced scattering also contributes significantly to the PF improvement.

\subsection{Lattice thermal conductivity}
\label{subsec: Lattice thermal conductivity}
\begin{figure}[h!]
    \centering
    \includegraphics[width=\linewidth]{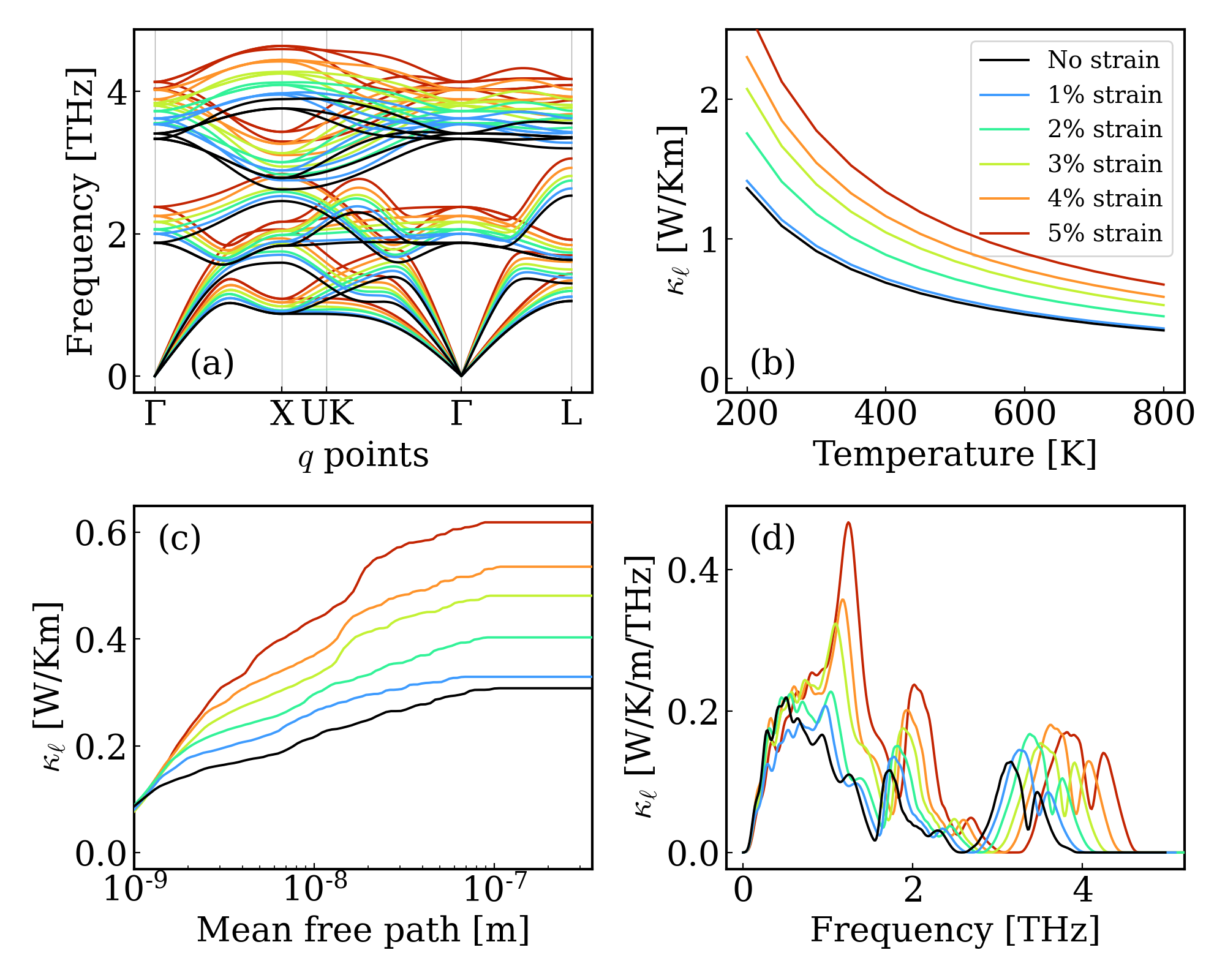}
    \caption{Phonon properties of CsK$_2$Sb at different compressive strains calculated with TDEP: (a) Phonon dispersion relation, (b) $\kappa_{\ell}$ as a function of temperature, (c) cumulative $\kappa_{\ell}$ as a function of the mean free path (MFP) ($T=800$ K), (d) spectral $\kappa_{\ell}$ as a function of the phonon frequency ($T=800$ K).
    \label{fig: ltc}}
\end{figure}

The calculated phonon properties of CsK$_2$Sb are shown in Fig.\ \ref{fig: ltc}. The phonon dispersion relation for $T=300$ K is given in Fig.\ \ref{fig: ltc}(a) with a compressive strain varying from 0 to 5{\%} (see legend in Fig.\ \ref{fig: ltc}(b).) It shows how the phonon frequencies increased as the unit cell was compressed. 

The resulting $\kappa_{\ell}$ is shown in Fig.\ \ref{fig: ltc}(b). The predicted lattice thermal conductivity of the unstrained structure was $\kappa_{\ell}=0.91$\;Wm$^{-1}$K$^{-1}$ at 300\;K, going down to 0.35\;Wm$^{-1}$K$^{-1}$ at 800\;K. Although the predicted $\kappa_{\ell}$ values were low, they are 3--6 times higher than previous predictions, \cite{Sharma_2023,Yuan_2022}, possibly due to the highly anharmonic effect captured by previous studies using a molecular dynamics-based $\kappa_{\ell}$ evaluation. $\kappa_\ell$ increased with increasing compressive strain, except when going from no strain to 1\;{\%} strain, for which there was virtually no change in $\kappa_{\ell}$.

This behavior can be understood from the cumulative and spectral $\kappa_{\ell}$ (Fig.\ \ref{fig: ltc}(c) and (d)).
Looking at the contribution from phonons with varying MFP (c), the ones with the lowest MFP contributed less to the total $\kappa_{\ell}$ when compressive strain was applied, while the opposite was the case for high MFPs. The case with 1{\%} strain stands out, where the highest MFP phonons contributed less than those of the strain-free material.\cite{LTC_note} Turning to the spectral $\kappa_{\ell}$ (d), strain reduced the contributions to $\kappa_{\ell}$ from phonons with low frequency ($<0.5$~THz), while contributions from the regions around 1 and 2\;THz increased. At 1{\%}, the reduction at low frequencies almost cancelled the increase at higher frequencies, while there was a significant net increase for larger strains.

\subsection{Figure of merit}
\label{subsec: Figure of merit}

\begin{figure}[h]
    \centering
    \includegraphics[width=\linewidth]{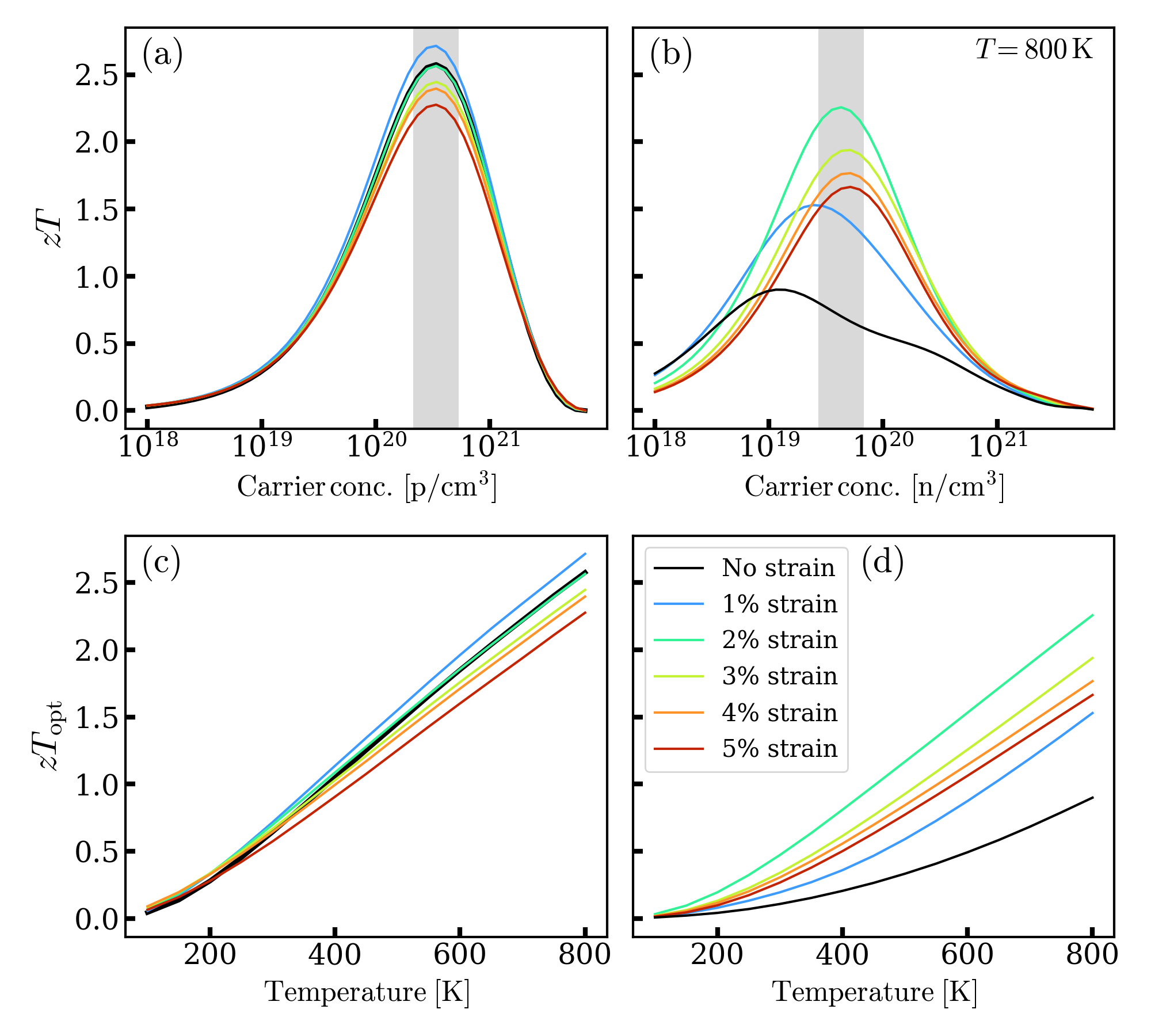}
    \caption{$zT$ of (a) p-type and (b) n-type doped CsK$_2$Sb at 800\;K at varying strain. (c) and (d) show the charge carrier-optimized $zT$ at different temperatures for p- and n-type, respectively.
    \label{fig: zT}}
\end{figure}

Figure \ref{fig: zT}(a) and (b) show $zT$ as a function of carrier concentration for p- and n-type CsK$_2$Sb at 800\;K. For the unstrained p-type material, the high PF combined with the low $\kappa =\kappa_\mathrm{e} + \kappa_\mathrm{\ell}$ resulted in a $zT$ of 2.6 at a carrier concentration of 3.39$\times10^{20}$ p/cm$^{-3}$. As the material was strained, flattening the valence band and increasing velocities, the $PF$ increased more than $\kappa$, resulting in a maximum $zT$ of 2.7 at 1\%\ strain. After this, the increase in $\kappa_\ell$ dominated and further strain only served to reduce the maximum $zT$. If $\kappa_\ell$ was kept constant at the 0\%\ strain value, the $zT$ would continue to increase, reaching a value of $zT= 3.3$ at 5\%\ strain, as shown in SM\cite{SM}.

For n-type, the low $S$ and relatively high $\kappa_\mathrm{e}$ resulting from the low DOS gave a $zT$ of 0.9 at 800\;K. With the alignment of the $\Gamma$ and X valleys under strain, a simultaneous increase of $S$ and decrease of $\kappa_\mathrm{e}$ led to a large increase of the PF and the $zT$, an effect well documented in literature\cite{Pei_2011,Zhang_2016,Li_2018,Zhu_2022}. With 1\%\ strain resulting in a $zT$ of 1.5 and 2\%\ strain resulting in $zT=$2.3, an almost three-fold increase from the unstrained structure could be observed. Further strain removed the valley alignment, thus decreasing the power factor. The accompanying increase in $\kappa_\ell$ decreased $zT$.

Figure~\ref{fig: zT}(c) and (d) show the maximal p- and n-type $zT$ as a function of $T$ for each strain. While only moderate changes can be seen with strain, the p-type material can achieve a $zT$ of 1 below 400\;K. For the n-type material, the optimal strain of 2\%\ resulted in a large increase in $zT$ at all temperatures, with $zT$ reaching just below 1 at 400\;K.

\section{Discussion}
\label{sec: Discussion}

TE materials with low-dimensional Fermi surfaces are attractive due to their high DOS near the band edge, which allows $S$ to be high at high carrier concentrations. A high DOS, however, gives many allowed scattering processes, which may increase scattering rates and thereby reduce $\mu$. Comparing the scattering rates of n- and p-type CsK$_2$Sb can elucidate the role of the DOS on the scattering rates, since the valence band has a much larger DOS than the conduction band. At the same time, the material parameters, i.e. scattering prefactors, are quite similar. In the case of ADP scattering, the DOS difference resulted in an order of magnitude larger scattering rates for p-type than for n-type. On the other hand, this difference was reduced to about a factor of 4 (2) for IMP (POP) scattering. The reduced DOS sensitivity can be explained by the $\mathbf{q}$-dependence of these scattering rates. Since POP often had the largest scattering rate, this result is very encouraging for TE materials with low-dimensional band structures since they extend over large parts of the Brillouin zone.

There is an interesting trade-off in the effect of strain on
$\kappa_\ell$ and $\tau_{\mathrm{eff}}$ related to the material parameters in Tab.~\ref{tab: Mat_prop} entering into the scattering coefficients. Reducing the volume increases elastic constants and thus ADP scattering through the reduced sound velocity, $c$ (Eq. \ref{eq: ADP}). Further, the ionic contribution to the dielectric tensor is reduced, thereby reducing POP scattering, but it also increases IMP scattering, see Eqs. \ref{eq: POP},\ref{eq: IMP}. In most materials, an increase in elastic constants under compressive strain will also increase $\kappa_\ell$\cite{Hofmeister_2007,Ravichandran_2019,Li_2022,Cherniushok_2024}, though exceptions do exist\cite{Lindsay_2015,Ravichandran_2019,Li_2022}. This inverse relationship between $\kappa_\ell$ and $\tau$ could be expected in other materials as well, where POP or ADP are the dominant scattering mechanisms.

The optimal p-type carrier concentrations, i.e. $\sim 10^{20}$ e/cm$^{-3}$ is more than an order of magnitude larger than that of the n-type, which should also be the case for other materials with low-dimensional electronic band structures\cite{Parker_2013,Dylla_2019,Brod_2021}. Recently, Lou et al. \cite{Lou_2024} argued that this could be an attractive feature as low optimal doping values are more vulnerable to doping variations such as intrinsic doping and impurities, which can increase the cost of synthesis and raw materials. On the other hand, high doping also requires high dopability. The fact that a high $zT$ can be achieved at finite strain for both p- and n-type is technologically attractive since thermomechanical properties would be similar, and the same metallization schemes and bonding materials might be used for both contacts. 

As discussed in the introduction, the hydrostatic stress needed to reach the larger strain values studied here is unrealistic in practical devices. In practice, one would need "chemical" pressure, i.e. alloying with smaller, isovalent elements\cite{Cherniushok_2024}. However, such isovalent substitution might also introduce significant changes to the band structure and material properties. Critically, it would also introduce additional electronic carrier scattering mechanisms. The overall effect might still be beneficial, as isovalent alloying can further decrease $\kappa_\ell$~\cite{Zevalkink_2012,Shuai_2016,Tranas_2022}.

CsK$_2$Sb is not the only material that displays a low-dimensional band structure. Examples of low-dimensional band structures similar to CsK$_2$Sb include Rb$_2$CsSb\cite{Yue_2022}, KCs$_2$Sb, CsK$_2$Bi, and CsK$_2$As\cite{She_2023}, and compounds in the A$_3$Sb class, Cs$_3$Sb~\cite{Kalarasse_2010}, Na$_3$Sb~\cite{Ettema_2000}, and K$_3$Sb~\cite{Liu_2024a,Kalarasse_2010}. Many of the A$_3$Sb compounds without low-dimensional band structures have also been predicted to exhibit excellent TE properties due to their ultra-low $\kappa_\ell$. A promising path forward could be to investigate the miscibility of the solid solutions of these compounds, which could provide the desired chemical pressure. This could give access to a continuous range of band structures and scattering mechanisms, providing good prospects for designing materials with outstanding TE properties.

\section{Conclusion and outlook}
\label{sec: Conclusion and outlook}

The thermoelectric properties of CsK$_2$Sb were in this study investigated with electron transport based on momentum-resolved scattering in \textsc{AMSET} and phonon properties with TDEP. For p-doped CsK$_2$Sb we showed that low-dimensional band structures lead to promising thermoelectric materials because of the combination of high electronic DOS near the Fermi level and high electron velocities. Crucially, we found that the long distance between states in $\mathbf{k}$-space results in low polar optical phonon and ionized impurity scattering rates; significantly lower than what would generally be expected with such a high DOS. Similar results could be expected in other materials with low-dimensional band structure, where $\mathbf{q}$-dependent POP or IMP scattering dominates. These results highlight the importance of moving beyond the CRTA or simple DOS-based scattering to correctly identify materials with a high potential as TE materials.

When strain was added to the structure, the electronic transport properties changed significantly. For the p-doped material, the Fermi surfaces could be optimized towards highly pronounced low-dimensionality at around 2{\%} compressive strain, which led to a significant increase of the power factor. The n-doped material had an even higher increase, due to valley alignment of the electronic band structure. This was partially counteracted by the lattice thermal conductivity, which increased when strain was added. Nevertheless, the strain-optimized p- and n-doped materials showed very promising TE properties, with a maximal figure of merit at 800 K of 2.7 and 2.3, respectively.

We believe that CsK$_2$Sb with repeatedly predicted superior TE properties merits further investigation, as do related antimonides rich in alkali metals. Further \textit{ab initio} calculations to explore the n- and p-type dopability and identify proper candidates for solid solutions should be performed. More advanced methods for determining the band structure and transport properties should also be employed. Ultimately, targeted synthesis and characterization to experimentally verify TE properties is needed.

\begin{acknowledgments}
The computations in this work were funded by The Norwegian e-infrastructure for research and education, Sigma2, through grants No. nn9711k and nn2615k. Ø.A.G., O.M.L, and K.B. gratefully acknowledge funding by the Research Council of Norway through the Allotherm project (Project No. 314778)  and G.J.S. by the Humboldt Foundation.
\end{acknowledgments}

\appendix

\section{Scattering matrix elements}
\label{app: a}
In the \textsc{AMSET} code\cite{Ganose_2021}, the matrix elements between state $|n\mathbf{k}\rangle$ and $\langle m \mathbf{k} +\mathbf{q}|$ for acoustic deformation potential scattering are calculated as
\begin{equation}
\begin{aligned}
    \label{eq: ADP}
    g_{nm}^{\mathrm{adp}} = \sqrt{k_\mathrm{B}T} \sum _{\mathbf{G}\neq -\mathbf{q}} 
    \left[ 
    \frac{\mathbf{\tilde{D}}_{n\mathbf{k}}:\mathbf{\hat{S}}_l}{c_l\sqrt{\rho}} 
    + \frac{\mathbf{\tilde{D}}_{n\mathbf{k}}:\mathbf{\hat{S}}_{t_1}}{c_{t_1}\sqrt{\rho}}
    + \frac{\mathbf{\tilde{D}}_{n\mathbf{k}}:\mathbf{\hat{S}}_{t_2}}{c_{t_2}\sqrt{\rho}}
    \right] \\
    \times \langle m \mathbf{k} +\mathbf{q}| e^{i(\mathbf{q}+\mathbf{G})\cdot \mathbf{r}}|n\mathbf{k} \rangle,
\end{aligned}
\end{equation}
where $\mathbf{\tilde{D}}_{n\mathbf{k}}$ is the velocity-adjusted deformation potential $\mathbf{D}_{n\mathbf{k}} + \mathbf{v}_{n\mathbf{k}} \otimes \mathbf{v}_{n\mathbf{k}}$, $\rho$ is the density, $\hat{\mathbf{S}}$ is unit strain, and $c$ is the velocity of sound. The subscripts $l$, $t_1$, and $t_2$ denotes the one longitudinal and two transverse directions, respectively.

The ionized impurity scattering matrix elements are given by
\begin{equation}
    \label{eq: IMP}
    g_{nm}^{\mathrm{imp}} = \sum _{\mathbf{G}\neq -\mathbf{q}} 
     \frac{n_{ii}^{1/2} Z e}{\mathbf{\hat{n}} \cdot \boldsymbol{\epsilon_s} \cdot \mathbf{\hat{n}}}
     \frac{\langle m \mathbf{k} +\mathbf{q}| e^{i(\mathbf{q}+\mathbf{G})\cdot \mathbf{r}}|n\mathbf{k} \rangle}{|\mathbf{q} + \mathbf{G}|^2 + \beta_\mathrm{s}^2 },
\end{equation}
where $n_\mathrm{ii} = (n_\mathrm{h}-n_\mathrm{e})/Z$ is the ionized impurity concentration, $Z$ is the unscreened charge of the impurity, and $e$ is the electron charge. $\beta_\mathrm{s}$ is the inverse screening length, given as
\begin{equation}
    \label{eq: free_carrier_screening}
    \beta_\mathrm{s}^2 = \frac{e^2}{\boldsymbol{\epsilon}_\mathrm{s}k_\mathrm{B}T} \int \frac{\mathrm{d}\varepsilon}{V}\mathrm{D}(\varepsilon) f(\varepsilon)(1 - f(\varepsilon))
\end{equation}
where $\boldsymbol{\epsilon}_\mathrm{s}$ is the static dielectric tensor, $V$ is the unit cell volume, $\mathrm{D}$ is the DOS, and $f$ is the Fermi-Dirac distribution function. 

The polar optical phonon scattering matrix element is given by
\begin{equation}
\begin{aligned}  
    \label{eq: POP}
    g_{nm}^{\mathrm{pop}} = \left[ \frac{\hbar\omega_{po}}{2} \right]^{1/2} \sum _{\mathbf{G}\neq -\mathbf{q}}
    \left(
    \frac{1}{\mathbf{\hat{n}} \cdot \boldsymbol{\epsilon _\infty} \cdot \mathbf{\hat{n}}}
     - \frac{1}{\mathbf{\hat{n}} \cdot \boldsymbol{\epsilon_s} \cdot \mathbf{\hat{n}}}
    \right)^{1/2} \\
    \times \frac{\langle m \mathbf{k} +\mathbf{q}| e^{i(\mathbf{q}+\mathbf{G})\cdot \mathbf{r}}|n\mathbf{k} \rangle}{\sqrt{|\mathbf{q} + \mathbf{G}|^2 + \beta_\infty^2}},
\end{aligned}
\end{equation}
where $\hat{\mathbf{n}}$ is the normalized vector of scattering, $\epsilon_\infty$ is the high frequency and $\epsilon_s$, the static dielectric tensors. $\beta_\infty ^2$ is defined as in Eq. \ref{eq: free_carrier_screening}, but uses $\boldsymbol{\epsilon}_\infty$ instead of $\boldsymbol{\epsilon}_\mathrm{s}$. An effective optical phonon frequency is obtained by the weighted average of the phonon frequencies $\omega_{\mathrm{po}}$, 
\begin{equation}
    \omega_\mathrm{po} =  \frac{\sum_\nu \omega_{\Gamma \nu}w_\nu}{\sum_\nu w_\nu}.
\end{equation}
Here, $\omega_{\Gamma v}$ is the frequency of phonon branch $\nu$ at the $\Gamma$-point and the weight is defined as
\begin{equation}
    \label{eq: pop_weight}
    w_\nu = \sum_\alpha \left[ \frac{1}{M_\alpha \omega_{\mathbf{q}\nu}}\right]^{1/2}\times [\mathbf{q}\cdot \mathbf{Z}_\alpha^* \cdot\mathbf{e}_{\alpha \nu}] \,,
\end{equation}
where $M$ is atomic mass, $\mathbf{Z}_\alpha^*$ is Born effective charge, $\mathbf{e}_{\alpha \nu}$ is a phonon eigenvector, and the index $\alpha$ runs over the atoms of the unit cell.

\bibliography{ref}

\begin{thebibliography}{89}%
\makeatletter
\providecommand \@ifxundefined [1]{%
 \@ifx{#1\undefined}
}%
\providecommand \@ifnum [1]{%
 \ifnum #1\expandafter \@firstoftwo
 \else \expandafter \@secondoftwo
 \fi
}%
\providecommand \@ifx [1]{%
 \ifx #1\expandafter \@firstoftwo
 \else \expandafter \@secondoftwo
 \fi
}%
\providecommand \natexlab [1]{#1}%
\providecommand \enquote  [1]{``#1''}%
\providecommand \bibnamefont  [1]{#1}%
\providecommand \bibfnamefont [1]{#1}%
\providecommand \citenamefont [1]{#1}%
\providecommand \href@noop [0]{\@secondoftwo}%
\providecommand \href [0]{\begingroup \@sanitize@url \@href}%
\providecommand \@href[1]{\@@startlink{#1}\@@href}%
\providecommand \@@href[1]{\endgroup#1\@@endlink}%
\providecommand \@sanitize@url [0]{\catcode `\\12\catcode `\$12\catcode `\&12\catcode `\#12\catcode `\^12\catcode `\_12\catcode `\%12\relax}%
\providecommand \@@startlink[1]{}%
\providecommand \@@endlink[0]{}%
\providecommand \url  [0]{\begingroup\@sanitize@url \@url }%
\providecommand \@url [1]{\endgroup\@href {#1}{\urlprefix }}%
\providecommand \urlprefix  [0]{URL }%
\providecommand \Eprint [0]{\href }%
\providecommand \doibase [0]{https://doi.org/}%
\providecommand \selectlanguage [0]{\@gobble}%
\providecommand \bibinfo  [0]{\@secondoftwo}%
\providecommand \bibfield  [0]{\@secondoftwo}%
\providecommand \translation [1]{[#1]}%
\providecommand \BibitemOpen [0]{}%
\providecommand \bibitemStop [0]{}%
\providecommand \bibitemNoStop [0]{.\EOS\space}%
\providecommand \EOS [0]{\spacefactor3000\relax}%
\providecommand \BibitemShut  [1]{\csname bibitem#1\endcsname}%
\let\auto@bib@innerbib\@empty
\bibitem [{\citenamefont {Snyder}\ and\ \citenamefont {Toberer}(2008)}]{Snyder_2008}%
  \BibitemOpen
  \bibfield  {author} {\bibinfo {author} {\bibfnamefont {G.~J.}\ \bibnamefont {Snyder}}\ and\ \bibinfo {author} {\bibfnamefont {E.~S.}\ \bibnamefont {Toberer}},\ }\bibfield  {title} {\bibinfo {title} {Complex thermoelectric materials},\ }\href {https://doi.org/10.1038/nmat2090} {\bibfield  {journal} {\bibinfo  {journal} {Nature Materials}\ }\textbf {\bibinfo {volume} {7}},\ \bibinfo {pages} {105} (\bibinfo {year} {2008})}\BibitemShut {NoStop}%
\bibitem [{\citenamefont {Vining}(2009)}]{Vining_2009}%
  \BibitemOpen
  \bibfield  {author} {\bibinfo {author} {\bibfnamefont {C.~B.}\ \bibnamefont {Vining}},\ }\bibfield  {title} {\bibinfo {title} {An inconvenient truth about thermoelectrics},\ }\href {https://doi.org/10.1038/nmat2361} {\bibfield  {journal} {\bibinfo  {journal} {Nature Materials}\ }\textbf {\bibinfo {volume} {8}},\ \bibinfo {pages} {83} (\bibinfo {year} {2009})}\BibitemShut {NoStop}%
\bibitem [{\citenamefont {Schwab}\ \emph {et~al.}(2022)\citenamefont {Schwab}, \citenamefont {Bernecker}, \citenamefont {Fischer}, \citenamefont {Seyed~Sadjjadi}, \citenamefont {Kober}, \citenamefont {Rinderknecht},\ and\ \citenamefont {Siefkes}}]{Schwab_2022}%
  \BibitemOpen
  \bibfield  {author} {\bibinfo {author} {\bibfnamefont {J.}~\bibnamefont {Schwab}}, \bibinfo {author} {\bibfnamefont {M.}~\bibnamefont {Bernecker}}, \bibinfo {author} {\bibfnamefont {S.}~\bibnamefont {Fischer}}, \bibinfo {author} {\bibfnamefont {B.}~\bibnamefont {Seyed~Sadjjadi}}, \bibinfo {author} {\bibfnamefont {M.}~\bibnamefont {Kober}}, \bibinfo {author} {\bibfnamefont {F.}~\bibnamefont {Rinderknecht}},\ and\ \bibinfo {author} {\bibfnamefont {T.}~\bibnamefont {Siefkes}},\ }\bibfield  {title} {\bibinfo {title} {Exergy {{Analysis}} of the {{Prevailing Residential Heating System}} and {{Derivation}} of {{Future CO2-Reduction Potential}}},\ }\href {https://doi.org/10.3390/en15103502} {\bibfield  {journal} {\bibinfo  {journal} {Energies}\ }\textbf {\bibinfo {volume} {15}},\ \bibinfo {pages} {3502} (\bibinfo {year} {2022})}\BibitemShut {NoStop}%
\bibitem [{\citenamefont {Bell}(2008)}]{Bell_2008}%
  \BibitemOpen
  \bibfield  {author} {\bibinfo {author} {\bibfnamefont {L.~E.}\ \bibnamefont {Bell}},\ }\bibfield  {title} {\bibinfo {title} {Cooling, {{Heating}}, {{Generating Power}}, and {{Recovering Waste Heat}} with {{Thermoelectric Systems}}},\ }\href {https://doi.org/10.1126/science.1158899} {\bibfield  {journal} {\bibinfo  {journal} {Science}\ }\textbf {\bibinfo {volume} {321}},\ \bibinfo {pages} {1457} (\bibinfo {year} {2008})}\BibitemShut {NoStop}%
\bibitem [{\citenamefont {Tran{\aa}s}\ \emph {et~al.}(2022)\citenamefont {Tran{\aa}s}, \citenamefont {L{\o}vvik},\ and\ \citenamefont {Berland}}]{Tranas_2022}%
  \BibitemOpen
  \bibfield  {author} {\bibinfo {author} {\bibfnamefont {R.}~\bibnamefont {Tran{\aa}s}}, \bibinfo {author} {\bibfnamefont {O.~M.}\ \bibnamefont {L{\o}vvik}},\ and\ \bibinfo {author} {\bibfnamefont {K.}~\bibnamefont {Berland}},\ }\bibfield  {title} {\bibinfo {title} {Attaining {{Low Lattice Thermal Conductivity}} in {{Half-Heusler Sublattice Solid Solutions}}: {{Which Substitution Site Is Most Effective}}?},\ }\href {https://doi.org/10.3390/electronicmat3010001} {\bibfield  {journal} {\bibinfo  {journal} {Electronic Materials}\ }\textbf {\bibinfo {volume} {3}},\ \bibinfo {pages} {1} (\bibinfo {year} {2022})}\BibitemShut {NoStop}%
\bibitem [{\citenamefont {Minnich}\ \emph {et~al.}(2009)\citenamefont {Minnich}, \citenamefont {Dresselhaus}, \citenamefont {Ren},\ and\ \citenamefont {Chen}}]{Minnich_2009}%
  \BibitemOpen
  \bibfield  {author} {\bibinfo {author} {\bibfnamefont {A.~J.}\ \bibnamefont {Minnich}}, \bibinfo {author} {\bibfnamefont {M.~S.}\ \bibnamefont {Dresselhaus}}, \bibinfo {author} {\bibfnamefont {Z.~F.}\ \bibnamefont {Ren}},\ and\ \bibinfo {author} {\bibfnamefont {G.}~\bibnamefont {Chen}},\ }\bibfield  {title} {\bibinfo {title} {Bulk nanostructured thermoelectric materials: Current research and future prospects},\ }\href {https://doi.org/10.1039/b822664b} {\bibfield  {journal} {\bibinfo  {journal} {Energy \& Environmental Science}\ }\textbf {\bibinfo {volume} {2}},\ \bibinfo {pages} {466} (\bibinfo {year} {2009})}\BibitemShut {NoStop}%
\bibitem [{\citenamefont {Zhang}\ \emph {et~al.}(2016)\citenamefont {Zhang}, \citenamefont {Song}, \citenamefont {Madsen}, \citenamefont {Fischer}, \citenamefont {Zhang}, \citenamefont {Shi},\ and\ \citenamefont {Iversen}}]{Zhang_2016}%
  \BibitemOpen
  \bibfield  {author} {\bibinfo {author} {\bibfnamefont {J.}~\bibnamefont {Zhang}}, \bibinfo {author} {\bibfnamefont {L.}~\bibnamefont {Song}}, \bibinfo {author} {\bibfnamefont {G.~K.~H.}\ \bibnamefont {Madsen}}, \bibinfo {author} {\bibfnamefont {K.~F.~F.}\ \bibnamefont {Fischer}}, \bibinfo {author} {\bibfnamefont {W.}~\bibnamefont {Zhang}}, \bibinfo {author} {\bibfnamefont {X.}~\bibnamefont {Shi}},\ and\ \bibinfo {author} {\bibfnamefont {B.~B.}\ \bibnamefont {Iversen}},\ }\bibfield  {title} {\bibinfo {title} {Designing high-performance layered thermoelectric materials through orbital engineering},\ }\href {https://doi.org/10.1038/ncomms10892} {\bibfield  {journal} {\bibinfo  {journal} {Nature Communications}\ }\textbf {\bibinfo {volume} {7}},\ \bibinfo {pages} {10892} (\bibinfo {year} {2016})}\BibitemShut {NoStop}%
\bibitem [{\citenamefont {Li}\ \emph {et~al.}(2018)\citenamefont {Li}, \citenamefont {Zhang}, \citenamefont {Wang}, \citenamefont {Liu}, \citenamefont {Yue}, \citenamefont {Lu},\ and\ \citenamefont {Zheng}}]{Li_2018}%
  \BibitemOpen
  \bibfield  {author} {\bibinfo {author} {\bibfnamefont {J.}~\bibnamefont {Li}}, \bibinfo {author} {\bibfnamefont {S.}~\bibnamefont {Zhang}}, \bibinfo {author} {\bibfnamefont {B.}~\bibnamefont {Wang}}, \bibinfo {author} {\bibfnamefont {S.}~\bibnamefont {Liu}}, \bibinfo {author} {\bibfnamefont {L.}~\bibnamefont {Yue}}, \bibinfo {author} {\bibfnamefont {G.}~\bibnamefont {Lu}},\ and\ \bibinfo {author} {\bibfnamefont {S.}~\bibnamefont {Zheng}},\ }\bibfield  {title} {\bibinfo {title} {Designing high-performance n-type {{Mg}}{\textsubscript{3}} {{Sb}}{\textsubscript{2}} -based thermoelectric materials through forming solid solutions and biaxial strain},\ }\href {https://doi.org/10.1039/C8TA07285J} {\bibfield  {journal} {\bibinfo  {journal} {Journal of Materials Chemistry A}\ }\textbf {\bibinfo {volume} {6}},\ \bibinfo {pages} {20454} (\bibinfo {year} {2018})}\BibitemShut {NoStop}%
\bibitem [{\citenamefont {Pei}\ \emph {et~al.}(2011)\citenamefont {Pei}, \citenamefont {Shi}, \citenamefont {LaLonde}, \citenamefont {Wang}, \citenamefont {Chen},\ and\ \citenamefont {Snyder}}]{Pei_2011}%
  \BibitemOpen
  \bibfield  {author} {\bibinfo {author} {\bibfnamefont {Y.}~\bibnamefont {Pei}}, \bibinfo {author} {\bibfnamefont {X.}~\bibnamefont {Shi}}, \bibinfo {author} {\bibfnamefont {A.}~\bibnamefont {LaLonde}}, \bibinfo {author} {\bibfnamefont {H.}~\bibnamefont {Wang}}, \bibinfo {author} {\bibfnamefont {L.}~\bibnamefont {Chen}},\ and\ \bibinfo {author} {\bibfnamefont {G.~J.}\ \bibnamefont {Snyder}},\ }\bibfield  {title} {\bibinfo {title} {Convergence of electronic bands for high performance bulk thermoelectrics},\ }\href {https://doi.org/10.1038/nature09996} {\bibfield  {journal} {\bibinfo  {journal} {Nature}\ }\textbf {\bibinfo {volume} {473}},\ \bibinfo {pages} {66} (\bibinfo {year} {2011})}\BibitemShut {NoStop}%
\bibitem [{\citenamefont {Zhu}\ \emph {et~al.}(2022)\citenamefont {Zhu}, \citenamefont {Wang}, \citenamefont {Hong}, \citenamefont {Hu}, \citenamefont {Ina}, \citenamefont {Zhan}, \citenamefont {Qin}, \citenamefont {Shi}, \citenamefont {Su}, \citenamefont {Gao},\ and\ \citenamefont {Zhao}}]{Zhu_2022}%
  \BibitemOpen
  \bibfield  {author} {\bibinfo {author} {\bibfnamefont {Y.}~\bibnamefont {Zhu}}, \bibinfo {author} {\bibfnamefont {D.}~\bibnamefont {Wang}}, \bibinfo {author} {\bibfnamefont {T.}~\bibnamefont {Hong}}, \bibinfo {author} {\bibfnamefont {L.}~\bibnamefont {Hu}}, \bibinfo {author} {\bibfnamefont {T.}~\bibnamefont {Ina}}, \bibinfo {author} {\bibfnamefont {S.}~\bibnamefont {Zhan}}, \bibinfo {author} {\bibfnamefont {B.}~\bibnamefont {Qin}}, \bibinfo {author} {\bibfnamefont {H.}~\bibnamefont {Shi}}, \bibinfo {author} {\bibfnamefont {L.}~\bibnamefont {Su}}, \bibinfo {author} {\bibfnamefont {X.}~\bibnamefont {Gao}},\ and\ \bibinfo {author} {\bibfnamefont {L.-D.}\ \bibnamefont {Zhao}},\ }\bibfield  {title} {\bibinfo {title} {Multiple valence bands convergence and strong phonon scattering lead to high thermoelectric performance in p-type {{PbSe}}},\ }\href {https://doi.org/10.1038/s41467-022-31939-4} {\bibfield  {journal} {\bibinfo  {journal} {Nature Communications}\ }\textbf {\bibinfo {volume} {13}},\ \bibinfo {pages}
  {4179} (\bibinfo {year} {2022})}\BibitemShut {NoStop}%
\bibitem [{\citenamefont {Bilc}\ \emph {et~al.}(2004)\citenamefont {Bilc}, \citenamefont {Mahanti}, \citenamefont {Quarez}, \citenamefont {Hsu}, \citenamefont {Pcionek},\ and\ \citenamefont {Kanatzidis}}]{Bilc_2004}%
  \BibitemOpen
  \bibfield  {author} {\bibinfo {author} {\bibfnamefont {D.}~\bibnamefont {Bilc}}, \bibinfo {author} {\bibfnamefont {S.~D.}\ \bibnamefont {Mahanti}}, \bibinfo {author} {\bibfnamefont {E.}~\bibnamefont {Quarez}}, \bibinfo {author} {\bibfnamefont {K.-F.}\ \bibnamefont {Hsu}}, \bibinfo {author} {\bibfnamefont {R.}~\bibnamefont {Pcionek}},\ and\ \bibinfo {author} {\bibfnamefont {M.~G.}\ \bibnamefont {Kanatzidis}},\ }\bibfield  {title} {\bibinfo {title} {Resonant {{States}} in the {{Electronic Structure}} of the {{High Performance Thermoelectrics A}} g {{P}} b m {{S}} b {{T}} e 2 + m : {{The Role}} of {{Ag-Sb Microstructures}}},\ }\href {https://doi.org/10.1103/PhysRevLett.93.146403} {\bibfield  {journal} {\bibinfo  {journal} {Physical Review Letters}\ }\textbf {\bibinfo {volume} {93}},\ \bibinfo {pages} {146403} (\bibinfo {year} {2004})}\BibitemShut {NoStop}%
\bibitem [{\citenamefont {Heremans}\ \emph {et~al.}(2008)\citenamefont {Heremans}, \citenamefont {Jovovic}, \citenamefont {Toberer}, \citenamefont {Saramat}, \citenamefont {Kurosaki}, \citenamefont {Charoenphakdee}, \citenamefont {Yamanaka},\ and\ \citenamefont {Snyder}}]{Heremans_2008}%
  \BibitemOpen
  \bibfield  {author} {\bibinfo {author} {\bibfnamefont {J.~P.}\ \bibnamefont {Heremans}}, \bibinfo {author} {\bibfnamefont {V.}~\bibnamefont {Jovovic}}, \bibinfo {author} {\bibfnamefont {E.~S.}\ \bibnamefont {Toberer}}, \bibinfo {author} {\bibfnamefont {A.}~\bibnamefont {Saramat}}, \bibinfo {author} {\bibfnamefont {K.}~\bibnamefont {Kurosaki}}, \bibinfo {author} {\bibfnamefont {A.}~\bibnamefont {Charoenphakdee}}, \bibinfo {author} {\bibfnamefont {S.}~\bibnamefont {Yamanaka}},\ and\ \bibinfo {author} {\bibfnamefont {G.~J.}\ \bibnamefont {Snyder}},\ }\bibfield  {title} {\bibinfo {title} {Enhancement of {{Thermoelectric Efficiency}} in {{PbTe}} by {{Distortion}} of the {{Electronic Density}} of {{States}}},\ }\href {https://doi.org/10.1126/science.1159725} {\bibfield  {journal} {\bibinfo  {journal} {Science}\ }\textbf {\bibinfo {volume} {321}},\ \bibinfo {pages} {554} (\bibinfo {year} {2008})}\BibitemShut {NoStop}%
\bibitem [{\citenamefont {Heremans}\ \emph {et~al.}(2012)\citenamefont {Heremans}, \citenamefont {Wiendlocha},\ and\ \citenamefont {Chamoire}}]{Heremans_2012}%
  \BibitemOpen
  \bibfield  {author} {\bibinfo {author} {\bibfnamefont {J.~P.}\ \bibnamefont {Heremans}}, \bibinfo {author} {\bibfnamefont {B.}~\bibnamefont {Wiendlocha}},\ and\ \bibinfo {author} {\bibfnamefont {A.~M.}\ \bibnamefont {Chamoire}},\ }\bibfield  {title} {\bibinfo {title} {Resonant levels in bulk thermoelectric semiconductors},\ }\href {https://doi.org/10.1039/C1EE02612G} {\bibfield  {journal} {\bibinfo  {journal} {Energy Environ. Sci.}\ }\textbf {\bibinfo {volume} {5}},\ \bibinfo {pages} {5510} (\bibinfo {year} {2012})}\BibitemShut {NoStop}%
\bibitem [{\citenamefont {Faleev}\ and\ \citenamefont {L{\'e}onard}(2008)}]{Faleev_2008}%
  \BibitemOpen
  \bibfield  {author} {\bibinfo {author} {\bibfnamefont {S.~V.}\ \bibnamefont {Faleev}}\ and\ \bibinfo {author} {\bibfnamefont {F.}~\bibnamefont {L{\'e}onard}},\ }\bibfield  {title} {\bibinfo {title} {Theory of enhancement of thermoelectric properties of materials with nanoinclusions},\ }\href {https://doi.org/10.1103/PhysRevB.77.214304} {\bibfield  {journal} {\bibinfo  {journal} {Physical Review B}\ }\textbf {\bibinfo {volume} {77}},\ \bibinfo {pages} {214304} (\bibinfo {year} {2008})}\BibitemShut {NoStop}%
\bibitem [{\citenamefont {Gayner}\ and\ \citenamefont {Amouyal}(2020)}]{Gayner_2020}%
  \BibitemOpen
  \bibfield  {author} {\bibinfo {author} {\bibfnamefont {C.}~\bibnamefont {Gayner}}\ and\ \bibinfo {author} {\bibfnamefont {Y.}~\bibnamefont {Amouyal}},\ }\bibfield  {title} {\bibinfo {title} {Energy {{Filtering}} of {{Charge Carriers}}: {{Current Trends}}, {{Challenges}}, and {{Prospects}} for {{Thermoelectric Materials}}},\ }\href {https://doi.org/10.1002/adfm.201901789} {\bibfield  {journal} {\bibinfo  {journal} {Advanced Functional Materials}\ }\textbf {\bibinfo {volume} {30}},\ \bibinfo {pages} {1901789} (\bibinfo {year} {2020})}\BibitemShut {NoStop}%
\bibitem [{\citenamefont {Lin}\ \emph {et~al.}(2020)\citenamefont {Lin}, \citenamefont {Wood}, \citenamefont {Imasato}, \citenamefont {Kuo}, \citenamefont {Lam}, \citenamefont {Mortazavi}, \citenamefont {Slade}, \citenamefont {Hodge}, \citenamefont {Xi}, \citenamefont {Kanatzidis}, \citenamefont {Clarke}, \citenamefont {Hersam},\ and\ \citenamefont {Snyder}}]{Lin_2020}%
  \BibitemOpen
  \bibfield  {author} {\bibinfo {author} {\bibfnamefont {Y.}~\bibnamefont {Lin}}, \bibinfo {author} {\bibfnamefont {M.}~\bibnamefont {Wood}}, \bibinfo {author} {\bibfnamefont {K.}~\bibnamefont {Imasato}}, \bibinfo {author} {\bibfnamefont {J.~J.}\ \bibnamefont {Kuo}}, \bibinfo {author} {\bibfnamefont {D.}~\bibnamefont {Lam}}, \bibinfo {author} {\bibfnamefont {A.~N.}\ \bibnamefont {Mortazavi}}, \bibinfo {author} {\bibfnamefont {T.~J.}\ \bibnamefont {Slade}}, \bibinfo {author} {\bibfnamefont {S.~A.}\ \bibnamefont {Hodge}}, \bibinfo {author} {\bibfnamefont {K.}~\bibnamefont {Xi}}, \bibinfo {author} {\bibfnamefont {M.~G.}\ \bibnamefont {Kanatzidis}}, \bibinfo {author} {\bibfnamefont {D.~R.}\ \bibnamefont {Clarke}}, \bibinfo {author} {\bibfnamefont {M.~C.}\ \bibnamefont {Hersam}},\ and\ \bibinfo {author} {\bibfnamefont {G.~J.}\ \bibnamefont {Snyder}},\ }\bibfield  {title} {\bibinfo {title} {Expression of interfacial {{Seebeck}} coefficient through grain boundary engineering with multi-layer graphene
  nanoplatelets},\ }\href {https://doi.org/10.1039/D0EE02490B} {\bibfield  {journal} {\bibinfo  {journal} {Energy \& Environmental Science}\ }\textbf {\bibinfo {volume} {13}},\ \bibinfo {pages} {4114} (\bibinfo {year} {2020})}\BibitemShut {NoStop}%
\bibitem [{\citenamefont {Hicks}\ and\ \citenamefont {Dresselhaus}(1993{\natexlab{a}})}]{Hicks_1993}%
  \BibitemOpen
  \bibfield  {author} {\bibinfo {author} {\bibfnamefont {L.~D.}\ \bibnamefont {Hicks}}\ and\ \bibinfo {author} {\bibfnamefont {M.~S.}\ \bibnamefont {Dresselhaus}},\ }\bibfield  {title} {\bibinfo {title} {Effect of quantum-well structures on the thermoelectric figure of merit},\ }\href {https://doi.org/10.1103/PhysRevB.47.12727} {\bibfield  {journal} {\bibinfo  {journal} {Physical Review B}\ }\textbf {\bibinfo {volume} {47}},\ \bibinfo {pages} {12727} (\bibinfo {year} {1993}{\natexlab{a}})}\BibitemShut {NoStop}%
\bibitem [{\citenamefont {Hicks}\ and\ \citenamefont {Dresselhaus}(1993{\natexlab{b}})}]{Hicks_1993a}%
  \BibitemOpen
  \bibfield  {author} {\bibinfo {author} {\bibfnamefont {L.~D.}\ \bibnamefont {Hicks}}\ and\ \bibinfo {author} {\bibfnamefont {M.~S.}\ \bibnamefont {Dresselhaus}},\ }\bibfield  {title} {\bibinfo {title} {Thermoelectric figure of merit of a one-dimensional conductor},\ }\href {https://doi.org/10.1103/PhysRevB.47.16631} {\bibfield  {journal} {\bibinfo  {journal} {Physical Review B}\ }\textbf {\bibinfo {volume} {47}},\ \bibinfo {pages} {16631} (\bibinfo {year} {1993}{\natexlab{b}})}\BibitemShut {NoStop}%
\bibitem [{\citenamefont {Cornett}\ and\ \citenamefont {Rabin}(2011{\natexlab{a}})}]{Cornett_2011}%
  \BibitemOpen
  \bibfield  {author} {\bibinfo {author} {\bibfnamefont {J.~E.}\ \bibnamefont {Cornett}}\ and\ \bibinfo {author} {\bibfnamefont {O.}~\bibnamefont {Rabin}},\ }\bibfield  {title} {\bibinfo {title} {Thermoelectric figure of merit calculations for semiconducting nanowires},\ }\href {https://doi.org/10.1063/1.3585659} {\bibfield  {journal} {\bibinfo  {journal} {Applied Physics Letters}\ }\textbf {\bibinfo {volume} {98}},\ \bibinfo {pages} {182104} (\bibinfo {year} {2011}{\natexlab{a}})}\BibitemShut {NoStop}%
\bibitem [{\citenamefont {Cornett}\ and\ \citenamefont {Rabin}(2011{\natexlab{b}})}]{Cornett_2011a}%
  \BibitemOpen
  \bibfield  {author} {\bibinfo {author} {\bibfnamefont {J.~E.}\ \bibnamefont {Cornett}}\ and\ \bibinfo {author} {\bibfnamefont {O.}~\bibnamefont {Rabin}},\ }\bibfield  {title} {\bibinfo {title} {Universal scaling relations for the thermoelectric power factor of semiconducting nanostructures},\ }\href {https://doi.org/10.1103/PhysRevB.84.205410} {\bibfield  {journal} {\bibinfo  {journal} {Physical Review B}\ }\textbf {\bibinfo {volume} {84}},\ \bibinfo {pages} {205410} (\bibinfo {year} {2011}{\natexlab{b}})}\BibitemShut {NoStop}%
\bibitem [{\citenamefont {Yang}\ \emph {et~al.}(2018)\citenamefont {Yang}, \citenamefont {Chen}, \citenamefont {Dargusch},\ and\ \citenamefont {Zou}}]{Yang_2018}%
  \BibitemOpen
  \bibfield  {author} {\bibinfo {author} {\bibfnamefont {L.}~\bibnamefont {Yang}}, \bibinfo {author} {\bibfnamefont {Z.-G.}\ \bibnamefont {Chen}}, \bibinfo {author} {\bibfnamefont {M.~S.}\ \bibnamefont {Dargusch}},\ and\ \bibinfo {author} {\bibfnamefont {J.}~\bibnamefont {Zou}},\ }\bibfield  {title} {\bibinfo {title} {High {{Performance Thermoelectric Materials}}: {{Progress}} and {{Their Applications}}},\ }\href {https://doi.org/10.1002/aenm.201701797} {\bibfield  {journal} {\bibinfo  {journal} {Advanced Energy Materials}\ }\textbf {\bibinfo {volume} {8}},\ \bibinfo {pages} {1701797} (\bibinfo {year} {2018})}\BibitemShut {NoStop}%
\bibitem [{\citenamefont {Parker}\ \emph {et~al.}(2013)\citenamefont {Parker}, \citenamefont {Chen},\ and\ \citenamefont {Singh}}]{Parker_2013}%
  \BibitemOpen
  \bibfield  {author} {\bibinfo {author} {\bibfnamefont {D.}~\bibnamefont {Parker}}, \bibinfo {author} {\bibfnamefont {X.}~\bibnamefont {Chen}},\ and\ \bibinfo {author} {\bibfnamefont {D.~J.}\ \bibnamefont {Singh}},\ }\bibfield  {title} {\bibinfo {title} {High {{Three-Dimensional Thermoelectric Performance}} from {{Low-Dimensional Bands}}},\ }\href {https://doi.org/10.1103/PhysRevLett.110.146601} {\bibfield  {journal} {\bibinfo  {journal} {Physical Review Letters}\ }\textbf {\bibinfo {volume} {110}},\ \bibinfo {pages} {146601} (\bibinfo {year} {2013})}\BibitemShut {NoStop}%
\bibitem [{\citenamefont {Bilc}\ \emph {et~al.}(2015)\citenamefont {Bilc}, \citenamefont {Hautier}, \citenamefont {Waroquiers}, \citenamefont {Rignanese},\ and\ \citenamefont {Ghosez}}]{Bilc_2015}%
  \BibitemOpen
  \bibfield  {author} {\bibinfo {author} {\bibfnamefont {D.~I.}\ \bibnamefont {Bilc}}, \bibinfo {author} {\bibfnamefont {G.}~\bibnamefont {Hautier}}, \bibinfo {author} {\bibfnamefont {D.}~\bibnamefont {Waroquiers}}, \bibinfo {author} {\bibfnamefont {G.-M.}\ \bibnamefont {Rignanese}},\ and\ \bibinfo {author} {\bibfnamefont {P.}~\bibnamefont {Ghosez}},\ }\bibfield  {title} {\bibinfo {title} {Low-{{Dimensional Transport}} and {{Large Thermoelectric Power Factors}} in {{Bulk Semiconductors}} by {{Band Engineering}} of {{Highly Directional Electronic States}}},\ }\href {https://doi.org/10.1103/PhysRevLett.114.136601} {\bibfield  {journal} {\bibinfo  {journal} {Physical Review Letters}\ }\textbf {\bibinfo {volume} {114}},\ \bibinfo {pages} {136601} (\bibinfo {year} {2015})}\BibitemShut {NoStop}%
\bibitem [{\citenamefont {Dylla}\ \emph {et~al.}(2019)\citenamefont {Dylla}, \citenamefont {Kang},\ and\ \citenamefont {Snyder}}]{Dylla_2019}%
  \BibitemOpen
  \bibfield  {author} {\bibinfo {author} {\bibfnamefont {M.~T.}\ \bibnamefont {Dylla}}, \bibinfo {author} {\bibfnamefont {S.~D.}\ \bibnamefont {Kang}},\ and\ \bibinfo {author} {\bibfnamefont {G.~J.}\ \bibnamefont {Snyder}},\ }\bibfield  {title} {\bibinfo {title} {Effect of {{Two}}-{{Dimensional Crystal Orbitals}} on {{Fermi Surfaces}} and {{Electron Transport}} in {{Three}}-{{Dimensional Perovskite Oxides}}},\ }\href {https://doi.org/10.1002/anie.201812230} {\bibfield  {journal} {\bibinfo  {journal} {Angewandte Chemie International Edition}\ }\textbf {\bibinfo {volume} {58}},\ \bibinfo {pages} {5503} (\bibinfo {year} {2019})}\BibitemShut {NoStop}%
\bibitem [{\citenamefont {Park}\ \emph {et~al.}(2021{\natexlab{a}})\citenamefont {Park}, \citenamefont {Xia}, \citenamefont {Ozoli{\c n}{\v s}},\ and\ \citenamefont {Jain}}]{Park_2021}%
  \BibitemOpen
  \bibfield  {author} {\bibinfo {author} {\bibfnamefont {J.}~\bibnamefont {Park}}, \bibinfo {author} {\bibfnamefont {Y.}~\bibnamefont {Xia}}, \bibinfo {author} {\bibfnamefont {V.}~\bibnamefont {Ozoli{\c n}{\v s}}},\ and\ \bibinfo {author} {\bibfnamefont {A.}~\bibnamefont {Jain}},\ }\bibfield  {title} {\bibinfo {title} {Optimal band structure for thermoelectrics with realistic scattering and bands},\ }\href {https://doi.org/10.1038/s41524-021-00512-w} {\bibfield  {journal} {\bibinfo  {journal} {npj Computational Materials}\ }\textbf {\bibinfo {volume} {7}},\ \bibinfo {pages} {43} (\bibinfo {year} {2021}{\natexlab{a}})}\BibitemShut {NoStop}%
\bibitem [{\citenamefont {Brod}\ and\ \citenamefont {Snyder}(2021)}]{Brod_2021}%
  \BibitemOpen
  \bibfield  {author} {\bibinfo {author} {\bibfnamefont {M.~K.}\ \bibnamefont {Brod}}\ and\ \bibinfo {author} {\bibfnamefont {G.~J.}\ \bibnamefont {Snyder}},\ }\bibfield  {title} {\bibinfo {title} {Orbital chemistry of high valence band convergence and low-dimensional topology in {{PbTe}}},\ }\href {https://doi.org/10.1039/D1TA01273H} {\bibfield  {journal} {\bibinfo  {journal} {Journal of Materials Chemistry A}\ }\textbf {\bibinfo {volume} {9}},\ \bibinfo {pages} {12119} (\bibinfo {year} {2021})}\BibitemShut {NoStop}%
\bibitem [{\citenamefont {Madsen}(2006)}]{Madsen_2006}%
  \BibitemOpen
  \bibfield  {author} {\bibinfo {author} {\bibfnamefont {G.~K.~H.}\ \bibnamefont {Madsen}},\ }\bibfield  {title} {\bibinfo {title} {Automated {{Search}} for {{New Thermoelectric Materials}}: {{The Case}} of {{LiZnSb}}},\ }\href {https://doi.org/10.1021/ja062526a} {\bibfield  {journal} {\bibinfo  {journal} {Journal of the American Chemical Society}\ }\textbf {\bibinfo {volume} {128}},\ \bibinfo {pages} {12140} (\bibinfo {year} {2006})}\BibitemShut {NoStop}%
\bibitem [{\citenamefont {Wang}\ \emph {et~al.}(2011)\citenamefont {Wang}, \citenamefont {Wang}, \citenamefont {Setyawan}, \citenamefont {Mingo},\ and\ \citenamefont {Curtarolo}}]{Wang_2011}%
  \BibitemOpen
  \bibfield  {author} {\bibinfo {author} {\bibfnamefont {S.}~\bibnamefont {Wang}}, \bibinfo {author} {\bibfnamefont {Z.}~\bibnamefont {Wang}}, \bibinfo {author} {\bibfnamefont {W.}~\bibnamefont {Setyawan}}, \bibinfo {author} {\bibfnamefont {N.}~\bibnamefont {Mingo}},\ and\ \bibinfo {author} {\bibfnamefont {S.}~\bibnamefont {Curtarolo}},\ }\bibfield  {title} {\bibinfo {title} {Assessing the {{Thermoelectric Properties}} of {{Sintered Compounds}} via {{High-Throughput}} {{{\emph{Ab-Initio}}}} {{Calculations}}},\ }\href {https://doi.org/10.1103/PhysRevX.1.021012} {\bibfield  {journal} {\bibinfo  {journal} {Physical Review X}\ }\textbf {\bibinfo {volume} {1}},\ \bibinfo {pages} {021012} (\bibinfo {year} {2011})}\BibitemShut {NoStop}%
\bibitem [{\citenamefont {Bhattacharya}\ and\ \citenamefont {Madsen}(2015)}]{Bhattacharya_2015}%
  \BibitemOpen
  \bibfield  {author} {\bibinfo {author} {\bibfnamefont {S.}~\bibnamefont {Bhattacharya}}\ and\ \bibinfo {author} {\bibfnamefont {G.~K.~H.}\ \bibnamefont {Madsen}},\ }\bibfield  {title} {\bibinfo {title} {High-throughput exploration of alloying as design strategy for thermoelectrics},\ }\href {https://doi.org/10.1103/PhysRevB.92.085205} {\bibfield  {journal} {\bibinfo  {journal} {Physical Review B}\ }\textbf {\bibinfo {volume} {92}},\ \bibinfo {pages} {085205} (\bibinfo {year} {2015})}\BibitemShut {NoStop}%
\bibitem [{\citenamefont {Ricci}\ \emph {et~al.}(2017)\citenamefont {Ricci}, \citenamefont {Chen}, \citenamefont {Aydemir}, \citenamefont {Snyder}, \citenamefont {Rignanese}, \citenamefont {Jain},\ and\ \citenamefont {Hautier}}]{Ricci_2017}%
  \BibitemOpen
  \bibfield  {author} {\bibinfo {author} {\bibfnamefont {F.}~\bibnamefont {Ricci}}, \bibinfo {author} {\bibfnamefont {W.}~\bibnamefont {Chen}}, \bibinfo {author} {\bibfnamefont {U.}~\bibnamefont {Aydemir}}, \bibinfo {author} {\bibfnamefont {G.~J.}\ \bibnamefont {Snyder}}, \bibinfo {author} {\bibfnamefont {G.-M.}\ \bibnamefont {Rignanese}}, \bibinfo {author} {\bibfnamefont {A.}~\bibnamefont {Jain}},\ and\ \bibinfo {author} {\bibfnamefont {G.}~\bibnamefont {Hautier}},\ }\bibfield  {title} {\bibinfo {title} {An ab initio electronic transport database for inorganic materials},\ }\href {https://doi.org/10.1038/sdata.2017.85} {\bibfield  {journal} {\bibinfo  {journal} {Scientific Data}\ }\textbf {\bibinfo {volume} {4}},\ \bibinfo {pages} {170085} (\bibinfo {year} {2017})}\BibitemShut {NoStop}%
\bibitem [{\citenamefont {Berland}\ \emph {et~al.}(2019)\citenamefont {Berland}, \citenamefont {Shulumba}, \citenamefont {Hellman}, \citenamefont {Persson},\ and\ \citenamefont {L{\o}vvik}}]{Berland_2019}%
  \BibitemOpen
  \bibfield  {author} {\bibinfo {author} {\bibfnamefont {K.}~\bibnamefont {Berland}}, \bibinfo {author} {\bibfnamefont {N.}~\bibnamefont {Shulumba}}, \bibinfo {author} {\bibfnamefont {O.}~\bibnamefont {Hellman}}, \bibinfo {author} {\bibfnamefont {C.}~\bibnamefont {Persson}},\ and\ \bibinfo {author} {\bibfnamefont {O.~M.}\ \bibnamefont {L{\o}vvik}},\ }\bibfield  {title} {\bibinfo {title} {Thermoelectric transport trends in group 4 half-{{Heusler}} alloys},\ }\href {https://doi.org/10.1063/1.5117288} {\bibfield  {journal} {\bibinfo  {journal} {Journal of Applied Physics}\ }\textbf {\bibinfo {volume} {126}},\ \bibinfo {pages} {145102} (\bibinfo {year} {2019})}\BibitemShut {NoStop}%
\bibitem [{\citenamefont {Berland}\ \emph {et~al.}(2021)\citenamefont {Berland}, \citenamefont {L{\o}vvik},\ and\ \citenamefont {Tran{\aa}s}}]{Berland_2021}%
  \BibitemOpen
  \bibfield  {author} {\bibinfo {author} {\bibfnamefont {K.}~\bibnamefont {Berland}}, \bibinfo {author} {\bibfnamefont {O.~M.}\ \bibnamefont {L{\o}vvik}},\ and\ \bibinfo {author} {\bibfnamefont {R.}~\bibnamefont {Tran{\aa}s}},\ }\bibfield  {title} {\bibinfo {title} {Discarded gems: {{Thermoelectric}} performance of materials with band gap emerging at the hybrid-functional level},\ }\href {https://doi.org/10.1063/5.0058685} {\bibfield  {journal} {\bibinfo  {journal} {Applied Physics Letters}\ }\textbf {\bibinfo {volume} {119}},\ \bibinfo {pages} {081902} (\bibinfo {year} {2021})}\BibitemShut {NoStop}%
\bibitem [{\citenamefont {Park}\ \emph {et~al.}(2019)\citenamefont {Park}, \citenamefont {Xia},\ and\ \citenamefont {Ozoli{\c n}{\v s}}}]{Park_2019}%
  \BibitemOpen
  \bibfield  {author} {\bibinfo {author} {\bibfnamefont {J.}~\bibnamefont {Park}}, \bibinfo {author} {\bibfnamefont {Y.}~\bibnamefont {Xia}},\ and\ \bibinfo {author} {\bibfnamefont {V.}~\bibnamefont {Ozoli{\c n}{\v s}}},\ }\bibfield  {title} {\bibinfo {title} {High {{Thermoelectric Power Factor}} and {{Efficiency}} from a {{Highly Dispersive Band}} in {{Ba}} 2 {{Bi Au}}},\ }\href {https://doi.org/10.1103/PhysRevApplied.11.014058} {\bibfield  {journal} {\bibinfo  {journal} {Physical Review Applied}\ }\textbf {\bibinfo {volume} {11}},\ \bibinfo {pages} {014058} (\bibinfo {year} {2019})}\BibitemShut {NoStop}%
\bibitem [{\citenamefont {Park}\ \emph {et~al.}(2021{\natexlab{b}})\citenamefont {Park}, \citenamefont {Dylla}, \citenamefont {Xia}, \citenamefont {Wood}, \citenamefont {Snyder},\ and\ \citenamefont {Jain}}]{Park_2021a}%
  \BibitemOpen
  \bibfield  {author} {\bibinfo {author} {\bibfnamefont {J.}~\bibnamefont {Park}}, \bibinfo {author} {\bibfnamefont {M.}~\bibnamefont {Dylla}}, \bibinfo {author} {\bibfnamefont {Y.}~\bibnamefont {Xia}}, \bibinfo {author} {\bibfnamefont {M.}~\bibnamefont {Wood}}, \bibinfo {author} {\bibfnamefont {G.~J.}\ \bibnamefont {Snyder}},\ and\ \bibinfo {author} {\bibfnamefont {A.}~\bibnamefont {Jain}},\ }\bibfield  {title} {\bibinfo {title} {When band convergence is not beneficial for thermoelectrics},\ }\href {https://doi.org/10.1038/s41467-021-23839-w} {\bibfield  {journal} {\bibinfo  {journal} {Nature Communications}\ }\textbf {\bibinfo {volume} {12}},\ \bibinfo {pages} {3425} (\bibinfo {year} {2021}{\natexlab{b}})}\BibitemShut {NoStop}%
\bibitem [{\citenamefont {Li}\ \emph {et~al.}(2024)\citenamefont {Li}, \citenamefont {Graziosi},\ and\ \citenamefont {Neophytou}}]{Li_2024}%
  \BibitemOpen
  \bibfield  {author} {\bibinfo {author} {\bibfnamefont {Z.}~\bibnamefont {Li}}, \bibinfo {author} {\bibfnamefont {P.}~\bibnamefont {Graziosi}},\ and\ \bibinfo {author} {\bibfnamefont {N.}~\bibnamefont {Neophytou}},\ }\bibfield  {title} {\bibinfo {title} {Efficient first-principles electronic transport approach to complex band structure materials: The case of n-type {{Mg3Sb2}}},\ }\href {https://doi.org/10.1038/s41524-023-01192-4} {\bibfield  {journal} {\bibinfo  {journal} {npj Computational Materials}\ }\textbf {\bibinfo {volume} {10}},\ \bibinfo {pages} {8} (\bibinfo {year} {2024})}\BibitemShut {NoStop}%
\bibitem [{\citenamefont {Ji}\ \emph {et~al.}(2022)\citenamefont {Ji}, \citenamefont {Tang}, \citenamefont {Yao}, \citenamefont {Yang}, \citenamefont {Jin}, \citenamefont {Zhang}, \citenamefont {Xi}, \citenamefont {Singh}, \citenamefont {Yang},\ and\ \citenamefont {Zhang}}]{Ji_2022}%
  \BibitemOpen
  \bibfield  {author} {\bibinfo {author} {\bibfnamefont {J.}~\bibnamefont {Ji}}, \bibinfo {author} {\bibfnamefont {Q.}~\bibnamefont {Tang}}, \bibinfo {author} {\bibfnamefont {M.}~\bibnamefont {Yao}}, \bibinfo {author} {\bibfnamefont {H.}~\bibnamefont {Yang}}, \bibinfo {author} {\bibfnamefont {Y.}~\bibnamefont {Jin}}, \bibinfo {author} {\bibfnamefont {Y.}~\bibnamefont {Zhang}}, \bibinfo {author} {\bibfnamefont {J.}~\bibnamefont {Xi}}, \bibinfo {author} {\bibfnamefont {D.~J.}\ \bibnamefont {Singh}}, \bibinfo {author} {\bibfnamefont {J.}~\bibnamefont {Yang}},\ and\ \bibinfo {author} {\bibfnamefont {W.}~\bibnamefont {Zhang}},\ }\bibfield  {title} {\bibinfo {title} {Functional-{{Unit-Based Material Design}}: {{Ultralow Thermal Conductivity}} in {{Thermoelectrics}} with {{Linear Triatomic Resonant Bonds}}},\ }\href {https://doi.org/10.1021/jacs.2c08062} {\bibfield  {journal} {\bibinfo  {journal} {Journal of the American Chemical Society}\ }\textbf {\bibinfo {volume} {144}},\ \bibinfo {pages} {18552} (\bibinfo
  {year} {2022})}\BibitemShut {NoStop}%
\bibitem [{\citenamefont {Ettema}\ and\ \citenamefont {De~Groot}(2000)}]{Ettema_2000}%
  \BibitemOpen
  \bibfield  {author} {\bibinfo {author} {\bibfnamefont {A.~R. H.~F.}\ \bibnamefont {Ettema}}\ and\ \bibinfo {author} {\bibfnamefont {R.~A.}\ \bibnamefont {De~Groot}},\ }\bibfield  {title} {\bibinfo {title} {Electronic structure of {{Na}} 3 {{Sb}} and {{Na}} 2 {{KSb}}},\ }\href {https://doi.org/10.1103/PhysRevB.61.10035} {\bibfield  {journal} {\bibinfo  {journal} {Physical Review B}\ }\textbf {\bibinfo {volume} {61}},\ \bibinfo {pages} {10035} (\bibinfo {year} {2000})}\BibitemShut {NoStop}%
\bibitem [{\citenamefont {Kalarasse}\ \emph {et~al.}(2010)\citenamefont {Kalarasse}, \citenamefont {Bennecer}, \citenamefont {Kalarasse},\ and\ \citenamefont {Djeroud}}]{Kalarasse_2010}%
  \BibitemOpen
  \bibfield  {author} {\bibinfo {author} {\bibfnamefont {L.}~\bibnamefont {Kalarasse}}, \bibinfo {author} {\bibfnamefont {B.}~\bibnamefont {Bennecer}}, \bibinfo {author} {\bibfnamefont {F.}~\bibnamefont {Kalarasse}},\ and\ \bibinfo {author} {\bibfnamefont {S.}~\bibnamefont {Djeroud}},\ }\bibfield  {title} {\bibinfo {title} {Pressure effect on the electronic and optical properties of the alkali antimonide semiconductors {{Cs3Sb}}, {{KCs2Sb}}, {{CsK2Sb}} and {{K3Sb}}: {{Ab}} initio study},\ }\href {https://doi.org/10.1016/j.jpcs.2010.09.007} {\bibfield  {journal} {\bibinfo  {journal} {Journal of Physics and Chemistry of Solids}\ }\textbf {\bibinfo {volume} {71}},\ \bibinfo {pages} {1732} (\bibinfo {year} {2010})}\BibitemShut {NoStop}%
\bibitem [{\citenamefont {Liu}\ \emph {et~al.}(2024)\citenamefont {Liu}, \citenamefont {Zhao}, \citenamefont {Ni},\ and\ \citenamefont {Dai}}]{Liu_2024a}%
  \BibitemOpen
  \bibfield  {author} {\bibinfo {author} {\bibfnamefont {P.}~\bibnamefont {Liu}}, \bibinfo {author} {\bibfnamefont {Y.}~\bibnamefont {Zhao}}, \bibinfo {author} {\bibfnamefont {J.}~\bibnamefont {Ni}},\ and\ \bibinfo {author} {\bibfnamefont {Z.}~\bibnamefont {Dai}},\ }\bibfield  {title} {\bibinfo {title} {Influence of anharmonicity on the thermoelectric properties of alkali antimonide compounds {{M}} 3 {{Sb}} ( {{M}} = {{Na}} , {{K}} , {{Rb}} , {{Cs}} )},\ }\href {https://doi.org/10.1103/PhysRevApplied.22.034049} {\bibfield  {journal} {\bibinfo  {journal} {Physical Review Applied}\ }\textbf {\bibinfo {volume} {22}},\ \bibinfo {pages} {034049} (\bibinfo {year} {2024})}\BibitemShut {NoStop}%
\bibitem [{\citenamefont {Sommer}(1963)}]{Sommer_1963}%
  \BibitemOpen
  \bibfield  {author} {\bibinfo {author} {\bibfnamefont {A.~H.}\ \bibnamefont {Sommer}},\ }\bibfield  {title} {\bibinfo {title} {A {{NEW ALKALI ANTIMONIDE PHOTOEMITTER WITH HIGH SENSITIVITY TO VISIBLE LIGHT}}},\ }\href {https://doi.org/10.1063/1.1753869} {\bibfield  {journal} {\bibinfo  {journal} {Applied Physics Letters}\ }\textbf {\bibinfo {volume} {3}},\ \bibinfo {pages} {62} (\bibinfo {year} {1963})}\BibitemShut {NoStop}%
\bibitem [{\citenamefont {Ghosh}\ and\ \citenamefont {Varma}(1978)}]{Ghosh_1978}%
  \BibitemOpen
  \bibfield  {author} {\bibinfo {author} {\bibfnamefont {C.}~\bibnamefont {Ghosh}}\ and\ \bibinfo {author} {\bibfnamefont {B.~P.}\ \bibnamefont {Varma}},\ }\bibfield  {title} {\bibinfo {title} {Preparation and study of properties of a few alkali antimonide photocathodes},\ }\href {https://doi.org/10.1063/1.325465} {\bibfield  {journal} {\bibinfo  {journal} {Journal of Applied Physics}\ }\textbf {\bibinfo {volume} {49}},\ \bibinfo {pages} {4549} (\bibinfo {year} {1978})}\BibitemShut {NoStop}%
\bibitem [{\citenamefont {Yue}\ \emph {et~al.}(2022)\citenamefont {Yue}, \citenamefont {Sui}, \citenamefont {Zhao}, \citenamefont {Ni}, \citenamefont {Meng},\ and\ \citenamefont {Dai}}]{Yue_2022}%
  \BibitemOpen
  \bibfield  {author} {\bibinfo {author} {\bibfnamefont {T.}~\bibnamefont {Yue}}, \bibinfo {author} {\bibfnamefont {P.}~\bibnamefont {Sui}}, \bibinfo {author} {\bibfnamefont {Y.}~\bibnamefont {Zhao}}, \bibinfo {author} {\bibfnamefont {J.}~\bibnamefont {Ni}}, \bibinfo {author} {\bibfnamefont {S.}~\bibnamefont {Meng}},\ and\ \bibinfo {author} {\bibfnamefont {Z.}~\bibnamefont {Dai}},\ }\bibfield  {title} {\bibinfo {title} {Theoretical prediction of mechanics, transport, and thermoelectric properties of full {{Heusler}} compounds {{Na}} 2 {{KSb}} and {{X}} 2 {{CsSb}} ( {{X}} = {{K}} , {{Rb}} )},\ }\href {https://doi.org/10.1103/PhysRevB.105.184304} {\bibfield  {journal} {\bibinfo  {journal} {Physical Review B}\ }\textbf {\bibinfo {volume} {105}},\ \bibinfo {pages} {184304} (\bibinfo {year} {2022})}\BibitemShut {NoStop}%
\bibitem [{\citenamefont {Yuan}\ \emph {et~al.}(2022)\citenamefont {Yuan}, \citenamefont {Zhang}, \citenamefont {Chang}, \citenamefont {Tang},\ and\ \citenamefont {Hu}}]{Yuan_2022}%
  \BibitemOpen
  \bibfield  {author} {\bibinfo {author} {\bibfnamefont {K.}~\bibnamefont {Yuan}}, \bibinfo {author} {\bibfnamefont {X.}~\bibnamefont {Zhang}}, \bibinfo {author} {\bibfnamefont {Z.}~\bibnamefont {Chang}}, \bibinfo {author} {\bibfnamefont {D.}~\bibnamefont {Tang}},\ and\ \bibinfo {author} {\bibfnamefont {M.}~\bibnamefont {Hu}},\ }\bibfield  {title} {\bibinfo {title} {Antibonding induced anharmonicity leading to ultralow lattice thermal conductivity and extraordinary thermoelectric performance in {{CsK}} {\textsubscript{2}} {{X}} ({{X}} = {{Sb}}, {{Bi}})},\ }\href {https://doi.org/10.1039/D2TC03356A} {\bibfield  {journal} {\bibinfo  {journal} {Journal of Materials Chemistry C}\ }\textbf {\bibinfo {volume} {10}},\ \bibinfo {pages} {15822} (\bibinfo {year} {2022})}\BibitemShut {NoStop}%
\bibitem [{\citenamefont {Singh}\ \emph {et~al.}(2022)\citenamefont {Singh}, \citenamefont {Singh}, \citenamefont {Zeeshan}, \citenamefont {Van Den~Brink},\ and\ \citenamefont {Kandpal}}]{Singh_2022}%
  \BibitemOpen
  \bibfield  {author} {\bibinfo {author} {\bibfnamefont {U.}~\bibnamefont {Singh}}, \bibinfo {author} {\bibfnamefont {S.}~\bibnamefont {Singh}}, \bibinfo {author} {\bibfnamefont {M.}~\bibnamefont {Zeeshan}}, \bibinfo {author} {\bibfnamefont {J.}~\bibnamefont {Van Den~Brink}},\ and\ \bibinfo {author} {\bibfnamefont {H.~C.}\ \bibnamefont {Kandpal}},\ }\bibfield  {title} {\bibinfo {title} {Low lattice thermal conductivity in alkali metal based {{Heusler}} alloys},\ }\href {https://doi.org/10.1103/PhysRevMaterials.6.125401} {\bibfield  {journal} {\bibinfo  {journal} {Physical Review Materials}\ }\textbf {\bibinfo {volume} {6}},\ \bibinfo {pages} {125401} (\bibinfo {year} {2022})}\BibitemShut {NoStop}%
\bibitem [{\citenamefont {Sharma}\ \emph {et~al.}(2023)\citenamefont {Sharma}, \citenamefont {Sajjad},\ and\ \citenamefont {Singh}}]{Sharma_2023}%
  \BibitemOpen
  \bibfield  {author} {\bibinfo {author} {\bibfnamefont {G.}~\bibnamefont {Sharma}}, \bibinfo {author} {\bibfnamefont {M.}~\bibnamefont {Sajjad}},\ and\ \bibinfo {author} {\bibfnamefont {N.}~\bibnamefont {Singh}},\ }\bibfield  {title} {\bibinfo {title} {Impressive {{Electronic}} and {{Thermal Transports}} in {{CsK}} {\textsubscript{2}} {{Sb}}: {{A Thermoelectric Perspective}}},\ }\href {https://doi.org/10.1021/acsaem.3c02024} {\bibfield  {journal} {\bibinfo  {journal} {ACS Applied Energy Materials}\ }\textbf {\bibinfo {volume} {6}},\ \bibinfo {pages} {11179} (\bibinfo {year} {2023})}\BibitemShut {NoStop}%
\bibitem [{\citenamefont {Ganose}\ \emph {et~al.}(2021{\natexlab{a}})\citenamefont {Ganose}, \citenamefont {Park}, \citenamefont {Faghaninia}, \citenamefont {{Woods-Robinson}}, \citenamefont {Persson},\ and\ \citenamefont {Jain}}]{Ganose_2021}%
  \BibitemOpen
  \bibfield  {author} {\bibinfo {author} {\bibfnamefont {A.~M.}\ \bibnamefont {Ganose}}, \bibinfo {author} {\bibfnamefont {J.}~\bibnamefont {Park}}, \bibinfo {author} {\bibfnamefont {A.}~\bibnamefont {Faghaninia}}, \bibinfo {author} {\bibfnamefont {R.}~\bibnamefont {{Woods-Robinson}}}, \bibinfo {author} {\bibfnamefont {K.~A.}\ \bibnamefont {Persson}},\ and\ \bibinfo {author} {\bibfnamefont {A.}~\bibnamefont {Jain}},\ }\bibfield  {title} {\bibinfo {title} {Efficient calculation of carrier scattering rates from first principles},\ }\href {https://doi.org/10.1038/s41467-021-22440-5} {\bibfield  {journal} {\bibinfo  {journal} {Nature Communications}\ }\textbf {\bibinfo {volume} {12}},\ \bibinfo {pages} {2222} (\bibinfo {year} {2021}{\natexlab{a}})}\BibitemShut {NoStop}%
\bibitem [{\citenamefont {Madsen}\ \emph {et~al.}(2018)\citenamefont {Madsen}, \citenamefont {Carrete},\ and\ \citenamefont {Verstraete}}]{Madsen_2018}%
  \BibitemOpen
  \bibfield  {author} {\bibinfo {author} {\bibfnamefont {G.~K.}\ \bibnamefont {Madsen}}, \bibinfo {author} {\bibfnamefont {J.}~\bibnamefont {Carrete}},\ and\ \bibinfo {author} {\bibfnamefont {M.~J.}\ \bibnamefont {Verstraete}},\ }\bibfield  {title} {\bibinfo {title} {{{BoltzTraP2}}, a program for interpolating band structures and calculating semi-classical transport coefficients},\ }\href {https://doi.org/10.1016/j.cpc.2018.05.010} {\bibfield  {journal} {\bibinfo  {journal} {Computer Physics Communications}\ }\textbf {\bibinfo {volume} {231}},\ \bibinfo {pages} {140} (\bibinfo {year} {2018})}\BibitemShut {NoStop}%
\bibitem [{\citenamefont {Kresse}\ and\ \citenamefont {Hafner}(1993)}]{Kresse_1993}%
  \BibitemOpen
  \bibfield  {author} {\bibinfo {author} {\bibfnamefont {G.}~\bibnamefont {Kresse}}\ and\ \bibinfo {author} {\bibfnamefont {J.}~\bibnamefont {Hafner}},\ }\bibfield  {title} {\bibinfo {title} {{\emph{Ab Initio}} molecular dynamics for liquid metals},\ }\href {https://doi.org/10.1103/PhysRevB.47.558} {\bibfield  {journal} {\bibinfo  {journal} {Physical Review B}\ }\textbf {\bibinfo {volume} {47}},\ \bibinfo {pages} {558} (\bibinfo {year} {1993})}\BibitemShut {NoStop}%
\bibitem [{\citenamefont {Kresse}\ and\ \citenamefont {Joubert}(1999)}]{Kresse_1999}%
  \BibitemOpen
  \bibfield  {author} {\bibinfo {author} {\bibfnamefont {G.}~\bibnamefont {Kresse}}\ and\ \bibinfo {author} {\bibfnamefont {D.}~\bibnamefont {Joubert}},\ }\bibfield  {title} {\bibinfo {title} {From ultrasoft pseudopotentials to the projector augmented-wave method},\ }\href {https://doi.org/10.1103/PhysRevB.59.1758} {\bibfield  {journal} {\bibinfo  {journal} {Physical Review B}\ }\textbf {\bibinfo {volume} {59}},\ \bibinfo {pages} {1758} (\bibinfo {year} {1999})}\BibitemShut {NoStop}%
\bibitem [{\citenamefont {Berland}\ and\ \citenamefont {Hyldgaard}(2014)}]{Berland_2014}%
  \BibitemOpen
  \bibfield  {author} {\bibinfo {author} {\bibfnamefont {K.}~\bibnamefont {Berland}}\ and\ \bibinfo {author} {\bibfnamefont {P.}~\bibnamefont {Hyldgaard}},\ }\bibfield  {title} {\bibinfo {title} {Exchange functional that tests the robustness of the plasmon description of the van der {{Waals}} density functional},\ }\href {https://doi.org/10.1103/PhysRevB.89.035412} {\bibfield  {journal} {\bibinfo  {journal} {Physical Review B}\ }\textbf {\bibinfo {volume} {89}},\ \bibinfo {pages} {035412} (\bibinfo {year} {2014})}\BibitemShut {NoStop}%
\bibitem [{\citenamefont {Berland}\ \emph {et~al.}(2014)\citenamefont {Berland}, \citenamefont {Arter}, \citenamefont {Cooper}, \citenamefont {Lee}, \citenamefont {Lundqvist}, \citenamefont {Schr{\"o}der}, \citenamefont {Thonhauser},\ and\ \citenamefont {Hyldgaard}}]{Berland_2014a}%
  \BibitemOpen
  \bibfield  {author} {\bibinfo {author} {\bibfnamefont {K.}~\bibnamefont {Berland}}, \bibinfo {author} {\bibfnamefont {C.~A.}\ \bibnamefont {Arter}}, \bibinfo {author} {\bibfnamefont {V.~R.}\ \bibnamefont {Cooper}}, \bibinfo {author} {\bibfnamefont {K.}~\bibnamefont {Lee}}, \bibinfo {author} {\bibfnamefont {B.~I.}\ \bibnamefont {Lundqvist}}, \bibinfo {author} {\bibfnamefont {E.}~\bibnamefont {Schr{\"o}der}}, \bibinfo {author} {\bibfnamefont {T.}~\bibnamefont {Thonhauser}},\ and\ \bibinfo {author} {\bibfnamefont {P.}~\bibnamefont {Hyldgaard}},\ }\bibfield  {title} {\bibinfo {title} {Van der {{Waals}} density functionals built upon the electron-gas tradition: {{Facing}} the challenge of competing interactions},\ }\href {https://doi.org/10.1063/1.4871731} {\bibfield  {journal} {\bibinfo  {journal} {The Journal of Chemical Physics}\ }\textbf {\bibinfo {volume} {140}},\ \bibinfo {pages} {18A539} (\bibinfo {year} {2014})}\BibitemShut {NoStop}%
\bibitem [{\citenamefont {Bj{\"o}rkman}(2014)}]{Bjorkman_2014}%
  \BibitemOpen
  \bibfield  {author} {\bibinfo {author} {\bibfnamefont {T.}~\bibnamefont {Bj{\"o}rkman}},\ }\bibfield  {title} {\bibinfo {title} {Testing several recent van der {{Waals}} density functionals for layered structures},\ }\href {https://doi.org/10.1063/1.4893329} {\bibfield  {journal} {\bibinfo  {journal} {The Journal of Chemical Physics}\ }\textbf {\bibinfo {volume} {141}},\ \bibinfo {pages} {074708} (\bibinfo {year} {2014})}\BibitemShut {NoStop}%
\bibitem [{\citenamefont {Tran}\ \emph {et~al.}(2019)\citenamefont {Tran}, \citenamefont {Kalantari}, \citenamefont {Traor{\'e}}, \citenamefont {Rocquefelte},\ and\ \citenamefont {Blaha}}]{Tran_2019}%
  \BibitemOpen
  \bibfield  {author} {\bibinfo {author} {\bibfnamefont {F.}~\bibnamefont {Tran}}, \bibinfo {author} {\bibfnamefont {L.}~\bibnamefont {Kalantari}}, \bibinfo {author} {\bibfnamefont {B.}~\bibnamefont {Traor{\'e}}}, \bibinfo {author} {\bibfnamefont {X.}~\bibnamefont {Rocquefelte}},\ and\ \bibinfo {author} {\bibfnamefont {P.}~\bibnamefont {Blaha}},\ }\bibfield  {title} {\bibinfo {title} {Nonlocal van der {{Waals}} functionals for solids: {{Choosing}} an appropriate one},\ }\href {https://doi.org/10.1103/PhysRevMaterials.3.063602} {\bibfield  {journal} {\bibinfo  {journal} {Physical Review Materials}\ }\textbf {\bibinfo {volume} {3}},\ \bibinfo {pages} {063602} (\bibinfo {year} {2019})}\BibitemShut {NoStop}%
\bibitem [{\citenamefont {Krukau}\ \emph {et~al.}(2006)\citenamefont {Krukau}, \citenamefont {Vydrov}, \citenamefont {Izmaylov},\ and\ \citenamefont {Scuseria}}]{Krukau_2006}%
  \BibitemOpen
  \bibfield  {author} {\bibinfo {author} {\bibfnamefont {A.~V.}\ \bibnamefont {Krukau}}, \bibinfo {author} {\bibfnamefont {O.~A.}\ \bibnamefont {Vydrov}}, \bibinfo {author} {\bibfnamefont {A.~F.}\ \bibnamefont {Izmaylov}},\ and\ \bibinfo {author} {\bibfnamefont {G.~E.}\ \bibnamefont {Scuseria}},\ }\bibfield  {title} {\bibinfo {title} {Influence of the exchange screening parameter on the performance of screened hybrid functionals},\ }\href {https://doi.org/10.1063/1.2404663} {\bibfield  {journal} {\bibinfo  {journal} {The Journal of Chemical Physics}\ }\textbf {\bibinfo {volume} {125}},\ \bibinfo {pages} {224106} (\bibinfo {year} {2006})}\BibitemShut {NoStop}%
\bibitem [{\citenamefont {Jain}\ \emph {et~al.}(2013)\citenamefont {Jain}, \citenamefont {Ong}, \citenamefont {Hautier}, \citenamefont {Chen}, \citenamefont {Richards}, \citenamefont {Dacek}, \citenamefont {Cholia}, \citenamefont {Gunter}, \citenamefont {Skinner}, \citenamefont {Ceder},\ and\ \citenamefont {Persson}}]{Jain_2013}%
  \BibitemOpen
  \bibfield  {author} {\bibinfo {author} {\bibfnamefont {A.}~\bibnamefont {Jain}}, \bibinfo {author} {\bibfnamefont {S.~P.}\ \bibnamefont {Ong}}, \bibinfo {author} {\bibfnamefont {G.}~\bibnamefont {Hautier}}, \bibinfo {author} {\bibfnamefont {W.}~\bibnamefont {Chen}}, \bibinfo {author} {\bibfnamefont {W.~D.}\ \bibnamefont {Richards}}, \bibinfo {author} {\bibfnamefont {S.}~\bibnamefont {Dacek}}, \bibinfo {author} {\bibfnamefont {S.}~\bibnamefont {Cholia}}, \bibinfo {author} {\bibfnamefont {D.}~\bibnamefont {Gunter}}, \bibinfo {author} {\bibfnamefont {D.}~\bibnamefont {Skinner}}, \bibinfo {author} {\bibfnamefont {G.}~\bibnamefont {Ceder}},\ and\ \bibinfo {author} {\bibfnamefont {K.~A.}\ \bibnamefont {Persson}},\ }\bibfield  {title} {\bibinfo {title} {Commentary: {{The Materials Project}}: {{A}} materials genome approach to accelerating materials innovation},\ }\href {https://doi.org/10.1063/1.4812323} {\bibfield  {journal} {\bibinfo  {journal} {APL Materials}\ }\textbf {\bibinfo {volume} {1}},\ \bibinfo {pages}
  {011002} (\bibinfo {year} {2013})}\BibitemShut {NoStop}%
\bibitem [{\citenamefont {Gajdo{\v s}}\ \emph {et~al.}(2006)\citenamefont {Gajdo{\v s}}, \citenamefont {Hummer}, \citenamefont {Kresse}, \citenamefont {Furthm{\"u}ller},\ and\ \citenamefont {Bechstedt}}]{Gajdos_2006}%
  \BibitemOpen
  \bibfield  {author} {\bibinfo {author} {\bibfnamefont {M.}~\bibnamefont {Gajdo{\v s}}}, \bibinfo {author} {\bibfnamefont {K.}~\bibnamefont {Hummer}}, \bibinfo {author} {\bibfnamefont {G.}~\bibnamefont {Kresse}}, \bibinfo {author} {\bibfnamefont {J.}~\bibnamefont {Furthm{\"u}ller}},\ and\ \bibinfo {author} {\bibfnamefont {F.}~\bibnamefont {Bechstedt}},\ }\bibfield  {title} {\bibinfo {title} {Linear optical properties in the projector-augmented wave methodology},\ }\href {https://doi.org/10.1103/PhysRevB.73.045112} {\bibfield  {journal} {\bibinfo  {journal} {Physical Review B}\ }\textbf {\bibinfo {volume} {73}},\ \bibinfo {pages} {045112} (\bibinfo {year} {2006})}\BibitemShut {NoStop}%
\bibitem [{\citenamefont {Le~Page}\ and\ \citenamefont {Saxe}(2002)}]{LePage_2002}%
  \BibitemOpen
  \bibfield  {author} {\bibinfo {author} {\bibfnamefont {Y.}~\bibnamefont {Le~Page}}\ and\ \bibinfo {author} {\bibfnamefont {P.}~\bibnamefont {Saxe}},\ }\bibfield  {title} {\bibinfo {title} {Symmetry-general least-squares extraction of elastic data for strained materials from {\emph{ab initio}} calculations of stress},\ }\href {https://doi.org/10.1103/PhysRevB.65.104104} {\bibfield  {journal} {\bibinfo  {journal} {Physical Review B}\ }\textbf {\bibinfo {volume} {65}},\ \bibinfo {pages} {104104} (\bibinfo {year} {2002})}\BibitemShut {NoStop}%
\bibitem [{\citenamefont {Shulumba}\ \emph {et~al.}(2017)\citenamefont {Shulumba}, \citenamefont {Hellman},\ and\ \citenamefont {Minnich}}]{Shulumba_2017}%
  \BibitemOpen
  \bibfield  {author} {\bibinfo {author} {\bibfnamefont {N.}~\bibnamefont {Shulumba}}, \bibinfo {author} {\bibfnamefont {O.}~\bibnamefont {Hellman}},\ and\ \bibinfo {author} {\bibfnamefont {A.~J.}\ \bibnamefont {Minnich}},\ }\bibfield  {title} {\bibinfo {title} {Intrinsic localized mode and low thermal conductivity of {{PbSe}}},\ }\href {https://doi.org/10.1103/PhysRevB.95.014302} {\bibfield  {journal} {\bibinfo  {journal} {Physical Review B}\ }\textbf {\bibinfo {volume} {95}},\ \bibinfo {pages} {014302} (\bibinfo {year} {2017})}\BibitemShut {NoStop}%
\bibitem [{\citenamefont {Hellman}\ \emph {et~al.}(2011)\citenamefont {Hellman}, \citenamefont {Abrikosov},\ and\ \citenamefont {Simak}}]{Hellman_2011}%
  \BibitemOpen
  \bibfield  {author} {\bibinfo {author} {\bibfnamefont {O.}~\bibnamefont {Hellman}}, \bibinfo {author} {\bibfnamefont {I.~A.}\ \bibnamefont {Abrikosov}},\ and\ \bibinfo {author} {\bibfnamefont {S.~I.}\ \bibnamefont {Simak}},\ }\bibfield  {title} {\bibinfo {title} {Lattice dynamics of anharmonic solids from first principles},\ }\href {https://doi.org/10.1103/PhysRevB.84.180301} {\bibfield  {journal} {\bibinfo  {journal} {Physical Review B}\ }\textbf {\bibinfo {volume} {84}},\ \bibinfo {pages} {180301} (\bibinfo {year} {2011})}\BibitemShut {NoStop}%
\bibitem [{\citenamefont {Knoop}\ \emph {et~al.}(2024)\citenamefont {Knoop}, \citenamefont {Shulumba}, \citenamefont {Castellano}, \citenamefont {Batista}, \citenamefont {Farris}, \citenamefont {Verstraete}, \citenamefont {Heine}, \citenamefont {Broido}, \citenamefont {Kim}, \citenamefont {Klarbring}, \citenamefont {Abrikosov}, \citenamefont {Simak},\ and\ \citenamefont {Hellman}}]{Knoop_2024}%
  \BibitemOpen
  \bibfield  {author} {\bibinfo {author} {\bibfnamefont {F.}~\bibnamefont {Knoop}}, \bibinfo {author} {\bibfnamefont {N.}~\bibnamefont {Shulumba}}, \bibinfo {author} {\bibfnamefont {A.}~\bibnamefont {Castellano}}, \bibinfo {author} {\bibfnamefont {J.~P.~A.}\ \bibnamefont {Batista}}, \bibinfo {author} {\bibfnamefont {R.}~\bibnamefont {Farris}}, \bibinfo {author} {\bibfnamefont {M.~J.}\ \bibnamefont {Verstraete}}, \bibinfo {author} {\bibfnamefont {M.}~\bibnamefont {Heine}}, \bibinfo {author} {\bibfnamefont {D.}~\bibnamefont {Broido}}, \bibinfo {author} {\bibfnamefont {D.~S.}\ \bibnamefont {Kim}}, \bibinfo {author} {\bibfnamefont {J.}~\bibnamefont {Klarbring}}, \bibinfo {author} {\bibfnamefont {I.~A.}\ \bibnamefont {Abrikosov}}, \bibinfo {author} {\bibfnamefont {S.~I.}\ \bibnamefont {Simak}},\ and\ \bibinfo {author} {\bibfnamefont {O.}~\bibnamefont {Hellman}},\ }\bibfield  {title} {\bibinfo {title} {{{TDEP}}: {{Temperature Dependent EffectivePotentials}}},\ }\href {https://doi.org/10.21105/joss.06150} {\bibfield
   {journal} {\bibinfo  {journal} {Journal of Open Source Software}\ }\textbf {\bibinfo {volume} {9}},\ \bibinfo {pages} {6150} (\bibinfo {year} {2024})}\BibitemShut {NoStop}%
\bibitem [{\citenamefont {McCarroll}(1965)}]{McCarroll_1965}%
  \BibitemOpen
  \bibfield  {author} {\bibinfo {author} {\bibfnamefont {W.}~\bibnamefont {McCarroll}},\ }\bibfield  {title} {\bibinfo {title} {Chemical and structural characteristics of the potassium-cesium-antimony photocathode},\ }\href {https://doi.org/10.1016/0022-3697(65)90087-9} {\bibfield  {journal} {\bibinfo  {journal} {Journal of Physics and Chemistry of Solids}\ }\textbf {\bibinfo {volume} {26}},\ \bibinfo {pages} {191} (\bibinfo {year} {1965})}\BibitemShut {NoStop}%
\bibitem [{SM()}]{SM}%
  \BibitemOpen
  \href@noop {} {}\bibinfo {note} {See Supplemental Material at [URL will be inserted by publisher] for equation for running average in Eq. S1, deformation potential of selected k-points in Table S1, band structure aligned at core level in Fig. S1, HSE06 band structure in Fig. S2, $zT$ under strain with constant $\kappa_\ell$ in Fig. S3, electron transport properties at 300 and 500\;K in Fig. S4--S7, and k-point and interpolation convergence of \textsc{AMSET} in Fig. S8--S11}\BibitemShut {NoStop}%
\bibitem [{\citenamefont {Wu}\ and\ \citenamefont {Ganose}(2023)}]{Wu_2023}%
  \BibitemOpen
  \bibfield  {author} {\bibinfo {author} {\bibfnamefont {R.}~\bibnamefont {Wu}}\ and\ \bibinfo {author} {\bibfnamefont {A.~M.}\ \bibnamefont {Ganose}},\ }\bibfield  {title} {\bibinfo {title} {Relativistic electronic structure and photovoltaic performance of {{K}}{\textsubscript{2}} {{CsSb}}},\ }\href {https://doi.org/10.1039/D3TA02061D} {\bibfield  {journal} {\bibinfo  {journal} {Journal of Materials Chemistry A}\ }\textbf {\bibinfo {volume} {11}},\ \bibinfo {pages} {21636} (\bibinfo {year} {2023})}\BibitemShut {NoStop}%
\bibitem [{\citenamefont {Cocchi}\ \emph {et~al.}(2019)\citenamefont {Cocchi}, \citenamefont {Mistry}, \citenamefont {Schmei{\ss}er}, \citenamefont {K{\"u}hn},\ and\ \citenamefont {Kamps}}]{Cocchi_2019}%
  \BibitemOpen
  \bibfield  {author} {\bibinfo {author} {\bibfnamefont {C.}~\bibnamefont {Cocchi}}, \bibinfo {author} {\bibfnamefont {S.}~\bibnamefont {Mistry}}, \bibinfo {author} {\bibfnamefont {M.}~\bibnamefont {Schmei{\ss}er}}, \bibinfo {author} {\bibfnamefont {J.}~\bibnamefont {K{\"u}hn}},\ and\ \bibinfo {author} {\bibfnamefont {T.}~\bibnamefont {Kamps}},\ }\bibfield  {title} {\bibinfo {title} {First-principles many-body study of the electronic and optical properties of {{CsK}}{\textsubscript{2}} {{Sb}}, a semiconducting material for ultra-bright electron sources},\ }\href {https://doi.org/10.1088/1361-648X/aaedee} {\bibfield  {journal} {\bibinfo  {journal} {Journal of Physics: Condensed Matter}\ }\textbf {\bibinfo {volume} {31}},\ \bibinfo {pages} {014002} (\bibinfo {year} {2019})}\BibitemShut {NoStop}%
\bibitem [{\citenamefont {Ganose}\ \emph {et~al.}(2021{\natexlab{b}})\citenamefont {Ganose}, \citenamefont {Searle}, \citenamefont {Jain},\ and\ \citenamefont {Griffin}}]{Ganose_2021a}%
  \BibitemOpen
  \bibfield  {author} {\bibinfo {author} {\bibfnamefont {A.}~\bibnamefont {Ganose}}, \bibinfo {author} {\bibfnamefont {A.}~\bibnamefont {Searle}}, \bibinfo {author} {\bibfnamefont {A.}~\bibnamefont {Jain}},\ and\ \bibinfo {author} {\bibfnamefont {S.}~\bibnamefont {Griffin}},\ }\bibfield  {title} {\bibinfo {title} {{{IFermi}}: {{A}} python library for {{Fermi}} surface generation and analysis},\ }\href {https://doi.org/10.21105/joss.03089} {\bibfield  {journal} {\bibinfo  {journal} {Journal of Open Source Software}\ }\textbf {\bibinfo {volume} {6}},\ \bibinfo {pages} {3089} (\bibinfo {year} {2021}{\natexlab{b}})}\BibitemShut {NoStop}%
\bibitem [{\citenamefont {Kane}(1957)}]{Kane_1957}%
  \BibitemOpen
  \bibfield  {author} {\bibinfo {author} {\bibfnamefont {E.~O.}\ \bibnamefont {Kane}},\ }\bibfield  {title} {\bibinfo {title} {Band structure of indium antimonide},\ }\href {https://doi.org/10.1016/0022-3697(57)90013-6} {\bibfield  {journal} {\bibinfo  {journal} {Journal of Physics and Chemistry of Solids}\ }\textbf {\bibinfo {volume} {1}},\ \bibinfo {pages} {249} (\bibinfo {year} {1957})}\BibitemShut {NoStop}%
\bibitem [{\citenamefont {Ziman}(1972)}]{Ziman_1972}%
  \BibitemOpen
  \bibfield  {author} {\bibinfo {author} {\bibfnamefont {J.~M.}\ \bibnamefont {Ziman}},\ }\href {https://doi.org/10.1017/CBO9781139644075} {\emph {\bibinfo {title} {Principles of the {{Theory}} of {{Solids}}}}},\ \bibinfo {edition} {2nd}\ ed.\ (\bibinfo  {publisher} {Cambridge University Press},\ \bibinfo {year} {1972})\BibitemShut {NoStop}%
\bibitem [{\citenamefont {Maeda}\ \emph {et~al.}(2015)\citenamefont {Maeda}, \citenamefont {Hattori}, \citenamefont {Chang}, \citenamefont {Arai},\ and\ \citenamefont {Kinoshita}}]{Maeda_2015}%
  \BibitemOpen
  \bibfield  {author} {\bibinfo {author} {\bibfnamefont {T.}~\bibnamefont {Maeda}}, \bibinfo {author} {\bibfnamefont {H.}~\bibnamefont {Hattori}}, \bibinfo {author} {\bibfnamefont {W.~H.}\ \bibnamefont {Chang}}, \bibinfo {author} {\bibfnamefont {Y.}~\bibnamefont {Arai}},\ and\ \bibinfo {author} {\bibfnamefont {K.}~\bibnamefont {Kinoshita}},\ }\bibfield  {title} {\bibinfo {title} {Hole {{Hall}} mobility of {{SiGe}} alloys grown by the traveling liquidus-zone method},\ }\href {https://doi.org/10.1063/1.4933330} {\bibfield  {journal} {\bibinfo  {journal} {Applied Physics Letters}\ }\textbf {\bibinfo {volume} {107}},\ \bibinfo {pages} {152104} (\bibinfo {year} {2015})}\BibitemShut {NoStop}%
\bibitem [{\citenamefont {Cha}\ \emph {et~al.}(2019)\citenamefont {Cha}, \citenamefont {Zhou}, \citenamefont {Cho}, \citenamefont {Park},\ and\ \citenamefont {Chung}}]{Cha_2019}%
  \BibitemOpen
  \bibfield  {author} {\bibinfo {author} {\bibfnamefont {J.}~\bibnamefont {Cha}}, \bibinfo {author} {\bibfnamefont {C.}~\bibnamefont {Zhou}}, \bibinfo {author} {\bibfnamefont {S.-P.}\ \bibnamefont {Cho}}, \bibinfo {author} {\bibfnamefont {S.~H.}\ \bibnamefont {Park}},\ and\ \bibinfo {author} {\bibfnamefont {I.}~\bibnamefont {Chung}},\ }\bibfield  {title} {\bibinfo {title} {Ultrahigh {{Power Factor}} and {{Electron Mobility}} in n-{{Type Bi}}{\textsubscript{2}} {{Te}}{\textsubscript{3}} -- {\emph{x}} \%{{Cu Stabilized}} under {{Excess Te Condition}}},\ }\href {https://doi.org/10.1021/acsami.9b10394} {\bibfield  {journal} {\bibinfo  {journal} {ACS Applied Materials \& Interfaces}\ }\textbf {\bibinfo {volume} {11}},\ \bibinfo {pages} {30999} (\bibinfo {year} {2019})}\BibitemShut {NoStop}%
\bibitem [{\citenamefont {Dughaish}(2002)}]{Dughaish_2002}%
  \BibitemOpen
  \bibfield  {author} {\bibinfo {author} {\bibfnamefont {Z.}~\bibnamefont {Dughaish}},\ }\bibfield  {title} {\bibinfo {title} {Lead telluride as a thermoelectric material for thermoelectric power generation},\ }\href {https://doi.org/10.1016/S0921-4526(02)01187-0} {\bibfield  {journal} {\bibinfo  {journal} {Physica B: Condensed Matter}\ }\textbf {\bibinfo {volume} {322}},\ \bibinfo {pages} {205} (\bibinfo {year} {2002})}\BibitemShut {NoStop}%
\bibitem [{\citenamefont {Kim}\ \emph {et~al.}(2015)\citenamefont {Kim}, \citenamefont {Gibbs}, \citenamefont {Tang}, \citenamefont {Wang},\ and\ \citenamefont {Snyder}}]{Kim_2015}%
  \BibitemOpen
  \bibfield  {author} {\bibinfo {author} {\bibfnamefont {H.-S.}\ \bibnamefont {Kim}}, \bibinfo {author} {\bibfnamefont {Z.~M.}\ \bibnamefont {Gibbs}}, \bibinfo {author} {\bibfnamefont {Y.}~\bibnamefont {Tang}}, \bibinfo {author} {\bibfnamefont {H.}~\bibnamefont {Wang}},\ and\ \bibinfo {author} {\bibfnamefont {G.~J.}\ \bibnamefont {Snyder}},\ }\bibfield  {title} {\bibinfo {title} {Characterization of {{Lorenz}} number with {{Seebeck}} coefficient measurement},\ }\href {https://doi.org/10.1063/1.4908244} {\bibfield  {journal} {\bibinfo  {journal} {APL Materials}\ }\textbf {\bibinfo {volume} {3}},\ \bibinfo {pages} {041506} (\bibinfo {year} {2015})}\BibitemShut {NoStop}%
\bibitem [{\citenamefont {{Flage-Larsen}}\ and\ \citenamefont {Prytz}(2011)}]{Flage-Larsen_2011}%
  \BibitemOpen
  \bibfield  {author} {\bibinfo {author} {\bibfnamefont {E.}~\bibnamefont {{Flage-Larsen}}}\ and\ \bibinfo {author} {\bibfnamefont {{\O}.}~\bibnamefont {Prytz}},\ }\bibfield  {title} {\bibinfo {title} {The {{Lorenz}} function: {{Its}} properties at optimum thermoelectric figure-of-merit},\ }\href {https://doi.org/10.1063/1.3656017} {\bibfield  {journal} {\bibinfo  {journal} {Applied Physics Letters}\ }\textbf {\bibinfo {volume} {99}},\ \bibinfo {pages} {202108} (\bibinfo {year} {2011})}\BibitemShut {NoStop}%
\bibitem [{\citenamefont {Zayachuk}(1997)}]{Zayachuk_1997b}%
  \BibitemOpen
  \bibfield  {author} {\bibinfo {author} {\bibfnamefont {D.~M.}\ \bibnamefont {Zayachuk}},\ }\bibfield  {title} {\bibinfo {title} {The dominant mechanisms of charge-carrier scattering in lead telluride},\ }\href {https://doi.org/10.1134/1.1187322} {\bibfield  {journal} {\bibinfo  {journal} {Semiconductors}\ }\textbf {\bibinfo {volume} {31}},\ \bibinfo {pages} {173} (\bibinfo {year} {1997})}\BibitemShut {NoStop}%
\bibitem [{\citenamefont {Cao}\ \emph {et~al.}(2018)\citenamefont {Cao}, \citenamefont {{Querales-Flores}}, \citenamefont {Murphy}, \citenamefont {Fahy},\ and\ \citenamefont {Savi{\'c}}}]{Cao_2018}%
  \BibitemOpen
  \bibfield  {author} {\bibinfo {author} {\bibfnamefont {J.}~\bibnamefont {Cao}}, \bibinfo {author} {\bibfnamefont {J.~D.}\ \bibnamefont {{Querales-Flores}}}, \bibinfo {author} {\bibfnamefont {A.~R.}\ \bibnamefont {Murphy}}, \bibinfo {author} {\bibfnamefont {S.}~\bibnamefont {Fahy}},\ and\ \bibinfo {author} {\bibfnamefont {I.}~\bibnamefont {Savi{\'c}}},\ }\bibfield  {title} {\bibinfo {title} {Dominant electron-phonon scattering mechanisms in n -type {{PbTe}} from first principles},\ }\href {https://doi.org/10.1103/PhysRevB.98.205202} {\bibfield  {journal} {\bibinfo  {journal} {Physical Review B}\ }\textbf {\bibinfo {volume} {98}},\ \bibinfo {pages} {205202} (\bibinfo {year} {2018})}\BibitemShut {NoStop}%
\bibitem [{\citenamefont {Hauble}\ \emph {et~al.}(2023)\citenamefont {Hauble}, \citenamefont {Toriyama}, \citenamefont {Bartling}, \citenamefont {{Abdel-Mageed}}, \citenamefont {Snyder},\ and\ \citenamefont {Kauzlarich}}]{Hauble_2023}%
  \BibitemOpen
  \bibfield  {author} {\bibinfo {author} {\bibfnamefont {A.~K.}\ \bibnamefont {Hauble}}, \bibinfo {author} {\bibfnamefont {M.~Y.}\ \bibnamefont {Toriyama}}, \bibinfo {author} {\bibfnamefont {S.}~\bibnamefont {Bartling}}, \bibinfo {author} {\bibfnamefont {A.~M.}\ \bibnamefont {{Abdel-Mageed}}}, \bibinfo {author} {\bibfnamefont {G.~J.}\ \bibnamefont {Snyder}},\ and\ \bibinfo {author} {\bibfnamefont {S.~M.}\ \bibnamefont {Kauzlarich}},\ }\bibfield  {title} {\bibinfo {title} {Experiment and {{Theory}} in {{Concert To Unravel}} the {{Remarkable Electronic Properties}} of {{Na-Doped Eu}} {\textsubscript{11}} {{Zn}} {\textsubscript{4}} {{Sn}} {\textsubscript{2}} {{As}} {\textsubscript{12}} : {{A Layered Zintl Phase}}},\ }\href {https://doi.org/10.1021/acs.chemmater.3c01509} {\bibfield  {journal} {\bibinfo  {journal} {Chemistry of Materials}\ }\textbf {\bibinfo {volume} {35}},\ \bibinfo {pages} {7719} (\bibinfo {year} {2023})}\BibitemShut {NoStop}%
\bibitem [{\citenamefont {Ziman}(1960)}]{Ziman_1960}%
  \BibitemOpen
  \bibfield  {author} {\bibinfo {author} {\bibfnamefont {J.~M.}\ \bibnamefont {Ziman}},\ }\href@noop {} {\emph {\bibinfo {title} {Electrons and Phonons: The Theory of Transport Phenomena in Solids}}}\ (\bibinfo  {publisher} {Clarendon Press},\ \bibinfo {address} {Oxford},\ \bibinfo {year} {1960})\BibitemShut {NoStop}%
\bibitem [{\citenamefont {Xu}\ and\ \citenamefont {Verstraete}(2014)}]{Xu_2014}%
  \BibitemOpen
  \bibfield  {author} {\bibinfo {author} {\bibfnamefont {B.}~\bibnamefont {Xu}}\ and\ \bibinfo {author} {\bibfnamefont {M.~J.}\ \bibnamefont {Verstraete}},\ }\bibfield  {title} {\bibinfo {title} {First {{Principles Explanation}} of the {{Positive Seebeck Coefficient}} of {{Lithium}}},\ }\href {https://doi.org/10.1103/PhysRevLett.112.196603} {\bibfield  {journal} {\bibinfo  {journal} {Physical Review Letters}\ }\textbf {\bibinfo {volume} {112}},\ \bibinfo {pages} {196603} (\bibinfo {year} {2014})}\BibitemShut {NoStop}%
\bibitem [{\citenamefont {Xia}\ \emph {et~al.}(2019)\citenamefont {Xia}, \citenamefont {Park}, \citenamefont {Zhou},\ and\ \citenamefont {Ozoli{\c n}{\v s}}}]{Xia_2019}%
  \BibitemOpen
  \bibfield  {author} {\bibinfo {author} {\bibfnamefont {Y.}~\bibnamefont {Xia}}, \bibinfo {author} {\bibfnamefont {J.}~\bibnamefont {Park}}, \bibinfo {author} {\bibfnamefont {F.}~\bibnamefont {Zhou}},\ and\ \bibinfo {author} {\bibfnamefont {V.}~\bibnamefont {Ozoli{\c n}{\v s}}},\ }\bibfield  {title} {\bibinfo {title} {High {{Thermoelectric Power Factor}} in {{Intermetallic Co Si Arising}} from {{Energy Filtering}} of {{Electrons}} by {{Phonon Scattering}}},\ }\href {https://doi.org/10.1103/PhysRevApplied.11.024017} {\bibfield  {journal} {\bibinfo  {journal} {Physical Review Applied}\ }\textbf {\bibinfo {volume} {11}},\ \bibinfo {pages} {024017} (\bibinfo {year} {2019})}\BibitemShut {NoStop}%
\bibitem [{\citenamefont {Graziosi}\ and\ \citenamefont {Neophytou}(2023)}]{Graziosi_2023}%
  \BibitemOpen
  \bibfield  {author} {\bibinfo {author} {\bibfnamefont {P.}~\bibnamefont {Graziosi}}\ and\ \bibinfo {author} {\bibfnamefont {N.}~\bibnamefont {Neophytou}},\ }\bibfield  {title} {\bibinfo {title} {The {{Role}} of {{Electronic Bandstructure Shape}} in {{Improving}} the {{Thermoelectric Power Factor}} of {{Complex Materials}}},\ }\href {https://doi.org/10.1021/acsaelm.3c00887} {\bibfield  {journal} {\bibinfo  {journal} {ACS Applied Electronic Materials}\ }\textbf {\bibinfo {volume} {6}},\ \bibinfo {pages} {2889} (\bibinfo {year} {2023})}\BibitemShut {NoStop}%
\bibitem [{LTC()}]{LTC_note}%
  \BibitemOpen
  \href@noop {} {}\bibinfo {note} {For technical reasons, the cumulative $\kappa_{\ell}$ was taken from the calculation using $25\times25\times25$ $q$-points and hence does not add up exactly to the extrapolated values at 300\;K in Fig.\ \ref{fig: ltc}(b).}\BibitemShut {Stop}%
\bibitem [{\citenamefont {Hofmeister}(2007)}]{Hofmeister_2007}%
  \BibitemOpen
  \bibfield  {author} {\bibinfo {author} {\bibfnamefont {A.~M.}\ \bibnamefont {Hofmeister}},\ }\bibfield  {title} {\bibinfo {title} {Pressure dependence of thermal transport properties},\ }\href {https://doi.org/10.1073/pnas.0610734104} {\bibfield  {journal} {\bibinfo  {journal} {Proceedings of the National Academy of Sciences}\ }\textbf {\bibinfo {volume} {104}},\ \bibinfo {pages} {9192} (\bibinfo {year} {2007})}\BibitemShut {NoStop}%
\bibitem [{\citenamefont {Ravichandran}\ and\ \citenamefont {Broido}(2019)}]{Ravichandran_2019}%
  \BibitemOpen
  \bibfield  {author} {\bibinfo {author} {\bibfnamefont {N.~K.}\ \bibnamefont {Ravichandran}}\ and\ \bibinfo {author} {\bibfnamefont {D.}~\bibnamefont {Broido}},\ }\bibfield  {title} {\bibinfo {title} {Non-monotonic pressure dependence of the thermal conductivity of boron arsenide},\ }\href {https://doi.org/10.1038/s41467-019-08713-0} {\bibfield  {journal} {\bibinfo  {journal} {Nature Communications}\ }\textbf {\bibinfo {volume} {10}},\ \bibinfo {pages} {827} (\bibinfo {year} {2019})}\BibitemShut {NoStop}%
\bibitem [{\citenamefont {Li}\ \emph {et~al.}(2022)\citenamefont {Li}, \citenamefont {Qin}, \citenamefont {Wu}, \citenamefont {Li}, \citenamefont {Kunz}, \citenamefont {Alatas}, \citenamefont {Kavner},\ and\ \citenamefont {Hu}}]{Li_2022}%
  \BibitemOpen
  \bibfield  {author} {\bibinfo {author} {\bibfnamefont {S.}~\bibnamefont {Li}}, \bibinfo {author} {\bibfnamefont {Z.}~\bibnamefont {Qin}}, \bibinfo {author} {\bibfnamefont {H.}~\bibnamefont {Wu}}, \bibinfo {author} {\bibfnamefont {M.}~\bibnamefont {Li}}, \bibinfo {author} {\bibfnamefont {M.}~\bibnamefont {Kunz}}, \bibinfo {author} {\bibfnamefont {A.}~\bibnamefont {Alatas}}, \bibinfo {author} {\bibfnamefont {A.}~\bibnamefont {Kavner}},\ and\ \bibinfo {author} {\bibfnamefont {Y.}~\bibnamefont {Hu}},\ }\bibfield  {title} {\bibinfo {title} {Anomalous thermal transport under high pressure in boron arsenide},\ }\href {https://doi.org/10.1038/s41586-022-05381-x} {\bibfield  {journal} {\bibinfo  {journal} {Nature}\ }\textbf {\bibinfo {volume} {612}},\ \bibinfo {pages} {459} (\bibinfo {year} {2022})}\BibitemShut {NoStop}%
\bibitem [{\citenamefont {Cherniushok}\ \emph {et~al.}(2024)\citenamefont {Cherniushok}, \citenamefont {Parashchuk}, \citenamefont {{Cardoso-Gil}}, \citenamefont {Grin},\ and\ \citenamefont {Wojciechowski}}]{Cherniushok_2024}%
  \BibitemOpen
  \bibfield  {author} {\bibinfo {author} {\bibfnamefont {O.}~\bibnamefont {Cherniushok}}, \bibinfo {author} {\bibfnamefont {T.}~\bibnamefont {Parashchuk}}, \bibinfo {author} {\bibfnamefont {R.}~\bibnamefont {{Cardoso-Gil}}}, \bibinfo {author} {\bibfnamefont {Y.}~\bibnamefont {Grin}},\ and\ \bibinfo {author} {\bibfnamefont {K.~T.}\ \bibnamefont {Wojciechowski}},\ }\bibfield  {title} {\bibinfo {title} {Controlled {{Phonon Transport}} via {{Chemical Bond Stretching}} and {{Defect Engineering}}: {{The Case Study}} of {{Filled}} {$\beta$}-{{Mn-Type Phases}}},\ }\href {https://doi.org/10.1021/acs.inorgchem.4c02562} {\bibfield  {journal} {\bibinfo  {journal} {Inorganic Chemistry}\ }\textbf {\bibinfo {volume} {63}},\ \bibinfo {pages} {18030} (\bibinfo {year} {2024})}\BibitemShut {NoStop}%
\bibitem [{\citenamefont {Lindsay}\ \emph {et~al.}(2015)\citenamefont {Lindsay}, \citenamefont {Broido}, \citenamefont {Carrete}, \citenamefont {Mingo},\ and\ \citenamefont {Reinecke}}]{Lindsay_2015}%
  \BibitemOpen
  \bibfield  {author} {\bibinfo {author} {\bibfnamefont {L.}~\bibnamefont {Lindsay}}, \bibinfo {author} {\bibfnamefont {D.~A.}\ \bibnamefont {Broido}}, \bibinfo {author} {\bibfnamefont {J.}~\bibnamefont {Carrete}}, \bibinfo {author} {\bibfnamefont {N.}~\bibnamefont {Mingo}},\ and\ \bibinfo {author} {\bibfnamefont {T.~L.}\ \bibnamefont {Reinecke}},\ }\bibfield  {title} {\bibinfo {title} {Anomalous pressure dependence of thermal conductivities of large mass ratio compounds},\ }\href {https://doi.org/10.1103/PhysRevB.91.121202} {\bibfield  {journal} {\bibinfo  {journal} {Physical Review B}\ }\textbf {\bibinfo {volume} {91}},\ \bibinfo {pages} {121202} (\bibinfo {year} {2015})}\BibitemShut {NoStop}%
\bibitem [{\citenamefont {Lou}\ \emph {et~al.}(2024)\citenamefont {Lou}, \citenamefont {Gao}, \citenamefont {Han}, \citenamefont {Liu}, \citenamefont {Fu},\ and\ \citenamefont {Zhu}}]{Lou_2024}%
  \BibitemOpen
  \bibfield  {author} {\bibinfo {author} {\bibfnamefont {Q.}~\bibnamefont {Lou}}, \bibinfo {author} {\bibfnamefont {Z.}~\bibnamefont {Gao}}, \bibinfo {author} {\bibfnamefont {S.}~\bibnamefont {Han}}, \bibinfo {author} {\bibfnamefont {F.}~\bibnamefont {Liu}}, \bibinfo {author} {\bibfnamefont {C.}~\bibnamefont {Fu}},\ and\ \bibinfo {author} {\bibfnamefont {T.}~\bibnamefont {Zhu}},\ }\bibfield  {title} {\bibinfo {title} {High {{Defect Tolerance}} in {{Heavy}}-{{Band Thermoelectrics}}},\ }\href {https://doi.org/10.1002/aenm.202402399} {\bibfield  {journal} {\bibinfo  {journal} {Advanced Energy Materials}\ ,\ \bibinfo {pages} {2402399}} (\bibinfo {year} {2024})}\BibitemShut {NoStop}%
\bibitem [{\citenamefont {Zevalkink}\ \emph {et~al.}(2012)\citenamefont {Zevalkink}, \citenamefont {Pomrehn}, \citenamefont {Johnson}, \citenamefont {Swallow}, \citenamefont {Gibbs},\ and\ \citenamefont {Snyder}}]{Zevalkink_2012}%
  \BibitemOpen
  \bibfield  {author} {\bibinfo {author} {\bibfnamefont {A.}~\bibnamefont {Zevalkink}}, \bibinfo {author} {\bibfnamefont {G.~S.}\ \bibnamefont {Pomrehn}}, \bibinfo {author} {\bibfnamefont {S.}~\bibnamefont {Johnson}}, \bibinfo {author} {\bibfnamefont {J.}~\bibnamefont {Swallow}}, \bibinfo {author} {\bibfnamefont {Z.~M.}\ \bibnamefont {Gibbs}},\ and\ \bibinfo {author} {\bibfnamefont {G.~J.}\ \bibnamefont {Snyder}},\ }\bibfield  {title} {\bibinfo {title} {Influence of the {{Triel Elements}} ( {{{\emph{M}}}} = {{Al}}, {{Ga}}, {{In}}) on the {{Transport Properties}} of {{Ca}}{\textsubscript{5}} {{{\emph{M}}}}{\textsubscript{2}} {{Sb}}{\textsubscript{6}} {{Zintl Compounds}}},\ }\href {https://doi.org/10.1021/cm300520w} {\bibfield  {journal} {\bibinfo  {journal} {Chemistry of Materials}\ }\textbf {\bibinfo {volume} {24}},\ \bibinfo {pages} {2091} (\bibinfo {year} {2012})}\BibitemShut {NoStop}%
\bibitem [{\citenamefont {Shuai}\ \emph {et~al.}(2016)\citenamefont {Shuai}, \citenamefont {Wang}, \citenamefont {Liu}, \citenamefont {Kim}, \citenamefont {Mao}, \citenamefont {Sui},\ and\ \citenamefont {Ren}}]{Shuai_2016}%
  \BibitemOpen
  \bibfield  {author} {\bibinfo {author} {\bibfnamefont {J.}~\bibnamefont {Shuai}}, \bibinfo {author} {\bibfnamefont {Y.}~\bibnamefont {Wang}}, \bibinfo {author} {\bibfnamefont {Z.}~\bibnamefont {Liu}}, \bibinfo {author} {\bibfnamefont {H.~S.}\ \bibnamefont {Kim}}, \bibinfo {author} {\bibfnamefont {J.}~\bibnamefont {Mao}}, \bibinfo {author} {\bibfnamefont {J.}~\bibnamefont {Sui}},\ and\ \bibinfo {author} {\bibfnamefont {Z.}~\bibnamefont {Ren}},\ }\bibfield  {title} {\bibinfo {title} {Enhancement of thermoelectric performance of phase pure {{Zintl}} compounds {{Ca1}}-{{Yb Zn2Sb2}}, {{Ca1}}-{{Eu Zn2Sb2}}, and {{Eu1}}-{{Yb Zn2Sb2}} by mechanical alloying and hot pressing},\ }\href {https://doi.org/10.1016/j.nanoen.2016.04.023} {\bibfield  {journal} {\bibinfo  {journal} {Nano Energy}\ }\textbf {\bibinfo {volume} {25}},\ \bibinfo {pages} {136} (\bibinfo {year} {2016})}\BibitemShut {NoStop}%
\bibitem [{\citenamefont {She}\ \emph {et~al.}(2023)\citenamefont {She}, \citenamefont {Sun}, \citenamefont {Zhao}, \citenamefont {Ni}, \citenamefont {Meng},\ and\ \citenamefont {Dai}}]{She_2023}%
  \BibitemOpen
  \bibfield  {author} {\bibinfo {author} {\bibfnamefont {A.}~\bibnamefont {She}}, \bibinfo {author} {\bibfnamefont {Y.}~\bibnamefont {Sun}}, \bibinfo {author} {\bibfnamefont {Y.}~\bibnamefont {Zhao}}, \bibinfo {author} {\bibfnamefont {J.}~\bibnamefont {Ni}}, \bibinfo {author} {\bibfnamefont {S.}~\bibnamefont {Meng}},\ and\ \bibinfo {author} {\bibfnamefont {Z.}~\bibnamefont {Dai}},\ }\bibfield  {title} {\bibinfo {title} {Transport and thermoelectric properties of strongly anharmonic {{Full-Heusler}} compounds {{CsK2M}} ({{M}}={{As}}, {{Bi}})},\ }\href {https://doi.org/10.1016/j.mtcomm.2022.105134} {\bibfield  {journal} {\bibinfo  {journal} {Materials Today Communications}\ }\textbf {\bibinfo {volume} {34}},\ \bibinfo {pages} {105134} (\bibinfo {year} {2023})}\BibitemShut {NoStop}%
\end{thebibliography}%

\end{document}


\title{Thermoelectric transport of strained CsK$_2$Sb: Role of momentum-dependent scattering in low-dimensional Fermi surfaces}

\author{Øven A. Grimenes}
\affiliation{
 Department of Mechanical Engineering and Technology Management, \\ Norwegian University of Life Sciences, NO-1432 Ås, Norway}

\author{G. Jeffrey Snyder}
\affiliation{Materials Science and Engineering, Northwestern University, Evanston, Illinois 60208, USA}

\author{Ole M. Løvvik}
\affiliation{SINTEF Sustainable Energy Technology, Forskningsveien 1, NO-0314 Oslo, Norway}

\author{Kristian Berland}
\affiliation{
 Department of Mechanical Engineering and Technology Management, \\ Norwegian University of Life Sciences, NO-1432 Ås, Norway}

\date{\today}

\maketitle

\section{Running average}

The running average electron velocity as a function of velocity is calculated as

\begin{equation}
    \langle |\mathbf{v}_{n\mathbf{k}}|\rangle(\varepsilon) = 
    \frac{
    \sum_{n\mathbf{k}} |\mathbf{v}_{n\mathbf{k}}|
    e^{-\frac{(\varepsilon-\varepsilon_{n\mathbf{k}})^2}{2 \sigma^2}}
    }
    {\sum_{n\mathbf{k}} 
    e^{-\frac{(\varepsilon-\varepsilon_{n\mathbf{k}})^2}{2 \sigma^2}}
    }\;,
\end{equation}
where $\mathbf{v}_{n\mathbf{k}}$ is the velocity and $\varepsilon_{n\mathbf{k}}$ is the energy at a given k-point $\mathbf{k}$ and band $n$. The standard deviation $\sigma$ was chosen to 0.01\;eV.

\newpage
\section{Deformation potential}
\begin{table}[h!]
  \caption{Unique elements of deformation potential for valence band and conduction bands at the $\Gamma$ and X points at various strain. The superscripts v and c indicate valence and conduction band at $\Gamma$ or X. The subscript indicates the tensor element of the deformation potential.}
\begin{tabular}{lllllll}

Strain             & 0\%    & 1\%        & 2\%       & 3\%    & 4\%    & 5\%\\ \hline

$D_{11}^{\mathrm{v}\Gamma}$ [eV] & 1.69 & 1.68 & 1.68 &  1.67 &  1.67 & 1.67  \\ 
$D_{12}^{\mathrm{v}\Gamma}$ [eV] & 1.10 & 1.14 & 1.18 & 1.24 & 1.29 & 1.34 \\ 
$D_{11}^{\mathrm{c}\Gamma}$ [eV] & 2.42 & 2.66 & 2.91 & 3.18 & 3.44 & 3.62 \\ 
$D_{12}^{\mathrm{c}\Gamma}$ [eV] & 0.06 & 0.06 & 0.07 & 0.06 & 0.03 & 0.06 \\ 
$D_{11}^{\mathrm{v}\mathrm{X}}$ [eV] & 0.59 & 0.54 & 0.50 & 0.44 & 0.39 & 0.34 \\ 
$D_{12}^{\mathrm{v}\mathrm{X}}$ [eV] & 0.06 & 0.06 & 0.06 & 0.06 & 0.09 & 0.08 \\ 
$D_{13}^{\mathrm{v}\mathrm{X}}$ [eV] & 0.04 & 0.04 & 0.05 & 0.04 & 0.01 & 0.02 \\ 
$D_{33}^{\mathrm{v}\mathrm{X}}$ [eV] & 2.67 & 2.68 & 2.70 & 2.71 & 2.73 & 2.75 \\ 
$D_{11}^{\mathrm{c}\mathrm{X}}$ [eV] & 1.19 & 1.25 & 1.33 & 1.42 &  1.51 & 1.60 \\ 
$D_{12}^{\mathrm{c}\mathrm{X}}$ [eV] & 0.10 & 0.08 & 0.09 & 0.08 & 0.05 & 0.06 \\ 
$D_{13}^{\mathrm{c}\mathrm{X}}$ [eV] & 0.02 & 0.02 & 0.03 & 0.02  & 0.01 & 0.02 \\ 
$D_{33}^{\mathrm{c}\mathrm{X}}$ [eV] & 2.43 & 2.38 & 2.34 & 2.28 & 2.21 & 2.14 \\ 

\end{tabular}
\end{table}

\newpage

\section{Band structure aligned at lowest energy}
\begin{figure*}[h!]
    \centering
    \includegraphics[width=\linewidth]{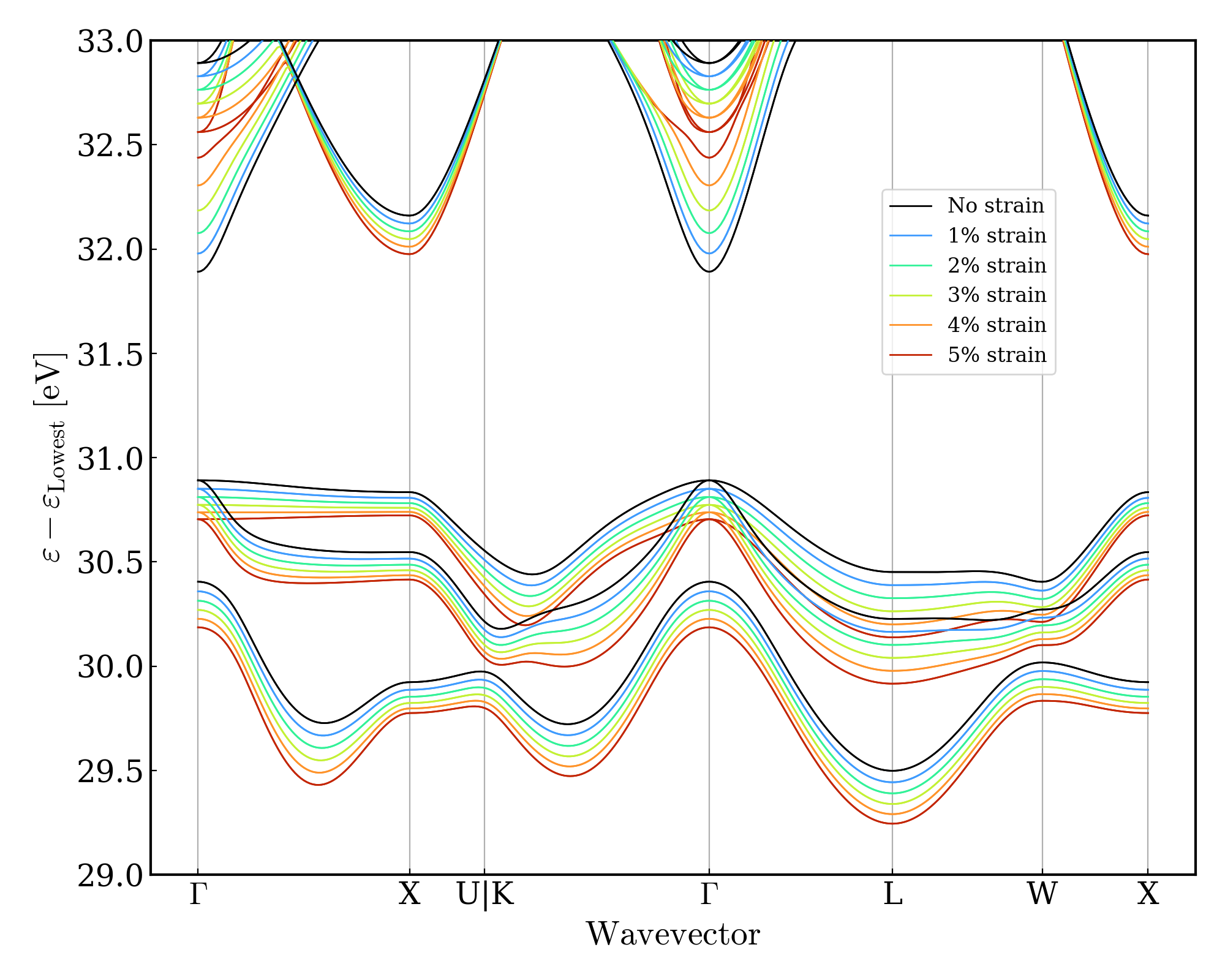}
    \caption{Band structures of CsK$_2$Sb at different strains. The lowest energy level of each band structure have been aligned.}
\end{figure*}

\newpage
\section{HSE06 band structure}
\begin{figure*}[h!]
    \centering
    \includegraphics[width=\linewidth]{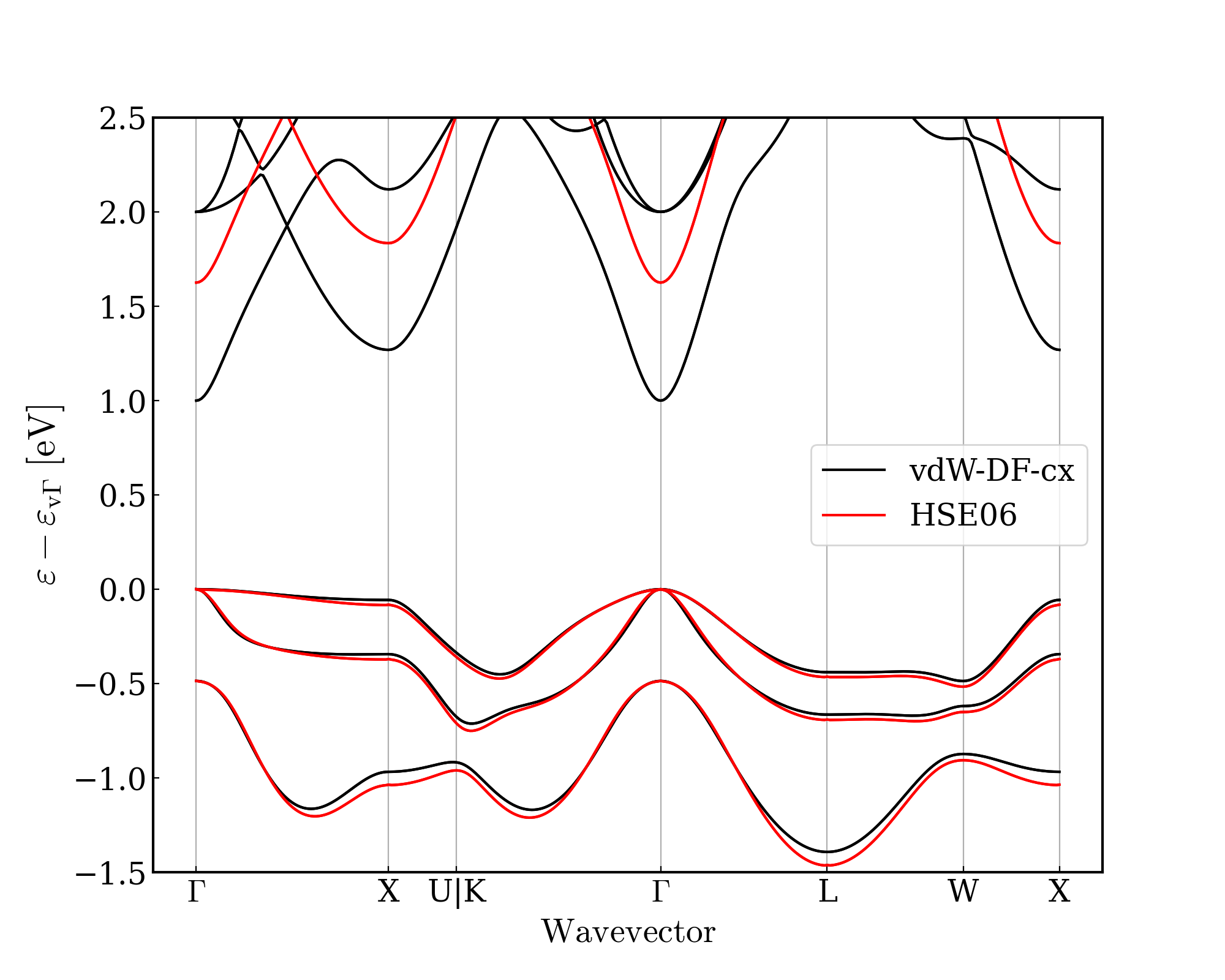}
    \caption{Band structure of CsK$_2$Sb plotted with hybrid functional HSE06 in red and vdW-DF-cx black.}
\end{figure*}

\newpage
\section{zT with constant $\kappa_\ell$}
\begin{figure*}[h!]
    \centering
    \includegraphics[width=\linewidth]{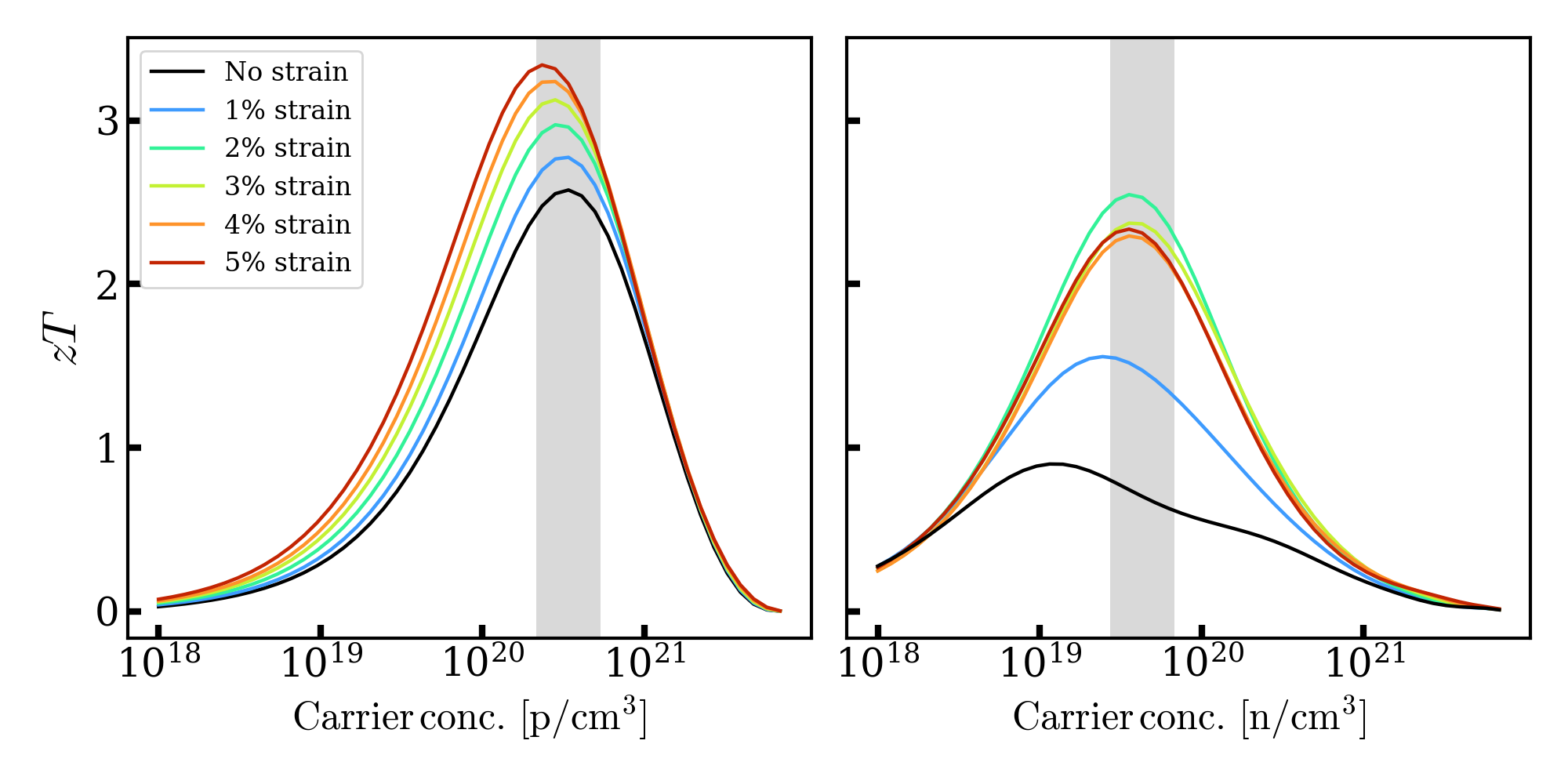}
    \caption{zT of CsK$_2$Sb at 800\;K with various strain, but constant $\kappa_\ell$ = 0.345 W/Km (equilibrium structure) for p-type (left) and n-type (right).}
\end{figure*}

\newpage
\section{Transport properties at 300\;K and 500\;K}

\begin{figure*}[h!]
    \centering
    \includegraphics[width=\linewidth]{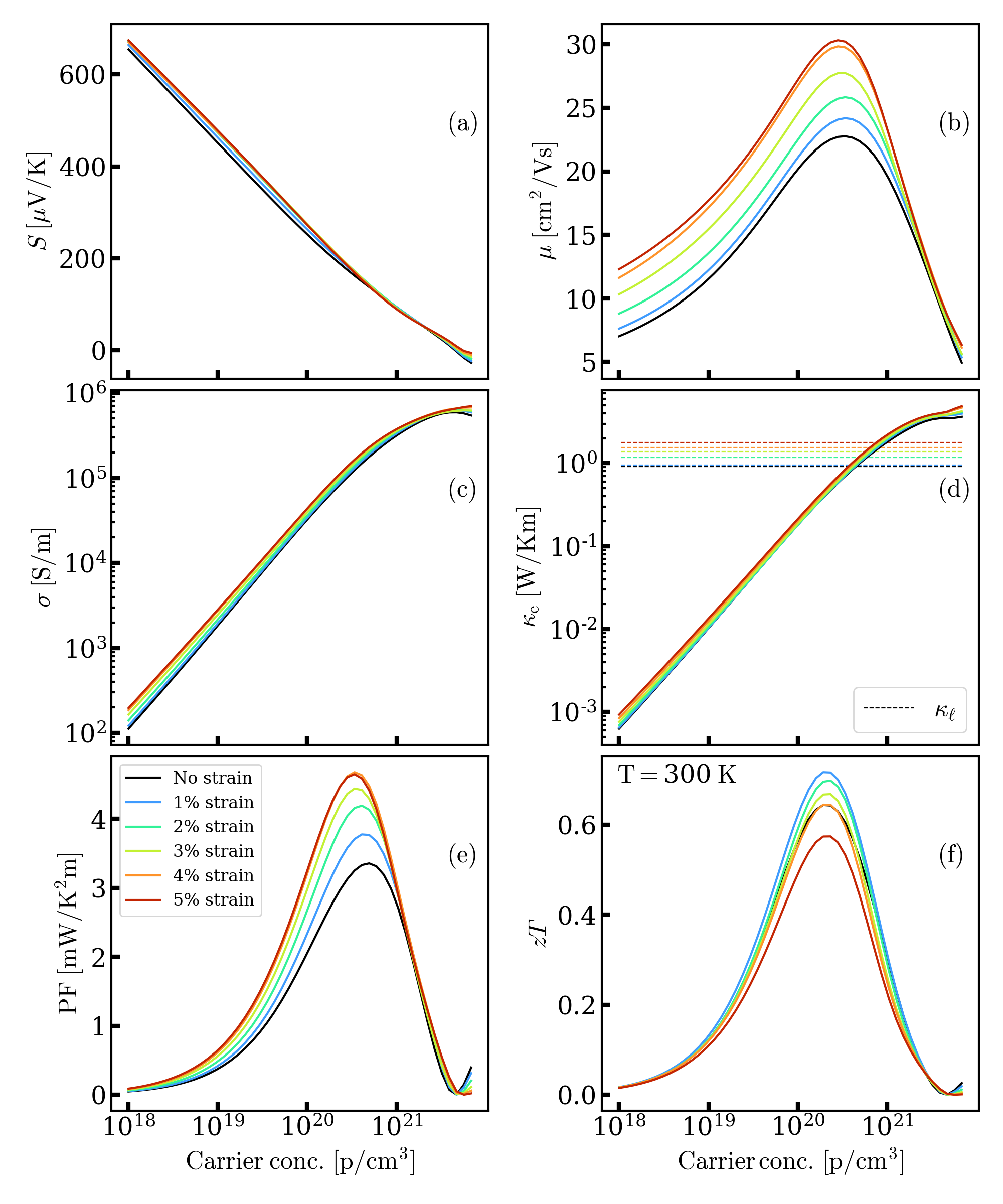}
    \caption{Seebeck coefficient ($S$), mobility ($\mu$), conductivity ($\sigma$), electron thermal conductivity ($\kappa_\ell$), power factor ($\mathrm{PF}$), and figure of merit ($zT$) of p-type CsK$_2$Sb at 300\;K.}
\end{figure*}

\begin{figure*}[h!]
    \centering
    \includegraphics[width=\linewidth]{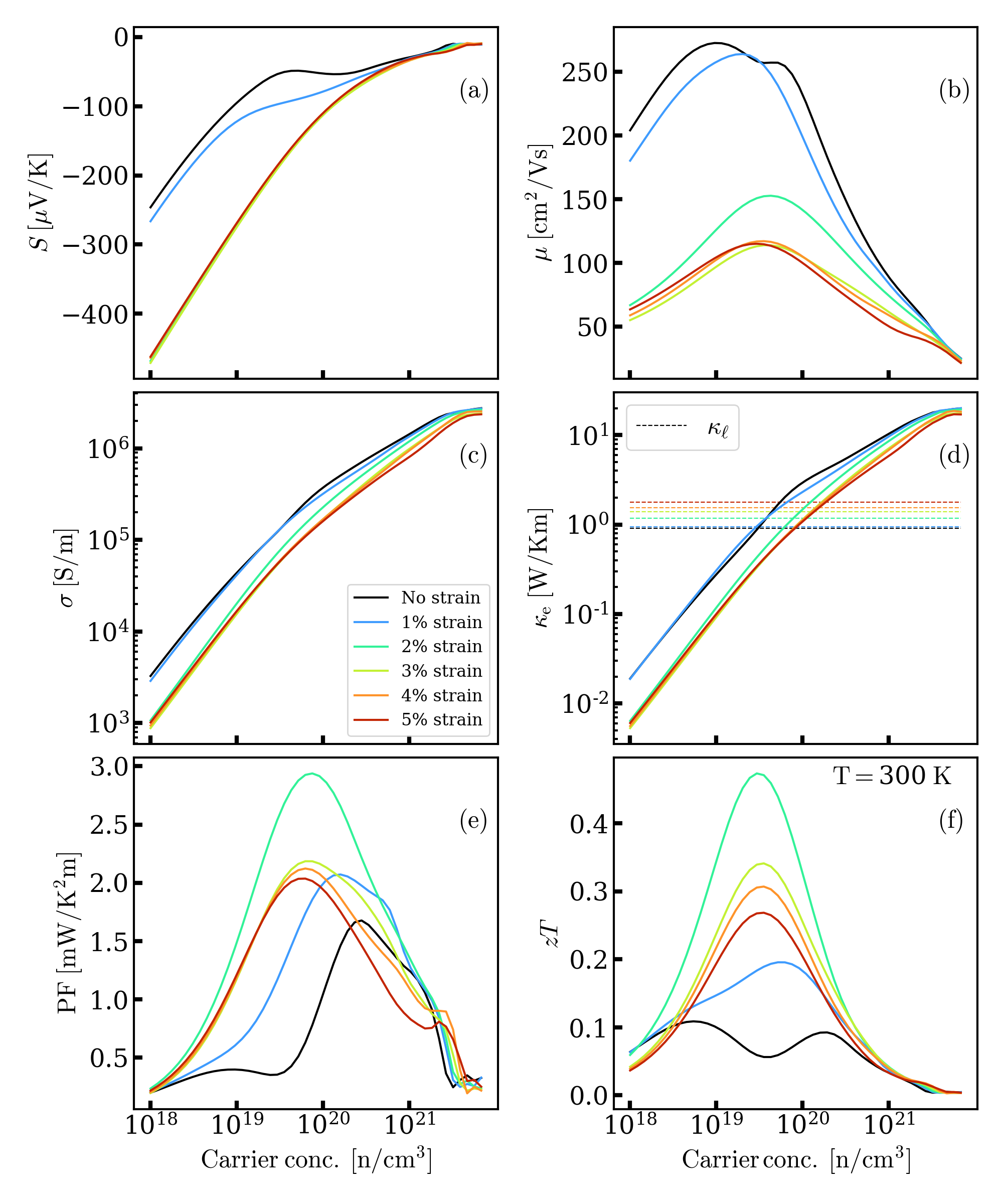}
    \caption{Electron transport properties of n-type CsK$_2$Sb at 300\;K.}
\end{figure*}

\newpage

\begin{figure*}[h!]
    \centering
    \includegraphics[width=\linewidth]{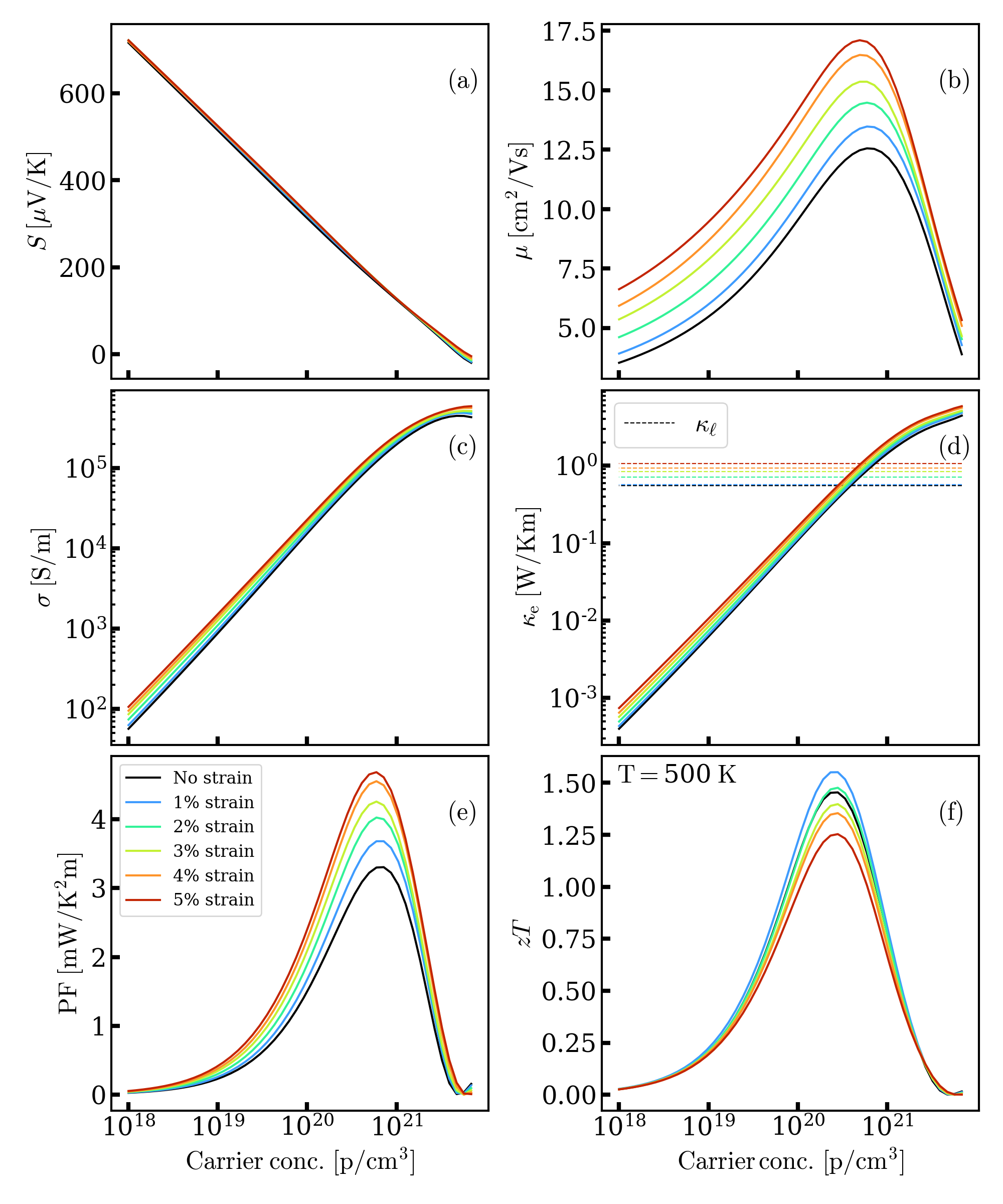}
    \caption{Electron transport properties of p-type CsK$_2$Sb at 500\;K.}
\end{figure*}

\newpage
\begin{figure*}[h!]
    \centering
    \includegraphics[width=\linewidth]{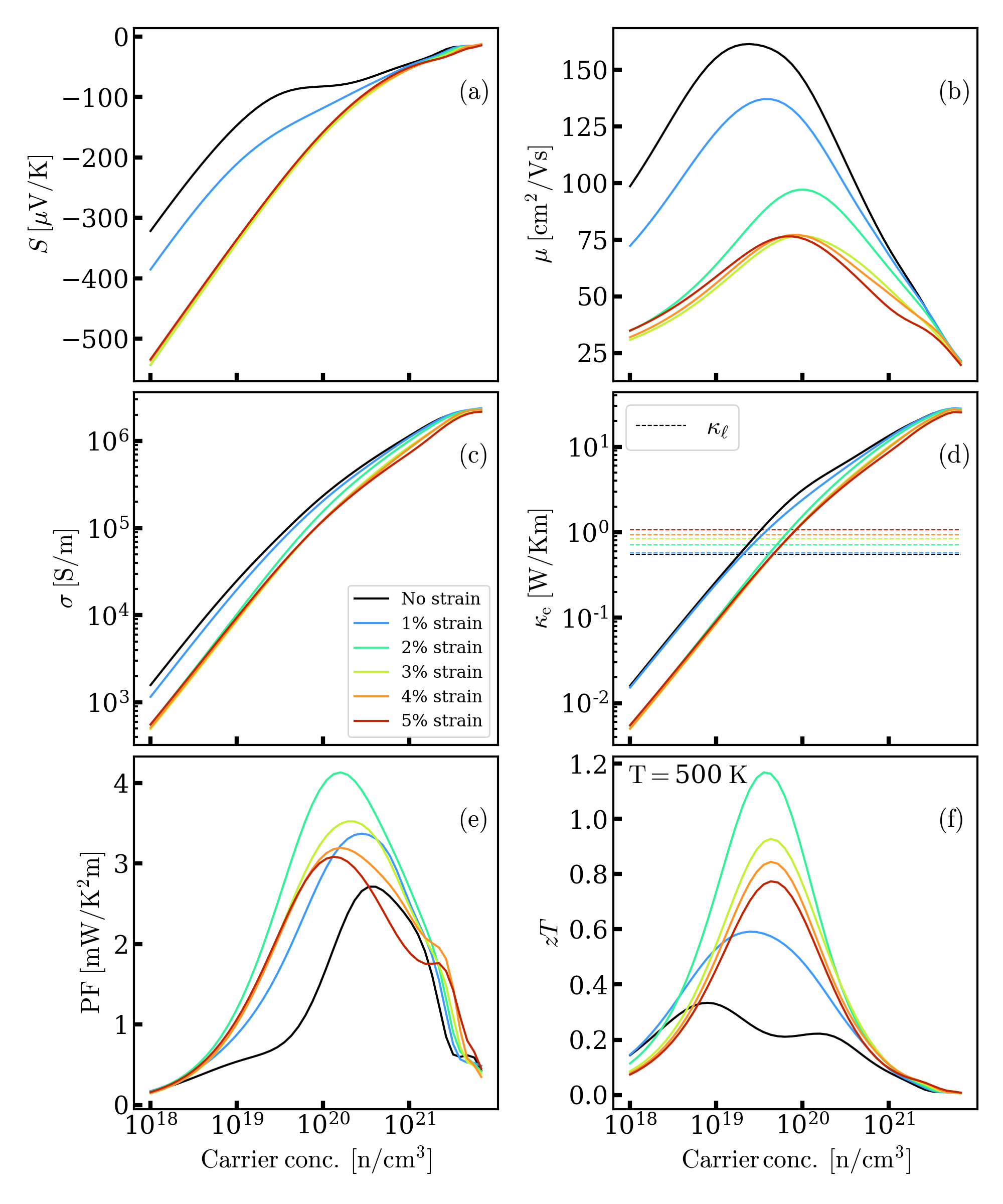}
    \caption{Electron transport properties of n-type CsK$_2$Sb at 500\;K.}
\end{figure*}

\newpage

\FloatBarrier

\section{k-point and interpolation factor convergence}
\begin{figure*}[h!]
    \centering
    \includegraphics[width=0.9\linewidth]{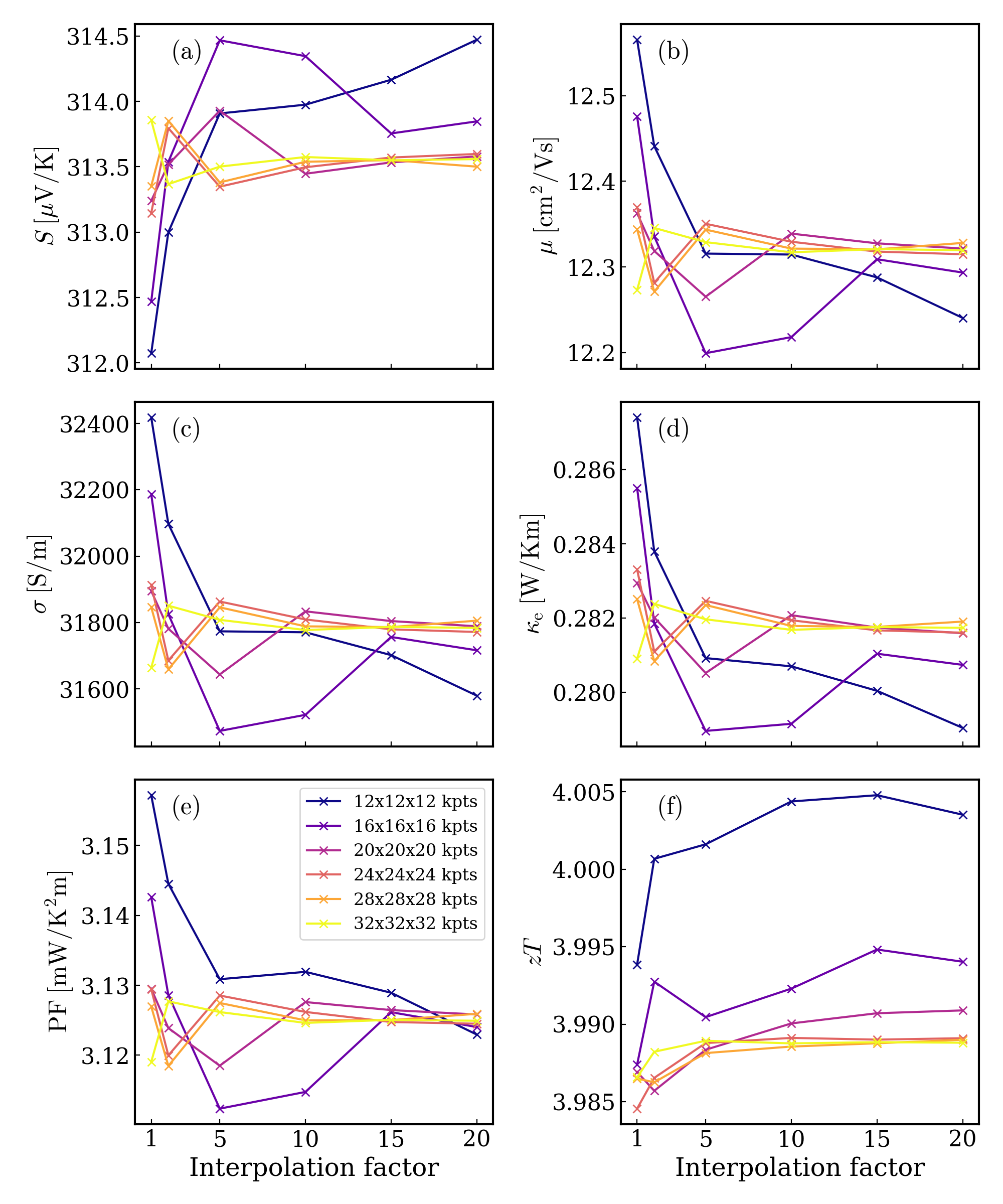}
    \caption{Constant relaxation time electronic transport properties of p-type CsK$_2$Sb with interpolation factor 1, 2, 5, 10, 15, and 20. Each line represents a density functional theory calculation with different k-point grid. The properties are at the carrier concentration resulting in maximum $zT$, $N = 1.61\times10^{20}$\;p/cm$^{3}$. A constant relaxation time of 10\;fs was used for this testing.}
\end{figure*}

\begin{figure*}[h!]
    \centering
    \includegraphics[width=\linewidth]{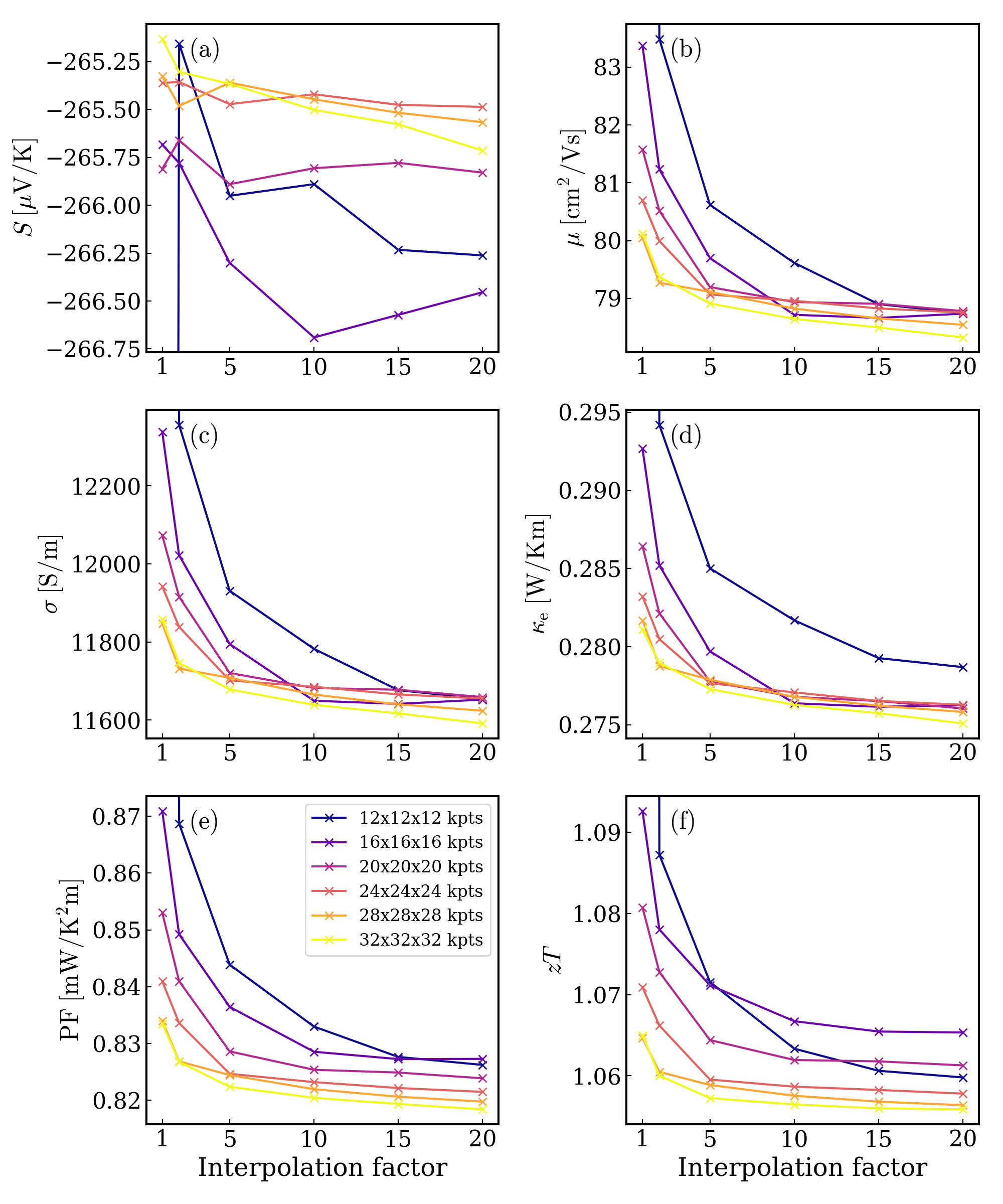}
    \caption{Constant relaxation time electronic transport properties of n-type CsK$_2$Sb with interpolation factor 1, 2, 5, 10, 15, and 20. Each line represents DFT calculations with different k-point grid. The properties are at the carrier concentration resulting in maximum $zT$, $N = 9.24\times 10^{18}$\;n/cm$^{3}$. A constant relaxation time of 10\;fs was used for this testing.}
\end{figure*}

\begin{figure*}[h!]
    \centering
    \includegraphics[width=\linewidth]{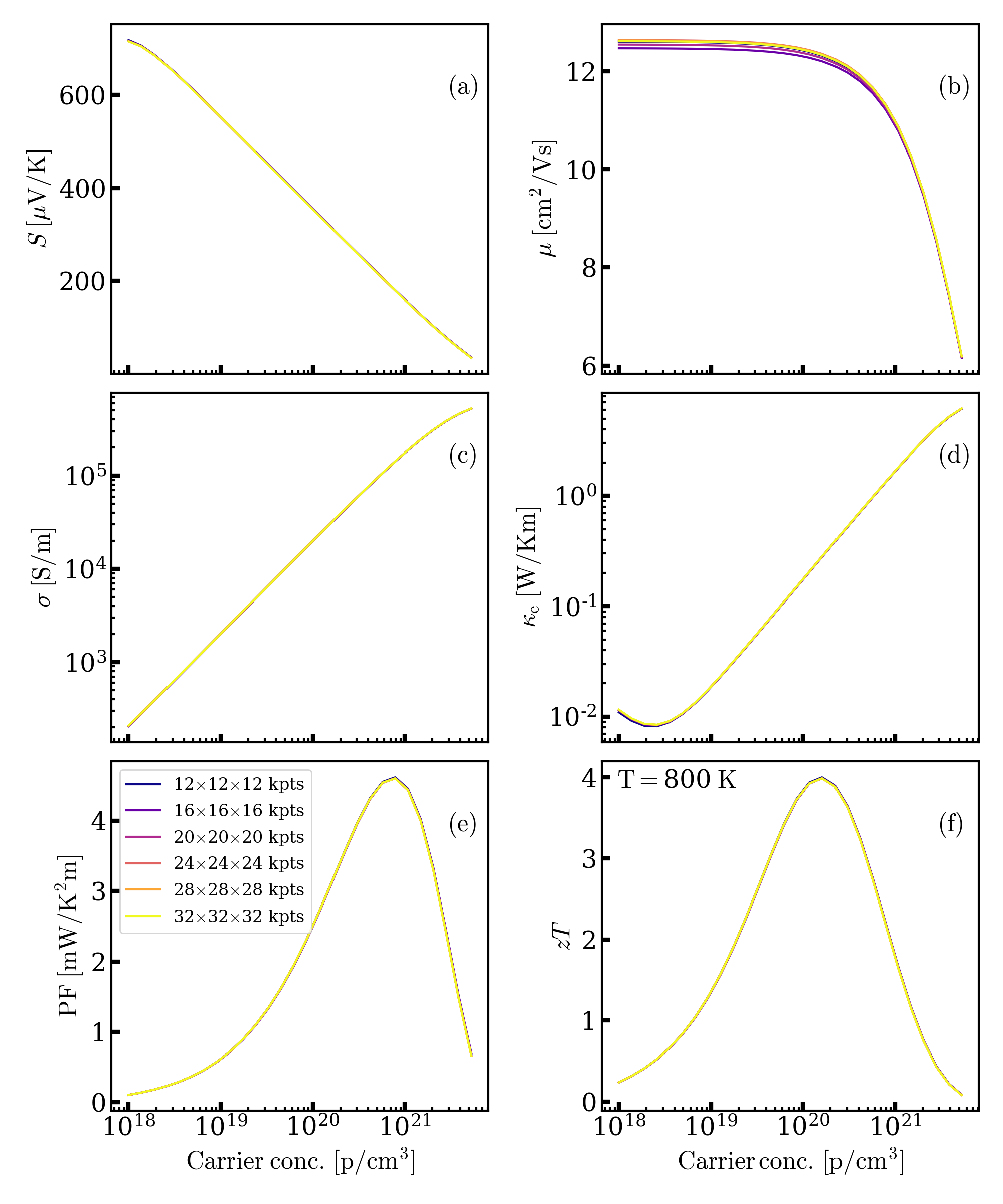}
    \caption{Constant relaxation time electronic transport properties of p-type CsK$_2$Sb with different k-points grid and an interpolation factor of 5. A relaxation time of 10\;fs was used for this testing.}
\end{figure*}

\begin{figure*}[h!]
    \centering
    \includegraphics[width=\linewidth]{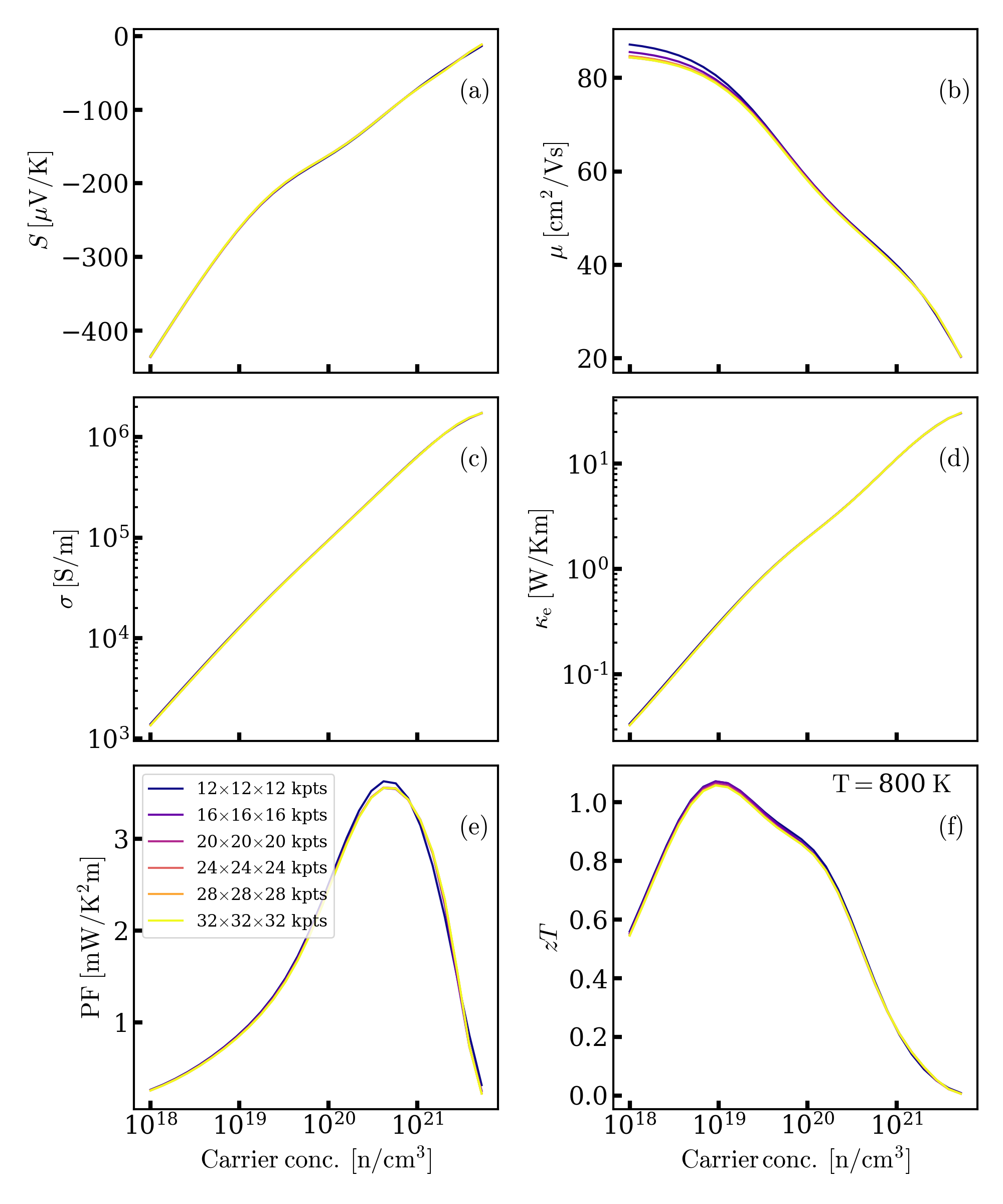}
    \caption{Constant relaxation time electronic transport properties of n-type CsK$_2$Sb with different k-points grid and an interpolation factor of 5. A relaxation time of 10\;fs was used for this testing.}
\end{figure*}